\documentclass{article} 
\usepackage{graphicx}
\usepackage{latexsym}
\usepackage{amsmath}
\usepackage{url}
\usepackage{comment}
\newcommand{\bookversion}{} 

\begin{document}

\ifdefined\bookversion
\title{Patterns and Interactions in Network Security\thanks{
\copyright~Pamela Zave and Jennifer Rexford, 2019, 2020.}}
\else
\title{Patterns and Interactions in Network Security}
\fi

\author{Pamela Zave and Jennifer Rexford}


\maketitle

\section{Introduction}
\label{sec:Introduction}

Today's Internet is not worthy of the trust society increasingly places in 
it. 
We hear every day about new security vulnerabilities and successful 
attacks, 
ranging from email viruses and Web sites overrun with unwanted traffic, 
to network outages, compromised user data, and downright espionage.  
These attacks are costly, leading to 
denial of service,
loss of revenue, 
identity theft, 
ransom demands,
subversion of the democratic process,
malfunctioning safety-critical equipment, and more.

Many successful security attacks use ``social engineering'' to prey
on naive users, for example by getting them to click on malicious 
hyperlinks.
Users are often guilty of using easily-guessed passwords,
or failing to reset a default password on a new device.
Application software is also the source of many security vulnerabilities,
due to bugs or poor programming practices.
Its complexity provides a big ``attack surface'' for adversaries
to probe for weaknesses.

Despite the prevalence of social engineering and vulnerable applications,
networks are an important part of the security landscape.
Networks make attacks on applications possible by
delivering unwanted traffic or leaking sensitive data. 
Network components and network services
are often the targets of attacks.  
Sometimes a network itself is the adversary, performing 
unethical surveillance or censoring communication.

Fortunately, networks can also be part of the solution, 
by blocking unwanted traffic, 
enabling anonymous communication,
circumventing censorship, or protecting both infrastructure and
applications from a range of known attacks.
And network protocols can protect users by 
authenticating and encrypting communications.

This article is intended as a concise tutorial on the very large subject
of security by and for networks,
specifically the mechanisms through which network security is achieved.
It is intended to be useful 
to all readers interested in networks, whether their
specialty is security or not.
Because the basic mechanisms have proven to be fairly stable over time,
we do not emphasize which particular attacks and defenses are
trending at the moment.
The details of well-motivated attacks or cost-effective
defenses change as
technology changes, and particular defenses might cycle in and out of
fashion.
Instead, to
achieve the goal of the paper, we derive our focus and organization
from two perspectives.

The first perspective is that, although mechanisms for network security
are extremely diverse, they are all instances of just a few patterns.
By emphasizing the patterns, we are able to cover more ground. 
We also aim to help the reader understand the big issues
and retain the most important facts.
\ifdefined\bookversion 

\else 
\fi
The second perspective comes from the observation that security
mechanisms interact in important ways, with each other and with
other aspects of networking.
\ifdefined\bookversion 
Although these interactions are not frequently discussed,
they deserve our attention.
\else 
These interactions deserve our attention.
\fi
To provide communication services that are secure and also fully
supportive of distributed applications, network designers must 
understand the consequences of their decisions on all aspects of
network architecture and services.

The boundaries of network security have been drawn by convention
over time, so 
\ifdefined\bookversion
\S\ref{sec:What-is-network-security} begins the tutorial by 
\else
we begin the tutorial by 
\fi
defining network security in two ways.
First, there is a practical classification of network security
attacks, based primarily on which agents are the attackers, defenders,
and potential victims.
The classification is based secondarily on defense mechanisms.
Second, we discuss 
how network security is related to information security and 
other forms of cyber-security, as well as the
gaps where no comprehensive defenses yet exist.

The four main sections of the tutorial cover the four major patterns 
for providing network security.
All agents can protect their own communications with 
{\it cryptographic protocols}
(\S\ref{sec:Cryptographic-protocols}),
which (among other benefits)
hide the data contents of packets.
\ifdefined\bookversion 
Networks can protect both themselves
\else 
Networks can protect themselves
\fi
and their users by {\it traffic filtering}
(\S\ref{sec:Traffic-filtering}).
Both users and networks can employ {\it dynamic resource allocation}
to overcome attacks (\S\ref{sec:Dynamic-resource-allocation}).
Although cryptographic protocols hide the data contents of
packets, they 
cannot hide packet headers, because the network
needs them to deliver the packets.
So when users need to hide packet headers from adversaries, 
which may
\ifdefined\bookversion 
include the network from which they are receiving service, 
\else 
include their own network,
\fi
they must resort to
{\it compound sessions and overlays}
(\S\ref{sec:Compound-sessions-and-overlays-for-security}).
The first three patterns will be familiar to anyone who has 
even dabbled in network security, while the importance of the
fourth pattern has not been sufficiently recognized. 

Between the definition of network security and the four major sections,
\S\ref{sec:A-model-of-networking} presents a new descriptive model of
networks and network services.
This model explains how network services are provided by
means of composition of many networks at many levels of abstraction,
where each network is self-contained in the sense of having---at least
potentially---all the
basic mechanisms of networking (such as routing, forwarding,
session protocols, and directories).
This model allows complete and precise descriptions of 
today's network architectures.
It is also necessary for recognition of the four patterns, because
the same patterns are reused in different networks in a compositional
architecture. 
The patterns are reusable precisely because the different
networks have fundamental similarities, despite the fact that they
may have different purposes, levels of abstraction, membership
scope, or geographical span.

In each of the four main sections, in addition to presenting a
security mechanism, we consider how the mechanism
interacts with other mechanisms within
its network and across composed networks.
This helps to determine where security could and should
be placed in a compositional network architecture.

\ifdefined\bookversion
\else
A version of this tutorial with more details, examples, commentary,
and references can be found on {\it arXiv} \cite{secur-arxiv}.
\fi

\section{What is network security?}
\label{sec:What-is-network-security}

Network security is a pragmatic subject with boundaries that have been
drawn by convention over time.
Although the focus of this tutorial is defense mechanisms, we must
have some idea of what kind of attacks they can defend against.

Classifying security attacks is extremely difficult because---by
their very nature---security attacks
are clever, they exploit gaps in standard
models, and they are always evolving.
In \S\ref{sec:A-practical-classification-of-network-security-attacks}
we present a practical classification scheme based on multiple factors.
It only covers known attacks, and there are some overlaps in the 
categories, but it does provide intuition that will be helpful for
understanding the defenses.

Of all the factors relevant to security attacks, the worst factor for
purposes of classification is real-world consequences (or, alternatively,
the motivations of attackers).
These consequences include financial loss, loss of time, loss of privacy,
loss of reputation, loss of political freedom, loss of physical
safety, and so on.
Often, these losses are intertwined, because one loss causes another.
Some attacks have no direct real-world consequences: their sole purpose
is to enable other, more damaging, attacks.

\begin{figure*}[hbt]
\begin{center}
\ifdefined\bookversion
\includegraphics[scale=0.60]{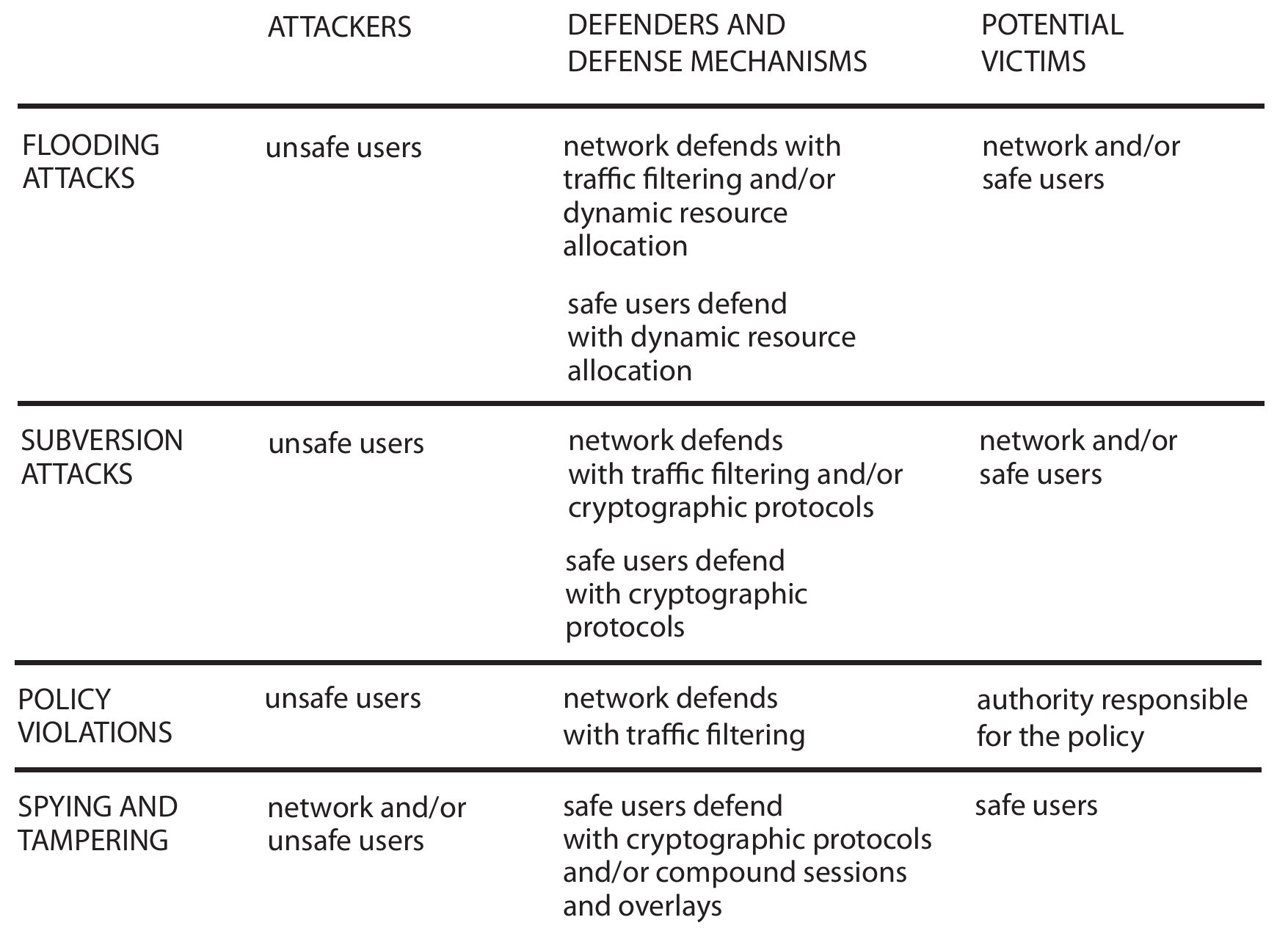} 
\else
\includegraphics[scale=0.50]{fig/attacktbl.pdf} 
\fi
\end{center}
\caption
{\small
{A practical classification of network security attacks.}
\normalsize}
\label{fig:attacktbl}
\end{figure*}

Our practical classification scheme, 
summarized in Figure~\ref{fig:attacktbl},
is based primarily on which agents are
the attackers, defenders, and potential victims.
\ifdefined\bookversion 
With one exception (see table), agents can be either:
\begin{itemize}
\item
{\it the network}, meaning the infrastructure machines provided by
the network operator to run the network,
\item 
{\it safe users,} meaning machines that use a network for 
communication and whose behavior is satisfactory according
to whatever rules or authorities
apply, or
\item
{\it unsafe users,} meaning machines that have access to a network
and whose behavior is unsatisfactory
because they have been programmed maliciously,
ignorantly, or erroneously.
\end{itemize}
\else 
With one exception (see table), agents can be either
{\it the network}, meaning the infrastructure machines provided by
the network operator to run the network,
{\it safe users,} meaning machines that use a network for 
communication and whose behavior is satisfactory according
to whatever rules or authorities
apply, or
{\it unsafe users,} meaning machines that have access to a network
and whose behavior is unsatisfactory
because they have been programmed maliciously,
ignorantly, or erroneously.
\fi
Classification is based secondarily on defense mechanisms;
these must be secondary to defenders because some mechanisms are only
available to some defenders.

Note that the network is usually a defender, but can be an attacker.
Even though traffic filtering is a possible defense for three attack
categories, as we will explain below, the details of filtering
against different attacks are quite different.

In \S\ref{sec:Relation-to-other-definitions-of-security}
we will discuss alternative definitions of security.
These include other kinds of cyber-security that complement
network security, attacks for which comprehensive defenses do not
yet exist, and alternative classification schemes.

\subsection{A practical classification of network security attacks}
\label{sec:A-practical-classification-of-network-security-attacks}

\subsubsection{Flooding attacks}
\label{sec:Flooding-attacks}

In a {\it flooding attack}, attackers
send floods of packets toward the victim, seeking to make it
unavailable by exhausting its resources.
Consequently,
flooding attacks are one type of {\it denial-of-service attack}
(see \S\ref{sec:The-information-security-triad}).
\ifdefined\bookversion 

\else 
\fi
The intended victims of flooding attacks vary.
If the victim is a public server or user machine, the attack might seek
to exhaust its compute-cycle, memory, or bandwidth resources.
Note that some public servers such as DNS servers are part of the
infrastructure of a network, 
so a flooding attack on a DNS server is an attempt to deny some network
services to a large number of users.
An attacker might also target some portion of a network, seeking to
exhaust the bandwidth of its links.
A bandwidth attack can make particular users unreachable, and can
also deny network service to many other users whose packets pass through
the congested portion of the network.
A bandwidth attack can also shift traffic to a less-secure part 
of a network, enabling other security attacks.

If an attacker simply sends as many packets as it can toward a victim,
the resources expended by the attacker may be similar to the resources
expended by the victim!
A {\it distributed denial-of-service attack} can be launched from
many coordinated machines, focusing the resources of many machines 
onto a smaller number of targets.
Alternatively, a flooding attack can employ some
form of {\it amplification,} in which the attacker's resources are
amplified to cause the victim to expend far more resources.
Here are some well-known forms of amplification:
\begin{itemize} 
\item
A ``botnet'' is formed by
penetrating large numbers (as in millions) of innocent-but-buggy
machines connected to the
Internet, and installing in them a particular kind of malware.
Subsequently the attacker sends a triggering
packet to each member of the botnet, causing it to launch a security
attack unbeknownst to the machine's owner.
This is another kind of distributed denial-of-service attack.
\item
An ``asymmetric attack'' sends requests to a server that
require it to expend significant compute or storage resources for
each request, so that a relatively small amount of traffic is
sufficient to launch a significant attack.
A typical IP example is a ``SYN flood,'' in which the victim receives
a flood of TCP SYN (session initiation) packets.
Each packet causes the server to do significant work and
allocate significant resources
such as buffer space.
Also in IP networks, attackers can flood DNS servers with queries
for random domain names (a ``random subdomain attack'').
These will force the servers to make many more queries, because they
will have no cached results to match them.
\ifdefined\bookversion 
In a Web-based application network, the attacker can send particular
HTTP requests that force a Web server to do a large amount of 
computation.
\else 
\fi
\item
An attacker can send many request packets to public servers, with the
intended victim's name as source name.
This ``reflection attack'' causes all the servers to send their responses
to the victim.
It amplifies work because responses (received by the victim) are
typically much longer than requests (sent by the attacker).
\ifdefined\bookversion 
\item
In an Ethernet network, a forwarder's response to receiving a packet
to an unknown destination is to ``flood'' the network with it,
which means (in this case) to send it out all links
so that eventually
all machines receive it and the designated destination responds to it.
An attacker can amplify any packet this way, simply by putting in
an unused destination name.
\else 
\fi
\end{itemize} 

Network infrastructure provides the principal defense against
flooding attacks, by filtering out attack packets
(\S\ref{sec:Traffic-filtering}).
Flooding attacks can also be countered by allocating additional
resources to handle peak loads
(\S\ref{sec:Dynamic-resource-allocation});
this is something that both network infrastructure and targeted users
can do.

If network infrastructure discovers where attack traffic is
coming from, defending against the attack becomes much easier.
For this reason, attackers employ various techniques to hide themselves,
for example:
\begin{itemize} 
\item
In an IP network, a sender can simply put a false source name in the
packet header, commonly called ``spoofing.''
In email applications, source email addresses are also easily
spoofed.
\item
With a botnet, even if bots use their true source names,
there may be too many of them to cut off.
The IP address of the master of the botnet remains hidden.
\item
An attacker can hide by putting a smaller-than-usual number
in IP packets' time-to-live fields, so that the packets are dropped
after they have done their damage in congesting the network, but before
they reach a place where measurements are collected or
defenses are deployed.
\end{itemize} 

Flooding attacks are a very serious problem in today's Internet.
There are businesses that generate them for small fees.
They target popular Web sites and (especially) DNS
\cite{dynAttack}.
The worst attacks are mounted by enterprises, albeit illegal ones,
that can draw on the same kind of professional knowledge, human resources,
and computer resources that legitimate businesses and governments have.
Such attackers will use many attacks and combinations of attacks
at once, and can continue them
over a long period of time.
According to industry reports, we are entering the era of flooding
attacks of terabits per second \cite{arbor}.

\subsubsection{Subversion attacks}
\label{sec:Subversion-attacks}

The purpose of a subversion attack on a network member
is to get the victim's machine to act as the
attacker wants it to, rather than as the owner of the machine wants.
\ifdefined\bookversion 
Here are some well-known examples of subversion attacks:
\else 
Here are some well-known examples:
\fi
\begin{itemize}
\item
The attacker sends malware
to infect or penetrate the machine.
The malware might be spyware or ransomware, capable of stealing or
damaging data stored in the machine.
The malware might turn the machine into a bot, so the botnet master
can exploit the machine's resources.
Or it might
attack the physical world through devices controlled by the machine.
\item
Port scanning is the process of trying TCP and UDP destination
ports on a range of IP addresses, to find pairs that
will accept a session initiation.
Port scanning does not in itself do much harm, but
it is gathering information to be used in launching other malware attacks.
This is because most malware targets a known vulnerability in a 
specific application program.
Scanning is less productive in IPv6, because the address space
is much larger, but specially focused scans may still succeed.
\item
The Border Gateway Protocol (BGP) 
is a control protocol through which IP networks exchange
routing information.
In ``BGP hijacking,''
an attacker uses BGP to insert false information,
telling routers to send packets
with certain destination names to the attacker rather than the
true destination.
The attacker may simply drop the redirected packets, denying service
to the victim.
The attacker can also respond to the packets as an impersonation of
the intended destination, for the purpose of
stealing commerce or secrets.
\item
Subversion attacks on directories also insert false information.
Higher-level names will then be mapped to the wrong lower-level
names, with the same consequences as route hijacking.
The directory protocol DNS (World-Wide Web name to IP address)
and the IPv4 directory protocol ARP (IP address to Ethernet address)
are subject to subversion attacks, as is the IPv6 replacement for
ARP, called Neighbor Discovery.
\item
Email spam and voice-over-IP robocalls can be considered subversion
attacks.
A networked device's owner wants the device for communicating with
acquaintances and chosen institutions.
These attacks force the device to present ads and other unsolicited
junk to the attention of its owner.
\end{itemize}

If a receiver of information knows the correct source of that
information, then 
both users and network components can protect themselves
from subversion by using cryptographic protocols.
With cryptographic authentication,
they know the identity of the agent with which they are communicating.

In other cases,
network infrastructure protects itself and
its users from subversion attacks by traffic filtering.
But filtering for
subversion attacks is significantly different from filtering for flooding
attacks because subversion requires two-way communication between
attacker and victim.
For example, if the victim is a server that communicates using TCP,
the attacker cannot send data to it until the initial TCP handshake
is completed.
This means that an attacker cannot hide by spoofing: 
if an attacker puts a false source name in its first packet to the
victim, it will never receive a reply to its SYN, and can never complete
the handshake.

\subsubsection{Policy violations}
\label{sec:Policy-violations}

Obviously, the default behavior of a network is to provide all
communication
services requested of it.
\ifdefined\bookversion 
These services should be provided according to explicit or implicit
agreements about quality and privacy.

\else 
\fi
On the other hand, the administrative authority of the network,
or other authorities such as governments, employers, and parents,
may have policies constraining network communication.
Specific communications that violate these policies are security attacks,
and the network defends against these attacks by tampering with
the communications (up to and including blocking them) or by spying
on them so that other enforcement actions become possible.
These defenses are exceptions to the default behavior of the network.
Examples of policy violations include:
\begin{itemize}
\item
Two users can communicate for the purpose of committing a crime.
This should be prevented, or in some cases recorded for
evidence in legal proceedings (``lawful intercept'').
Similarly, the communications of suspected individuals can be
monitored for surveillance and investigation.
\item
Saboteurs can attempt to access the control system of a power grid.
\item
A minor can attempt to access a Web site that
violates parental controls.
\item
A network may consider certain voice or video applications to 
take up more bandwidth than individual users are entitled to,
and rate-limit them to minimize their effects on overall performance.
\item
Operators of enterprise networks know which employees are using
which machines for which purposes.
Often they configure their networks to prevent unnecessary
communications, which may be attacks, and can be blocked without
harm even if they are only mistakes.
For example, machines used by engineers should not have access to
the enterprise's personnel database.
\end{itemize}

Network infrastructure
defends against policy violations by traffic filtering.
As indicated above, violating packets can simply be discarded,
but they can also be recorded, tampered with, or rate-limited.

Traffic filtering for policy enforcement is different from traffic
filtering against flooding and subversion attacks because the
filtering is so specific.
There is often a specified target whose
communications are being monitored.
Flooding and subversion attacks, in contrast, usually have unknown
sources, and their victims are often opportunistic.

\subsubsection{Spying and tampering}
\label{sec:Spying-and-tampering}

The victims of spying and tampering are network users, who want
their communications to be private, and want the network to be a
transparent and effective medium of communication.
The attackers in spying and tampering can be unsafe users, or they
can also be the infrastructure machines of the network itself.
Note that tampering is different from subversion because, in
subversion, one endpoint of the communication is the attacker.
In a tampering attack, the communication has two innocent endpoints,
and the attacker is causing what one endpoint receives to differ from
what the other endpoint sent.

When the attacker is the network, a spying or tampering attack is
the exact dual of a policy violation---both the users and the network
are doing exactly the same thing, and the only difference is 
which party we consider good or bad.
Judgments of which behaviors are good or bad emerge from
social debates involving legal, commercial, political, 
and ethical considerations.
These debates should not be constrained by technology.
Rather, the goal of technical experts should be to have the knowledge
to implement whatever decisions emerge from these debates \cite{tussle}.

Examples of spying and tampering include:
\begin{itemize}
\item
Some governments censor the Internet usage of their citizens.
Even if networks in their countries are privately owned, the governments
can insist that network providers enforce their policies.
\item
Some governments use surveillance of network usage 
\ifdefined\bookversion 
as a tool in
\else 
for
\fi
repression of or retaliation against political dissidents.
\item
By observing
the searches and Web accesses of a network user, 
an attacker can learn a great deal about the user's personal life.
\item
\ifdefined\bookversion 
Network infrastructure 
\else 
Networks
\fi
can insert into the paths of user sessions
middleboxes that insert ads or alter search results.
\end{itemize}
\ifdefined\bookversion 

The ``Great Cannon'' is an example of a tampering attack that does
no direct harm (to the endpoints of a targeted communication), but 
instead enables another attack.
The attacker is a network, as the Great Cannon appears to share many
resources with the Great Firewall of China.
The Cannon replaces ads being fetched from Chinese servers (by machines
outside China) with scripts that cause the recipients to access the
servers of designated victims.
At high volumes, this results in a distributed denial-of-service attack
on the victims \cite{greatCannon}.
\else 
\fi

Network users have two possible defenses against spying and
tampering.
The first is the use of cryptographic
protocols (\S\ref{sec:Cryptographic-protocols}), which conceal
the data in transmitted packets.
The second is the use of compound sessions and overlays
(\S\ref{sec:Compound-sessions-and-overlays-for-security}),
which seek to hide packets so that even their headers, sizes, and
timing cannot be observed.

\subsection{Relation to other definitions of security}
\label{sec:Relation-to-other-definitions-of-security}

\subsubsection{The information-security triad}
\label{sec:The-information-security-triad}

Governments, enterprises, and other 
institutions have broad concerns about information security.
These concerns are articulated by
the well-known ``information-security (CIA) triad,'' consisting of the
properties of {\it confidentiality} (secrecy, privacy, access control),
{\it integrity} (the information is valid or uncorrupted or
has correct provenance
information), and {\it availability} (information can be read or written
whenever needed).

These broad concerns about privacy include 
insider attacks and theft of physical storage media.
The broad concerns about integrity and availability include 
natural disasters and even military attacks
that might affect data centers.
If the opposite of availability is denial-of-service, we can see
that {\it denial-of-service attack} is also an extremely broad category.

Although the goals of the CIA triad have a great deal of overlap 
with the goals of network security, 
the classification scheme of 
\S\ref{sec:A-practical-classification-of-network-security-attacks}
is far more focused.
It is confined to threats incurred by operating a network or being
connected to one,
and it is closely tied to specific defense mechanisms within networks.

\subsubsection{Complementary forms of security}
\label{sec:Complementary-forms-of-security}

For network users, network security is a first line of defense against
subversion attacks;
a major goal is to keep subversion packets from being delivered
to user machines.
If the packets do arrive, then security measures in operating
systems and applications must take over.
Many applications and most operating systems now have well-developed
security measures of their own.
However, old operating systems,
real-time operating systems, and Internet of Things devices (which are
highly resource-constrained)
tend to have far fewer security mechanisms built in. 
For these endpoints,
network defenses against subversion remain important.

Another subfield of security research and practice concerns
``trust management,''
which is technology aimed at deciding which agents should have permission
to access which resources or perform which operations, 
based on the credentials and attributes of
the agent, and on the permission policies applicable to the object
(see, for example, \cite{trustSurvey,delegationLogic}).
Trust management is a decision-making
component of most forms of security, including
network security.
\ifdefined\bookversion 
Distributed trust-management
systems also rely on network security, for instance to communicate
secret information safely among nodes of the system.
\else 
\fi

Most security experts would probably agree that the human side of
security is the most important and the hardest to deal with.
In an ideal world, all institutions would have sophisticated
cyber-security policies, and enforce them.
These policies would prevent (among other problems)
insider attacks in which employees with access
to code deliberately put bugs or backdoors in it.
All people using computers would keep their software updated,
choose 
hard-to-guess passwords, and change default passwords immediately.
(Botnets are heavily populated with Internet of Things devices
such as baby monitors, because they come with factory-installed
passwords, and their naive owners do not change them.)
No one would be fooled by 
``phishing'' attacks, which imitate a legitimate email so
that the recipient clicks on a malicious hyperlink embedded in it.
And on and on.

\subsubsection{Threats with inadequate defenses}

Personal data privacy is a form of security that is much discussed
in today's world.
Individuals are concerned about the massive amounts of personal
data that is 
collected about them by Web sites, search engines, and other applications.
This data is extremely valuable for selling advertising, and can also
be used for worse purposes.
Individual users can protect their privacy to some extent
by using anti-spying defenses to achieve anonymity.
Anonymously, they can email and participate in social media.
At some point, however, full participation in electronic commerce
and institutional services almost forces people
to disclose their identities \cite{vertesi}.

Finally, 
there is the growing threat of side-channel attacks.
Network infrastructure monitors
traffic to filter out flooding attacks,
subversion attacks, and policy violations.
Attackers also observe and analyze network traffic, for the purpose
of spying and tampering.
What are the characteristics of network traffic to be observed and
analyzed,
in addition to principal header fields and packet contents (which
are explicitly intended and known to carry information)?

The timing and sizes of packets can be observed.
Pseudo-random header fields, intended merely to group or distinguish
packets, might be carrying secret codes.
Optional header fields might reveal the configuration of the machine
or software version that produced it.
If the observer has access to the machine that sent the packet,
it might be able to observe processor timing, 
power consumption, or usage of shared resources as the packet is prepared.
Such access is possible if the machine is a stolen mobile
device, or if multiple tenants share a 
physical machine in a cloud.

All of these characteristics are usually incidental, but they can be
controlled by the sender to signal information to a knowledgeable
observer that is invisible to other observers.
This is known as a ``covert channel.''
Incidental characteristics can also be analyzed by
an adversarial observer,
to gain information despite the intentions of the sender.
This is known as ``side-channel'' information \cite{sidechannels}.
Extracting side-channel
information from packet timing and sizes
is becoming more common, both for (good) filtering and
(bad) spying, because the expanding use of cryptography has hidden
much explicit information \cite{trafficanalysis}.
At present defenses against side-channel spying are patchy and
experimental.

\section{A model of networking}
\label{sec:A-model-of-networking}

To find the patterns underlying network security mechanisms,
and to understand how these patterns interact with each other and
with other aspects of network architecture, we must be able to describe
today's networks in a way that is somewhat abstract and yet very
precise.
The ``classic'' Internet architecture
\cite{philo} and the OSI reference model \cite{osi} have not kept
up with the Internet's evolution since the early 1990s.
For a better way to describe networks, we will use
the compositional model of networking
introduced in \cite{cacm}.
\ifdefined\bookversion 

\else 
\fi
In this section we give a brief overview of the
compositional model, covering the structures and aspects that will
be used in the rest of the tutorial.
Although the model uses familiar terms, be aware that when they have
definitions within the model, it is these precise and specific definitions 
that apply.

\subsection{Components of a network}
\label{sec:Components-of-a-network}

\begin{figure*}[hbt]
\begin{center}
\ifdefined\bookversion
\includegraphics[scale=0.60]{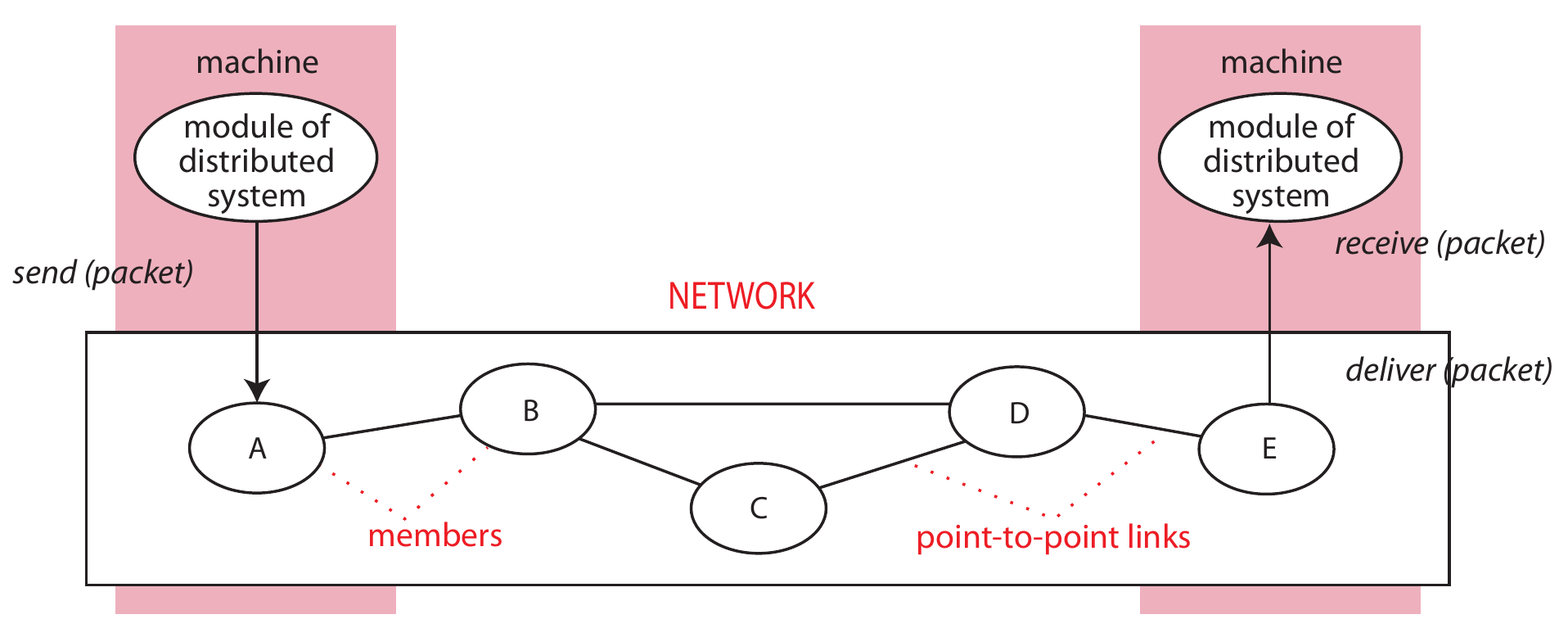} 
\else
\includegraphics[scale=0.45]{fig/oneNetwork.pdf} 
\fi
\end{center}
\caption
{\small
{Components of a single network, and its user interface.} 
\normalsize}
\label{fig:oneNetwork}
\end{figure*}

The components of a network are {\it members} and {\it links}.
A {\it member} of a network is a software and/or hardware module running
on a computing {\it machine,} and participating in the network.
As a participant, the member implements some subset of the 
network's protocols.
A network member usually has a unique {\it name} in the namespace of
the network.
For example, Figure~\ref{fig:oneNetwork} shows five members of
a network with unique names {\it A, B, etc.}

In the compositional
model a network always has a single {\it administrative authority}, 
or alternatively {\it network operator},
which is a person or organization responsible for the network.
The operator provides and administers resources for the network, in the
form of links, members, and additional resources on the members' machines.
The operator is expected to protect the network's resources and ensure
that users of the network enjoy the promised communication services.
It is convenient to partition the members of a network into 
{\it infrastructure members} administered by the operator
to provide services,
and {\it user members} belonging to the network for the purpose
\ifdefined\bookversion 
of employing its services.\footnote{In the case of peer-to-peer 
networks, each member contributes only its own resources, and there may
be no central operator or administrator.
For these networks, the authority is the cooperative
agreement among members.}
\else 
of employing its services.
\fi

\ifdefined\bookversion 
A network member can send or receive digital units called {\it packets} 
on one or more {\it links} of the network.
A {\it link} is a communication channel.
Most physical links are wires, optical fibers, or radio frequencies.
Wires and optical fibers are usually used as
{\it point-to-point links}, with two
endpoints and transmission in one or both directions.
Radio frequencies are {\it broadcast links}, on which any member with
suitable hardware and within
radio range can send or receive.
In wired networks, buses (used in older Ethernets and cable networks)
are also broadcast links, so packets can be sent
and received by any machine connected to them.

Broadcast links are mostly ignored in this tutorial.
The reason is that most links are virtual rather than physical (see 
\S\ref{sec:Composition-of-networks}), and the layering
mechanism that creates
virtual links is usually applied to make broadcast physical links
appear as point-to-point virtual links in higher layers of a network
architecture.
\else 
A network member can send or receive digital units called {\it packets} 
on one or more {\it links} of the network.
A {\it link} is a communication channel.
In this version of the tutorial we only consider point-to-point
links, as broadcast links are not fundamentally different for security. 
\fi

A {\it public} network allows any machine to host a network member and
connect to the network, while a {\it private} network allows only
authorized members.
\ifdefined\bookversion 
One of the two common authorization mechanisms is cryptographic
protocols (\S\ref{sec:Cryptographic-protocols}).
The other authorization mechanism is physical security,
in which intruders are denied physical access to the links of
the network.
\else 
The two common authorization mechanisms are cryptographic
protocols (\S\ref{sec:Cryptographic-protocols}) and physical security,
in which intruders are denied physical access to the links of
the network.
\fi

\subsection{Functions of a network}
\label{sec:Functions-of-a-network}

As shown in Figure~\ref{fig:oneNetwork}, a network enables
modules of a distributed system on different machines to communicate.
We say that a network provides one or more {\it communication services}.
A particular instance or usage of a communication service is called
a {\it session}.
Like a link, a session is also a communication channel for
digital packets.
The minimum semantics of a session is that it is a group of packets
that the users of the service regard as belonging together.

\begin{figure*}[hbt]
\begin{center}
\ifdefined\bookversion
\includegraphics[scale=0.60]{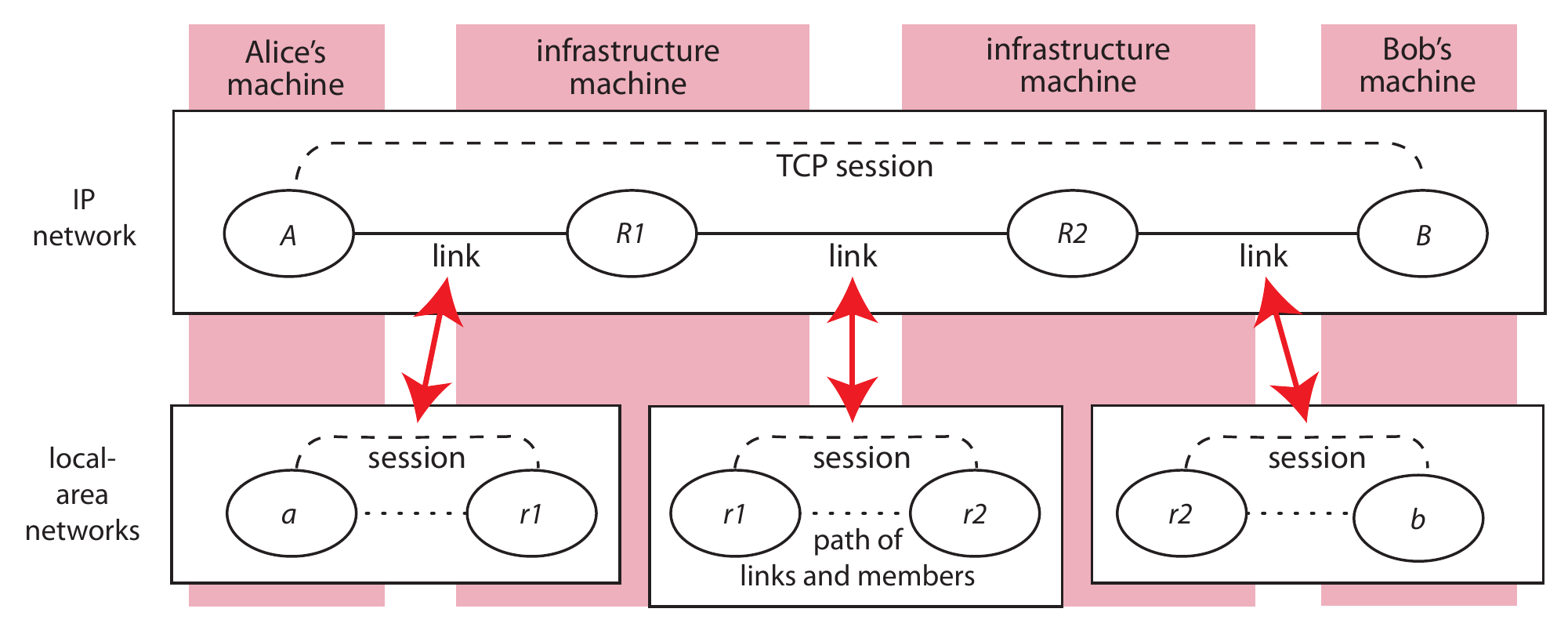} 
\else
\includegraphics[scale=0.50]{fig/layering.pdf} 
\fi
\end{center}
\caption
{\small
{The IP networks of the Internet are layered on many
\normalsize}
local-area networks.}
\label{fig:layering}
\end{figure*}

\ifdefined\bookversion 
Communication services can be specified to have a wide variety of
properties, which the network operator is obligated to enforce.
There are two major mechanisms in networks for satisfying service
specifications.
\else 
There are two major mechanisms for providing network services.
\fi
The first is {\it routing} and {\it forwarding}.
{\it Forwarding} is the mechanism that extends the reach of the
network beyond individual links to paths of links;
in forwarding, a member receives a packet on an incoming link,
and sends it out on an outgoing link to get it closer to its
destination.
A {\it forwarder} is an infrastructure member whose primary purpose
is forwarding.
Figure~\ref{fig:layering} shows a path through an IP network between user
members $A$ on Alice's machine and $B$ on Bob's machine.
In the figure, {\it R1} and {\it R2} are ``IP routers,'' i.e.,
forwarders.
All these names are ``IP addresses.''

{\it Routing} is the control mechanism that controls forwarding by
populating {\it forwarding tables} in forwarders.
Forwarders consult their tables to know where to forward packets.
Routing and forwarding can be extended beyond minimum requirements
of reachability to perform services such as broadcasting and
steering packets through {\it middleboxes}---members
that perform various packet-processing
functions related to security, efficiency, or interoperability.

\ifdefined\bookversion 
Routing and forwarding work together by means of a network's
{\it forwarding protocol}, which is a set of rules
governing the format of packets transmitted through the network.
Each packet has a {\it header part} and a payload or {\it data part.}
A header usually includes a {\it source name} indicating which member
originally sent the packet, and a {\it destination name} indicating
which member is intended to ultimately receive the packet.
Entries in a forwarding table match fields in a packet header.
\else 
\fi

The other major mechanism for satisfying service specifications
is {\it session protocols.}
A {\it session protocol} is a set of rules governing packet formats,
higher-level semantic units,
and participant behavior during a session.
In Figure~\ref{fig:layering}, 
the session shown in the IP network uses the TCP
session protocol. 
Following the rules of
TCP, the session endpoints maintain state and send extra packets to
provide reliable, ordered data delivery
despite the facts that IP links are not perfectly reliable, and different
packets of a session may be routed on different paths.
UDP (another IP session protocol) is much simpler and implements
fewer services, but it does define port numbers that can
be used to group related packets.

\subsection{Composition of networks}
\label{sec:Composition-of-networks}

We have defined networks as self-contained modules with members, links,
routing, forwarding, and session protocols.
In today's Internet, there are many networks,
each of which may be specialized according to its purpose, membership
scope, geographical span, and level of abstraction.
A network architecture is a flexible
composition of these networks, and thus called a
``compositional network architecture'' \cite{cacm}.

There are two composition operators on networks, the first being
{\it layering.}
The model
defines layering precisely: one network is {\it layered on} another
network if a link in the overlay network
is implemented by a session in the underlay network.
For example, each IP link in Figure~\ref{fig:layering}
\ifdefined\bookversion 
is implemented by a session in a local-area network, as indicated by
the bold arrows.
\else 
is implemented by a session in a local-area network
(see bold arrows).
\fi
Members of different networks on the same machine communicate through
the operating system and/or hardware of the machine.
IP packets sent on an IP link are actually encapsulated in Ethernet
headers and transported through local-area networks
as the data parts of Ethernet packets.
Since the implementation of an overlay link always 
consists of digital logic, whether in
hardware or software, an overlay link is always virtual,
regardless of whether the links in the underlay are physical or virtual.
\ifdefined\bookversion 
Note that the
\else 
The
\fi
IP network in Figure~\ref{fig:layering} plays the same role as the
distributed system in Figure~\ref{fig:oneNetwork}.

As Figure~\ref{fig:layering} shows, almost all networked machines host
members of at least two networks, and some host many more.
We use the term {\it member} rather than {\it node} because the
latter is too similar in connotation to {\it machine}.
The figure shows how layering extends the reach of the local-area
networks, each of which is isolated.
A local-area network only implements an IP link, but the IP network
can reach machines over paths that are concatenations of links.

The second composition operator on networks is {\it bridging}.
{\it Bridging} simply means that two particular networks share some 
links, so they can forward packets to each other.
If the designs of bridged networks are sufficiently homogeneous,
in particular if they share session protocols, then sessions can
cross network boundaries.
In the Internet, many IP networks are bridged together in this way.
These networks differ in their operators/administrative authorities,
but not their basic design.

The definition of layering in compositional network architecture
is very different from the older notion of
layering in networks found in the ``classic'' Internet architecture
\cite{philo} and OSI reference model \cite{osi}.
In the new model, each layer is a complete network, so IP 
routing/forwarding and IP session protocols belong to the same
network/layer.
In the new model, an architecture has as many layers as needed,
which often includes multiple IP networks layered on top of one
another.
We use the compositional model
in this tutorial because it allows comprehensive yet
precise descriptions of how the Internet actually works today \cite{cacm}.
It is also necessary for recognition of the four patterns, because
the same patterns are reused in different networks in a compositional
architecture. 
\ifdefined\bookversion 
And layering of networks over networks and bridged sets of networks is
especially important because it makes it possible to reason rigorously
about networks from the bottom up:
properties proved of
an implementing session are automatically true of the implemented link.
\else 
\fi

\ifdefined\bookversion 
\subsection{Other forms of composition}
\label{sec:Other-forms-of-composition}

Network composition is the most important form of composition in
network architecture.
Nevertheless, two other forms of composition, both found inside
individual networks, are relevant to security.

\subsubsection{Protocol composition}
\label{sec:Protocol-composition}

{\it Control packets} are used by a protocol to synchronize the
endpoints and share specific parameters.
{\it Data packets} contain the substance being communicated.
Although a session protocol may have only one of these packet types,
many protocols have both, or mix control information and data in a
single packet.

Within a network, session protocols can be composed, so that the same
session benefits from the services implemented by multiple protocols.
When two session protocols $P$ and $Q$ are composed, one of them is
{\it embedded in} the other.
If $P$ is embedded in $Q$, for instance, most packets in the session
will have the format shown in Figure~\ref{fig:composed-packet}, in which
the $P$ header and data are encapsulated in $Q$ data
(the figure also shows optional footers, which are required by
some protocols).
In addition, the session may include control packets of $Q$ that are
independent of $P$ and have no encapsulated $P$ packets.

\begin{figure*}[hbt]
\begin{center}
\ifdefined\bookversion
\includegraphics[scale=0.60]{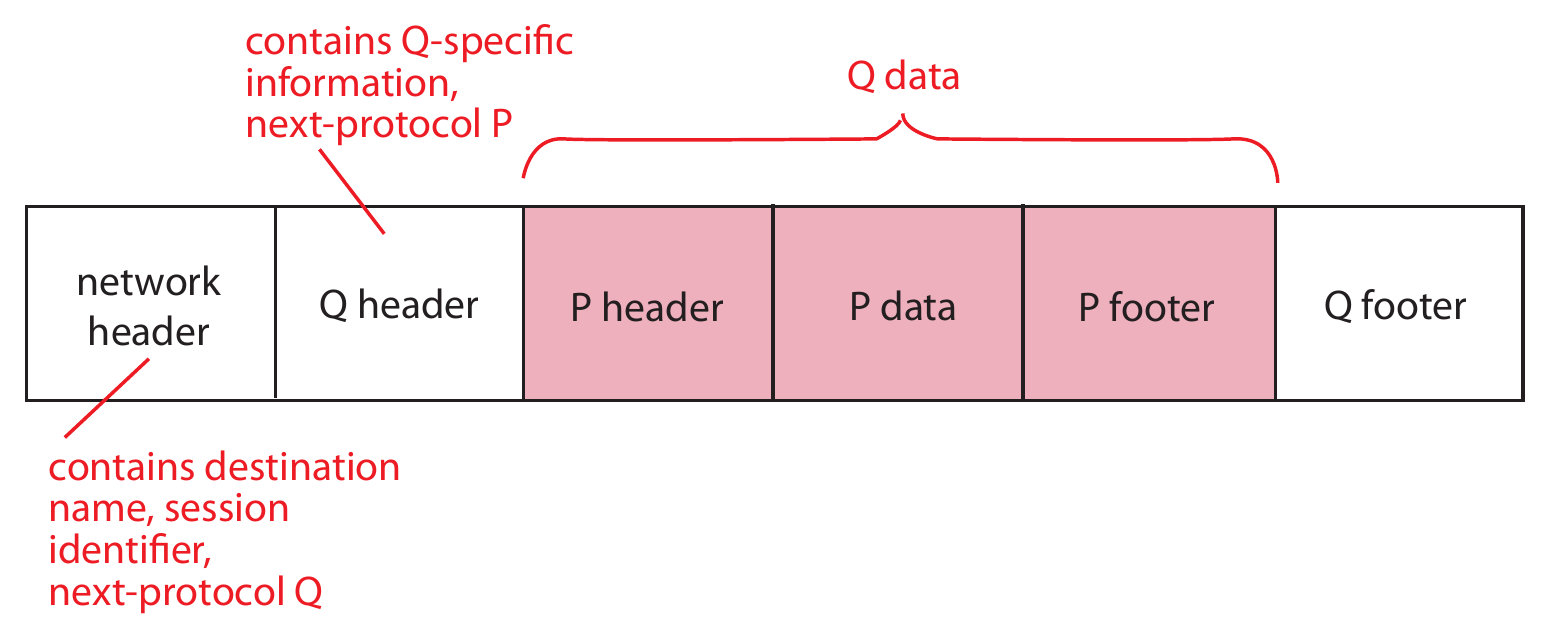} 
\else
\includegraphics[scale=0.50]{fig/composed-packet.pdf} 
\fi
\end{center}
\caption
{\small
{Packets of a network with session protocol $P$ embedded in
session protocol $Q$.  The shaded part is all specific to $P$.
Protocol footers are optional.}
\normalsize}
\label{fig:composed-packet}
\end{figure*}

The figure shows an ideal packet format in which
the network header of a packet
includes the destination name and session identifier
for the entire session, so that all packets of the session will
be easily identifiable to the forwarders.
Each network or protocol header names the type of the next header,
if any, so that session protocols can be composed freely.
Unfortunately, not all packet formats are so cleanly designed.

\subsubsection{Compound sessions}
\label{sec:Compound-sessions}

A user member initiating a session to some far endpoint can insert
another user member into the session path as a middlebox.
To do this, the initiating user must give the name of the middlebox
as the destination name of its outgoing packets, as shown in
Figure~\ref{fig:joinbox}.
The middlebox must learn the initiator's intended far endpoint, 
for example by getting it from some other field of the session-initiation
packet.
Then the middlebox changes the headers of the packets it receives
(source becomes its own name, destination becomes the
initiator's intended) and sends them out.
A middlebox that behaves in this way is called a {\it proxy}.
Each proxy accepts a session, initiates another
session with a different header, remembers the association
between the two sessions, and relays packets between them.
A {\it compound session} is a chain
of {\it simple sessions} composed
by proxies in this way.

\begin{figure*}[hbt]
\begin{center}
\ifdefined\bookversion
\includegraphics[scale=0.60]{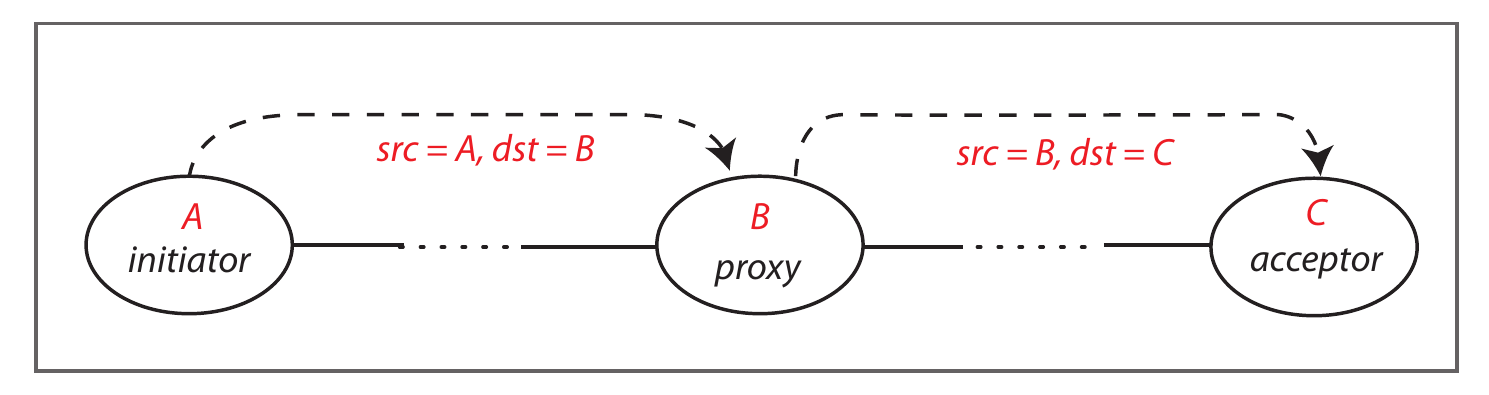} 
\else
\includegraphics[scale=0.50]{fig/joinbox.pdf} 
\fi
\end{center}
\caption
{\small
{A compound session with two simple sessions.}
\normalsize}
\label{fig:joinbox}
\end{figure*}

A compound session can have more than one proxy (as an example
of how to do this,
the session-initiation packet can contain a list of proxies to
visit, ending with the final destination).
Because of the names in forward packet headers, 
return packets naturally pass
through the same proxies in reverse order,
and have their headers re-translated in reverse order.

The principal
security significance of compound sessions is that each simple
session has a different header, so compound sessions can be employed
by users to obscure header information (making them
complementary to cryptography).
In Figure~\ref{fig:joinbox}, an observer between $A$ and $B$ cannot
observe the true acceptor of the compound session, at least from
packet headers alone,
and an observer between $B$ and $C$ cannot observe the true initiator
of the compound session.
\else 
\fi

\section{Cryptographic protocols}
\label{sec:Cryptographic-protocols}

Cryptographic protocols are incorporated into the session
protocols of a network.
Cryptographic protocols are executed by the endpoints of a point-to-point
session, so that the session will have {\it (data) integrity} and
{\it (data) confidentiality}.
These are the same terms used in the information-security triad,
but in this context they have a much more specific meaning.
Confidentiality means that no party except a designated receiver
can read the packets sent.
Integrity means that no third party can insert, modify, or replay
packets of the session, so that the packets received by a designated
receiver
are the exact packets sent by the designated sender,
and if the sender sends a distinguished
packet $m$ times, the receiver receives it at most $m$ times.

Cryptographic protocols can also achieve
{\it endpoint authentication}, which means that 
either session endpoint can be sure of the other endpoint's
identity.
Confidentiality should be reinforced by the property of
{\it forward secrecy}, which means that even if an encrypted session
is recorded by an attacker, and the attacker learns the secrets
of one of its endpoints at some later time, the attacker still cannot
decrypt and read the recorded packets.

User members of a network use cryptographic protocols to protect themselves
against spying and tampering attacks.
Infrastructure members, also, defend 
network operations against spying and tampering with cryptographic
protocols.

It is important that cryptographic protocols are designed for
the most hostile environments.
For example, in accepted proof systems (such as 
\cite{canetti,NRL}),
the baseline model of a security protocol allows an adversary to
control all communication channels between the endpoints (and other
agents they might query),
examining, storing, deleting, injecting, or altering any packets that the
adversary wishes.
Because cryptographic protocols are designed (and proved mathematically)
with such conservative assumptions, users trust them even
when they can trust nothing about the layers of
networking between endpoints.

\S\ref{sec:Trust-and-identity} begins our discussion of
cryptographic protocols by introducing the central concept of
{\it identity}.
The foundation for all cryptographic protocols is public-key
cryptography (\S\ref{sec:Public-key-cryptography-and-its-uses}),
because it provides some crucial functions and supports others.
In \S\ref{sec:Three-IP-cryptographic-protocols} we return to
the properties of data integrity and confidentiality.
Finally, in
\S\ref{sec:Interactions-between-cryptographic-protocols} 
we discuss architectural interactions with cryptographic protocols
that are relatively independent of other security patterns.

In \S\ref{sec:Trust-and-identity} through
\S\ref{sec:Three-IP-cryptographic-protocols} the context 
will be a single network of any kind.
The discussion also covers a set of 
similar bridged networks
all at the same level of the layering hierarchy, for example
the bridged IP networks of the Internet.
\S\ref{sec:Interactions-between-cryptographic-protocols} broadens
the context, as it includes how cryptographic protocols interact
with composition of networks by layering.

\subsection{Trust and identity}
\label{sec:Trust-and-identity}

Security requirements are based on which network members do and do not
``trust'' each other.
Of course a network member is a software or hardware module; it
cannot trust in any ordinary sense of the word, 
and has no legal responsibility
that it can be trusted to fulfill.
For the purpose of establishing trust, a network member that is an
endpoint of a session has an {\it identity}.
\ifdefined\bookversion 
This identity is given to the other endpoint of the session in answer
\else 
This identity is the answer
\fi
to the question, ``With whom am I communicating?''

This role implies that an identity should have meaning in the world
outside the network.
Often it is closely associated with a legal person---a person
or organization---who is legally responsible for the network member.
The identity is usually 
the source of the data that the network member sends
during the session.

Identities are related to layering, because layering allows a machine
to have different names---one in the namespace of each network it
participates in---at the same time.
For example, 
\ifdefined\bookversion 
in Figure 6, 
\else 
in Figure 4, 
\fi
each machine is participating in a higher-level Web-based application
network and a lower-level IP network bridged with other IP networks.
The dynamic sessions and links in the figure are formed as follows.
The client's browser at the upper level instructs its IP member $C$
to contact {\it bigbank.com}.
When there is layering of networks,
a {\it directory} is often used to find where an overlay member is
{\it attached to} an underlay network.
$C$ looks up {\it bigbank.com} in the DNS directory, 
and finds it is located
on the same machine as IP member $S$.
At the lower level, $C$ initiates a TCP session to $S$.
When the TCP session (and dynamic link) are ready, the browser 
initiates a request/response HTTP session over it.

\begin{figure*}[hbt]
\begin{center}
\ifdefined\bookversion 
\includegraphics[scale=0.60]{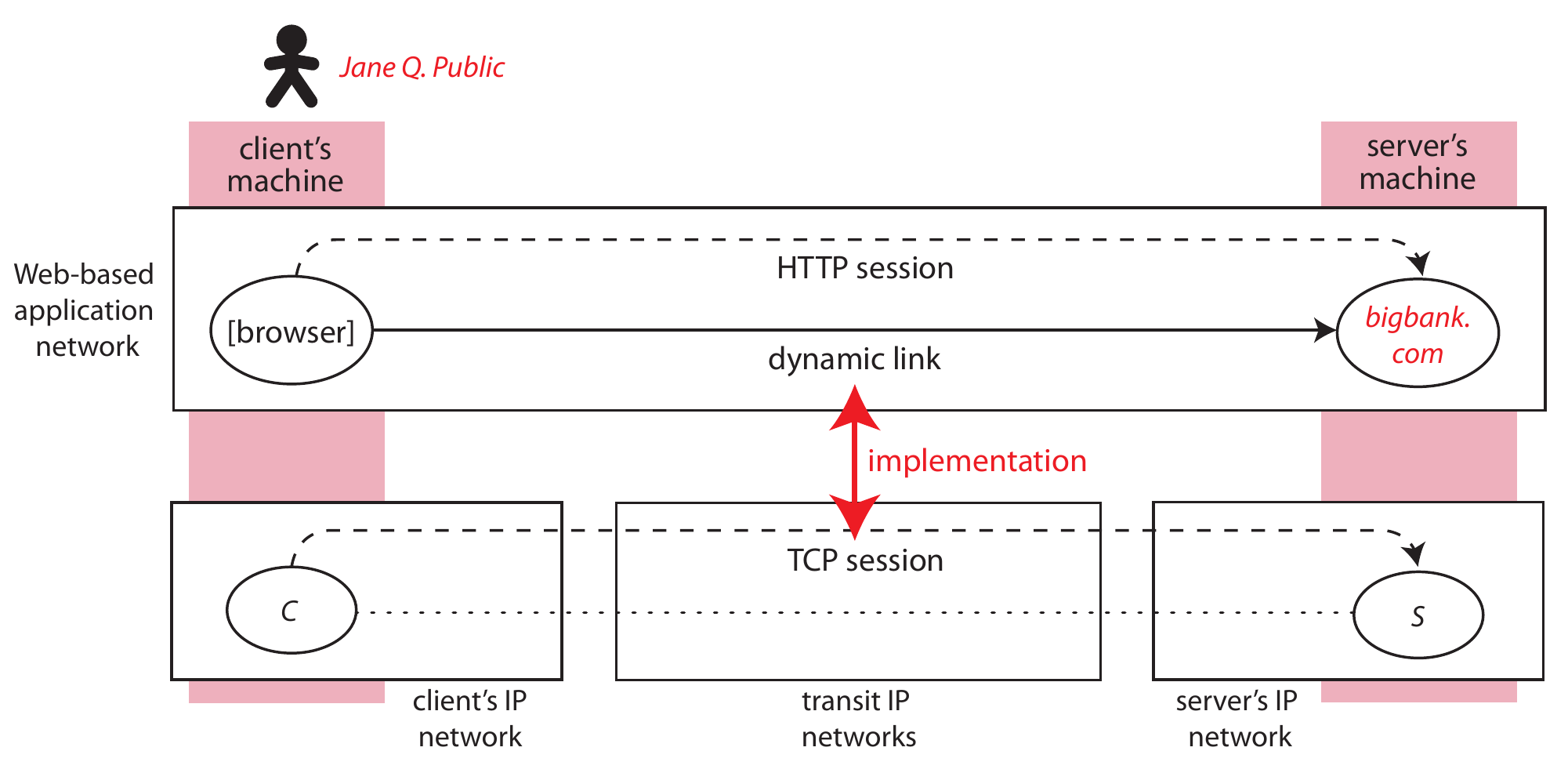} 
\else 
\includegraphics[scale=0.50]{fig/identity.pdf} 
\fi
\end{center}
{\caption
{Member names and identities in a Web application.}
\normalsize}
\label{fig:identity}
\end{figure*}

If the two endpoints of the TCP session need to authenticate each
other (as they should, for a banking transaction), what identities do
they give as their own?
The general answer is that each gives its member name or the 
name of a higher-level network member that is using it.
Either IP interface could give its IP name, but it would not be a
very good identifier---too transient, or with too little meaning in
the outside world.
Instead, the server's IP interface $S$ will be known by
its public Web name {\it bigbank.com}.
The client's machine does
not have a name in the application network, because
the browser only initiates sessions and never accepts them.
However, the user of the browser is a person named
{\it Jane Q. Public}, 
whose clicks and keystrokes provide input to the browser.
The browser will send {\it Jane Q. Public} as its identity,
and we can imagine this identity as a member of an 
even-higher-level distributed financial system.

For endpoint authentication, a member must have access to a secret
associated with the identity it provides.
One kind of secret, useful when the two endpoints have an
ongoing relationship, is a password.
The server {\it bigbank.com} knows Jane's password, and she can
type it into the browser when requested.

\begin{figure*}[hbt]
\begin{center}
\ifdefined\bookversion
\includegraphics[scale=0.60]{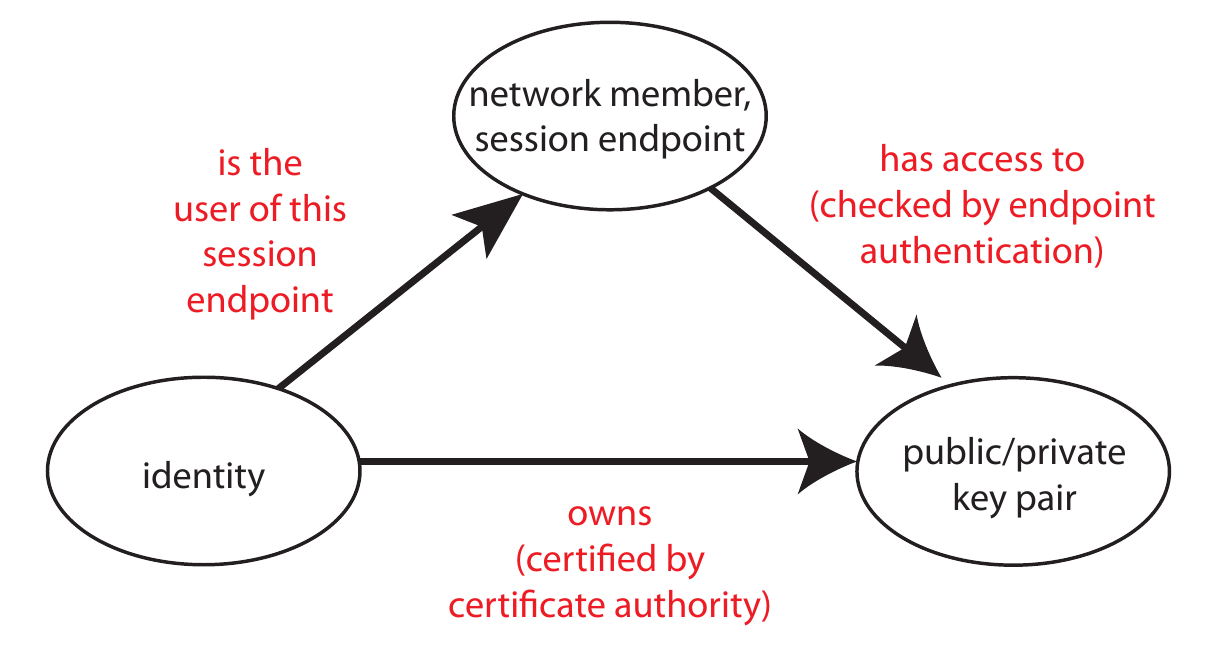} 
\else
\includegraphics[scale=0.50]{fig/identitytriangle.pdf} 
\fi
\end{center}
\caption
{\small
{Relationships among identification entities.}
\normalsize}
\label{fig:identitytriangle}
\end{figure*}

For the important cryptographic protocols, however,
the secret is always a public/private
cryptographic key pair (see next section).
The relationships among the important entities are shown in 
Figure~\ref{fig:identitytriangle}.
The identity is responsible for the packets sent by the network 
member,
and the network member has access to
the public key and its paired private key.

A ``certificate authority'' is trusted to ascertain that a particular
public key belongs to a particular identity; it issues a certificate
to that effect and signs it digitally.
Thus when an endpoint receives a certificate, it can trust the identity
that goes with the key (at least, 
as well as it trusts the certificate authority).
As indicated above, identities found in certificates include names of
legal persons, domain names, and IP addresses.

\ifdefined\bookversion 
It should be noted that 
trust between communicating endpoints is not necessarily simple or
absolute. 
For instance,
two endpoints may be communicating to negotiate a
contract, and (because they do not trust each other completely)
need to communicate through a third party trusted by both.
A trusted broker can ensure, for example, that both parties sign
the exact same contract
\cite{end-to-end2}.
\else 
\fi

\subsection{Public-key cryptography and its uses}
\label{sec:Public-key-cryptography-and-its-uses}

In public-key cryptography,
an identity generates and owns a coordinated
pair of keys, one public and one kept private and secret.
The important properties of these keys are that (i) it is extremely
difficult to compute the private key from the public key, 
and (ii) plaintext
encrypted with the public key can be decrypted with the private key,
and vice-versa.
Today's public-key cryptography is descended from the Diffie-Hellman
Key Exchange protocol and the RSA algorithm (named for its inventors
Ron Rivest, Adi Shamir, and Leonard Adleman).
\ifdefined\bookversion 

At present a key must be at least 2048 bits to be considered secure
(the minimum size is expected to increase in the future).
A public key is a pair of unsigned integers $(n,e)$.
The corresponding private key is a pair $(n,d)$.
To be encrypted, a packet must be divided into chunks such that each
chunk has an integer representation less than $n$.
If $m$ is such a chunk, then the public-key encryption of $m$ is
$m^e\;\mbox{mod}\;n$, and the private-key encryption of $m$ is  
$m^d\;\mbox{mod}\;n$.
The point of this isolated detail is to show why   
public-key cryptography is computationally expensive (think of what
big numbers the exponents are!), which is an
important factor in design of cryptographic protocols.
\else 
At present a key must be at least 2048 bits to be considered secure, and
the minimum size is expected to increase in the future.
\fi

\subsubsection{Endpoint authentication}
\label{sec:Endpoint-authentication}

A simple challenge protocol is sufficient to determine that an endpoint
has access to a public/private key pair.
Suppose that an endpoint $B$ is engaged in a session with endpoint $A$,
and wants to check its identity's claim to own public key
$K^+$.
$B$ can make sure of this by sending a
{\it nonce} (a random number used only once in its context) $n$.
$A$ is supposed to reply with $K^-(n)$, which
is $n$ encrypted using the private key $K^-$ that goes with public
key $K^+$.
$B$ then decrypts the reply with $K^+$.
If the result is $n$, then $B$ has authenticated that the other endpoint
indeed has access to public key $K^+$ and its private key $K^-$.

In practice $B$ may not know the public key ahead of time.
In a typical client/server protocol, the client needs to authenticate
the server, but the server does not authenticate the client.
The client $B$ might send its nonce to $A$, and $A$ might reply
with both its certificate and $K^-(n)$.
From the certificate, $B$ gets $K^+$.
The client should validate the certificate as well as the
encrypted nonce,
including checking that the identity in the certificate
is the identity expected,
checking that the certificate has not expired,
and checking that it has been signed by a legitimate certificate
authority.
Some client software validates certificates poorly or not al all,
causing it to be dubbed
``the most dangerous code in the world'' \cite{dangerous}.

A server can delegate its identity to another trusted network member,
by giving the delegate its certificate and keys.
For example, ``content-delivery networks'' host Web content on behalf
of other enterprises.
Content-delivery servers are trusted delegates of their customers,
and each such server can have many delegated identities.

\ifdefined\bookversion 
As mentioned in \S\ref{sec:Trust-and-identity}, IP names (addresses)
are not very good identities, because they are often assigned
transiently, and are never mnemonic. 
As a result, the names of IP members cannot be authenticated, leading
directly to the prominence of spoofing in a variety of security attacks.
The Accountable Internet \cite{aip} is a proposal
based on the alternative principle that Internet names should be
the persistent identities of Internet members, and that they should
be ``self-certifying.''
This means that 
any other member communicating with a member can authenticate
its name, even without trusting a certificate authority.
This is important in a global network, because there are no
certificate authorities
that are trusted by all countries \cite{clark-book}.

Clearly this could be achieved if the name of a member were its public
key, but public keys are too long for network names.
The Accountable Internet solves this problem by using as a member's
name a 144-bit {\it cryptographic hash} of its public key.
A cryptographic hash is computed by
a function $H$ from a digital message $m$ (of any length) to a
fixed-length bit string.
Its important property is that, given a hash $H(m)$, it is
extremely difficult to compute a different message $m'$ such that
$H(m) = H(m')$.
In AIP, having validated that a member has 
public key $K^+$, a validator completes the job by computing
$H(K^+)$ and checking that it is the same as the member's name.

In the Accountable Internet Protocol (AIP),
endpoint authentication is not implemented in user endpoints by session
protocols, as is usual;
rather it is part of routing and forwarding, and is implemented in
AIP forwarders.
The costs are considerable and everyone connected to the Internet must
bear them, which is why AIP is a radical proposal.
The Accountable Internet's
counter-argument 
would be that endpoint authentication is
essential for network security, so everyone needs it all the time.
\else 
\fi

\subsubsection{Digital signatures}
\label{sec:Digital-signatures}

A digital signature transmitted with a document can be checked to
verify that the document came from a specific identity, and has not
been modified in transit.
The simplest digital signature of a document $m$ would be
$K^-(m)$, i.e., the document itself encrypted with the private key of
the signer.
The recipient decrypts the signature with the public key of the signer.
If the result is $m$, then the signature and document are verified.

\ifdefined\bookversion 
Because public-key encryption is computationally expensive, encrypting
whole documents would be very inefficient.
In practice a (short) 
cryptographic hash $H(m)$ of the document is encrypted
with the private key
and used as a digital signature.
To verify the signature, the recipient both decrypts the signature with
the public key, and computes the same hash function on the plaintext
document.
Verification is successful if both computed values are the same.
\else 
Because public-key encryption is computationally expensive, encrypting
whole documents would be very inefficient.
Instead a {\it cryptographic hash} is used.
The hash function 
$H$ is computed from a digital message $m$ (of any length) to a
fixed-length bit string.
Its important property is that, given a hash $H(m)$, it is
extremely difficult to compute a different message $m'$ such that
$H(m) = H(m')$.
So a (short) 
cryptographic hash $H(m)$ of the document can be encrypted
with the private key
and used as a digital signature.
To verify the signature, the recipient both encrypts the signature with
the public key, and computes the same hash function on the plaintext
document.
Verification is successful if they match.
\fi

If a client is interested in the identity of a server only to obtain
its authentic data, then receiving data signed by the server is just as
good as receiving data directly from the server.
\ifdefined\bookversion 
This kind of delegation is used in Named Data Networking \cite{ndn14}.
\else 
\fi

\subsubsection{Key exchange}
\label{sec:Key-exchange}

\begin{figure*}[hbt]
\begin{center}
\ifdefined\bookversion
\includegraphics[scale=0.60]{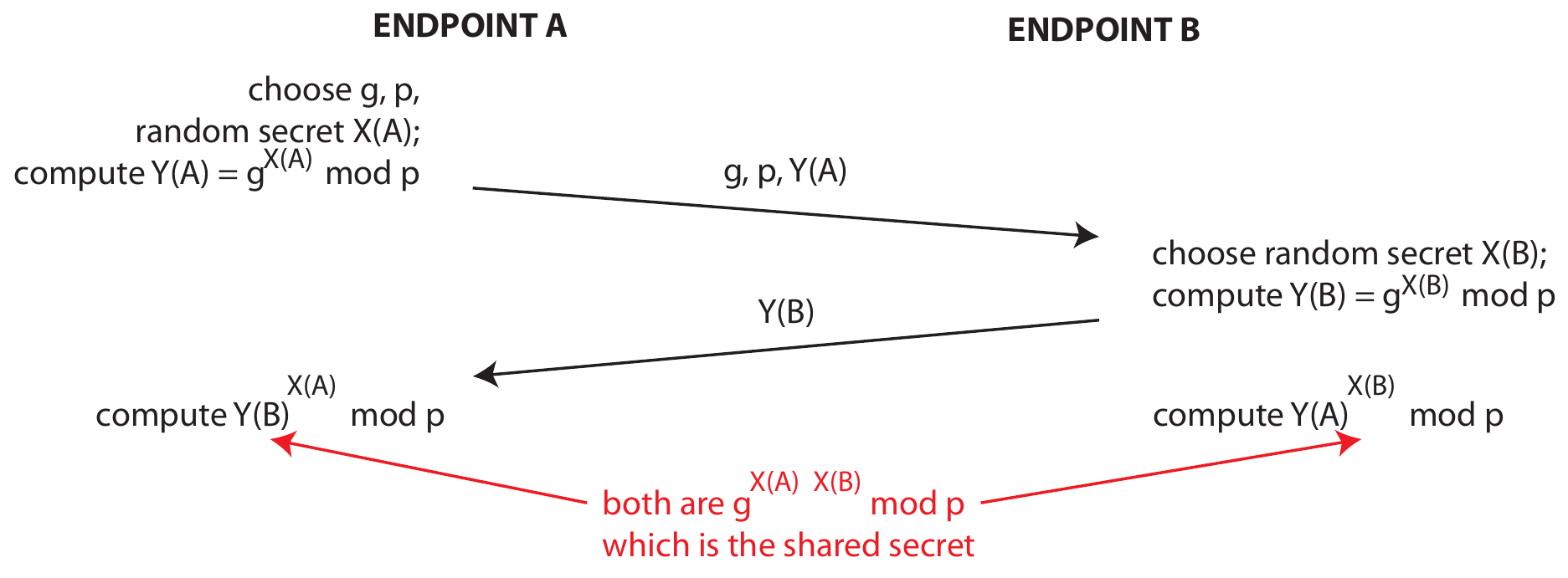} 
\else
\includegraphics[scale=0.50]{fig/dh.pdf} 
\fi
\end{center}
\caption
{\small
{Diffie-Hellman key exchange.  $g$ is a small number such as 
2 or 3, while $p$, $X(A)$, and $X(B)$ are large integers.}
\normalsize}
\label{fig:dh}
\end{figure*}

Because public-key cryptography is computationally expensive, it is used 
only to encrypt small amounts of data.
For encrypting the entire data stream being transmitted on a link, 
{\it symmetric-key cryptography}, which is much more efficient, is used.
As the name implies,
symmetric-key cryptography requires that both endpoints have the same
secret key, which is used to both encrypt and decrypt the data.
\ifdefined\bookversion 

This raises the problem of ``key exchange,'' or how to distribute
secret keys securely over insecure channels.
The basic solution to the problem of key exchange is the Diffie-Hellman
algorithm, shown in Figure~\ref{fig:dh}.
\else 
This raises the problem of ``key exchange,'' or how to distribute
secret keys securely over insecure channels.
The basic solution to the problem is the Diffie-Hellman
algorithm, shown in Figure~\ref{fig:dh}.
\fi

\begin{figure*}[hbt]
\begin{center}
\ifdefined\bookversion
\includegraphics[scale=0.55]{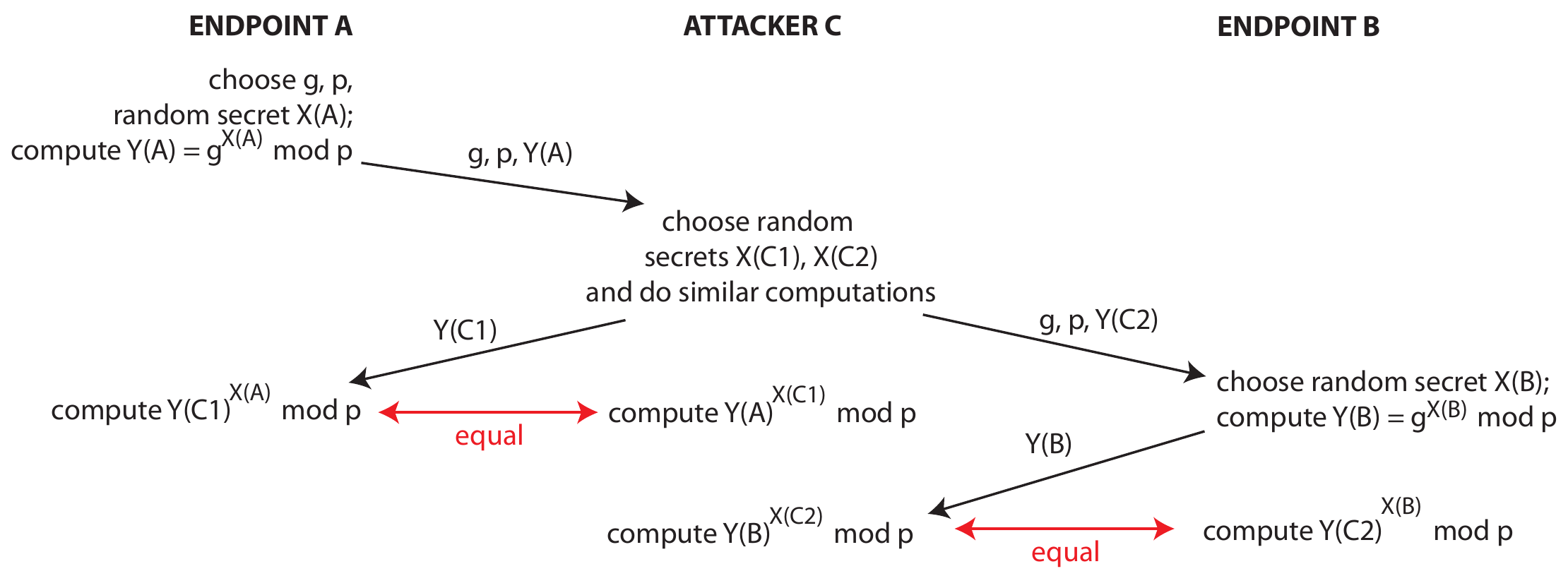} 
\else
\includegraphics[scale=0.50]{fig/dhattack.pdf} 
\fi
\end{center}
\caption
{\small
{A man-in-the-middle attack on Diffie-Hellman key exchange.}
\normalsize}
\label{fig:dhattack}
\end{figure*}

Unfortunately, the basic algorithm is vulnerable to a ``man-in-the-middle''
attack, which refers to any attack carried out by an adversary able
to intercept packets on a link.
The adversary can
read, absorb, inject, or alter any packet transmitted on
the link; the attacker can also ``replay'' packets by storing them
and retransmitting them later.
Figure~\ref{fig:dhattack} shows how such an attack would work.
The adversary simply engages in a separate key exchange with each
of the two endpoints.
After the key exchange the adversary can relay packets transparently
between $A$ and $B$ by decrypting with one key and encrypting with the
other; it can also read the packets and manipulate them in any way
whatsoever.

Fortunately, the solution to this problem is straightforward. 
$A$ and $B$ must have identities and public/private key pairs,
and must authenticate each other before the key exchange.
Then protocol packets must
bear the sender's digital signature.
Even if the attacker can read $Y(A)$ and $Y(B)$, it can do nothing
with them.

\subsection{Three IP cryptographic protocols}
\label{sec:Three-IP-cryptographic-protocols}

This section provides an overview of security in
the three most important cryptographic protocols in the IP suite:
\begin{itemize}
\item
Transport Layer Security (TLS) is the
successor to Secure Sockets Layer, and is an extension of TCP.
Two versions of TLS, 1.2 and 1.3, are in widespread use.
\item
Quic \cite{Quic} is a new protocol proposed as an alternative to TLS.
Its security mechanisms are similar to TLS 1.3.
\item
``IPsec'' refers to a family of related IP protocols, comprising
the Authentication Header and Encapsulating Security Payload (ESP)
protocols, each of which can be used in ``transport mode'' or
``tunnel mode.''
ESP is more useful than Authentication
Header, so only ESP will be discussed here.
\end{itemize}
These protocols provide endpoint authentication,
data integrity, data confidentiality, and forward secrecy.
They have interesting differences, and the differences are significant
for their use in compositional network architectures.

\ifdefined\bookversion 
\subsubsection{Protocol embeddings}
\label{sec:Protocol-embeddings}

TLS is composed with (embedded in) TCP 
(recall \S\ref{sec:Protocol-composition}).
If the Uniform Resource Locator (URL) of a Web site begins 
with {\tt https://}, then its clients should
make requests of it using IP protocol TCP and destination port 443,
signifying the use of TLS embedded in TCP.
Figure~\ref{fig:tlsesppackets} shows packet formats
for TLS, ESP in transport mode, and ESP in tunnel mode.

\begin{figure*}[hbt]
\begin{center}
\includegraphics[scale=0.60]{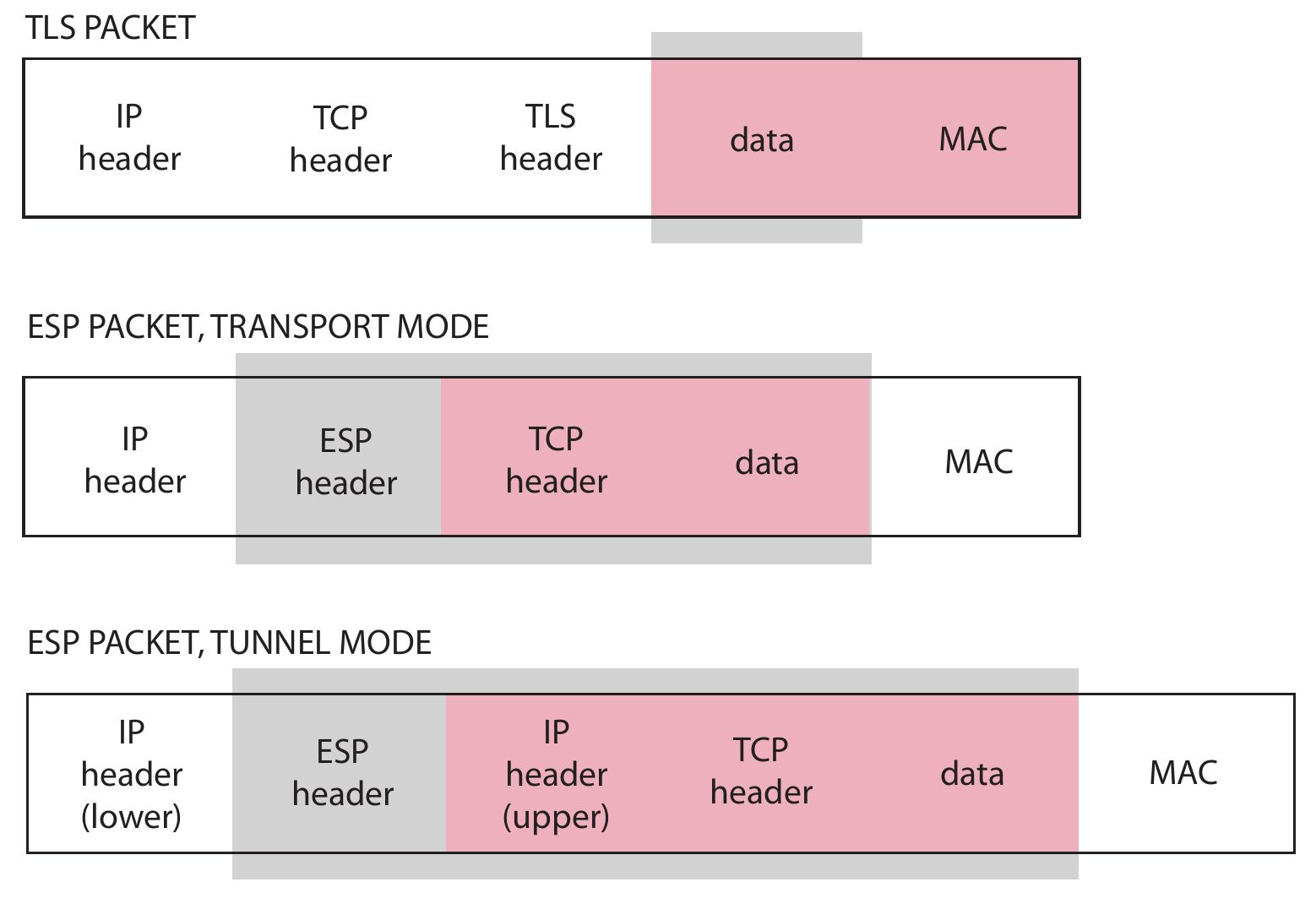} 
\end{center}
\caption
{\small
{Packet formats for cryptographic protocols, slightly simplified.  
The Message Authentication Code (MAC) is a footer that
assists in message authentication.
Pink parts of a packet are encrypted, while gray parts are authenticated.}
\normalsize}
\label{fig:tlsesppackets}
\end{figure*}

When ESP is used in composition with TCP in transport mode, 
TCP is simply embedded in ESP.
In contrast, ESP in composition with TCP in tunnel mode is an instance
of layering (recall \S\ref{sec:Composition-of-networks}).
An entire overlay packet with IP/TCP headers and data is encapsulated
in the data part of an underlay IP/ESP packet.
So the important distinction between ESP transport mode (session-protocol
composition with TCP) and ESP tunnel mode (layering composition with
TCP) is that in tunnel mode there is an upper IP header with a 
completely different destination than in the lower IP header.
Intuitively, the upper destination is the ultimate destination of the
TCP session, while the lower destination is the next hop in the session
path (see \S\ref{sec:Layering4}).

Quic is embedded in UDP, and also uses destination port 443.
When a client accesses an {\tt https://} Web site for the first time,
it should use TLS.
If responses carry an ``I support Quic'' code, subsequent requests
from that client to that server should use Quic, with TLS as a fallback
in case of problems.
\else 
\subsubsection{Protocol embeddings}
\label{sec:Protocol-embeddings}

Within a network, session protocols can be composed, so that the same
session benefits from the services implemented by multiple protocols.
When two session protocols $P$ and $Q$ are composed, one of them is
{\it embedded in} the other.

TLS is composed with (embedded in) TCP.
If the Uniform Resource Locator (URL) of a Web site begins 
with {\tt https://}, then its clients should
make requests of it using IP protocol TCP and destination port 443,
signifying the use of TLS embedded in TCP.
Figure~\ref{fig:tlsesppackets} shows packet formats
for TLS, ESP in transport mode, and ESP in tunnel mode.
Fields are labeled to show which headers belong to embedded and
embedding protocols.

\begin{figure*}[hbt]
\begin{center}
\includegraphics[scale=0.50]{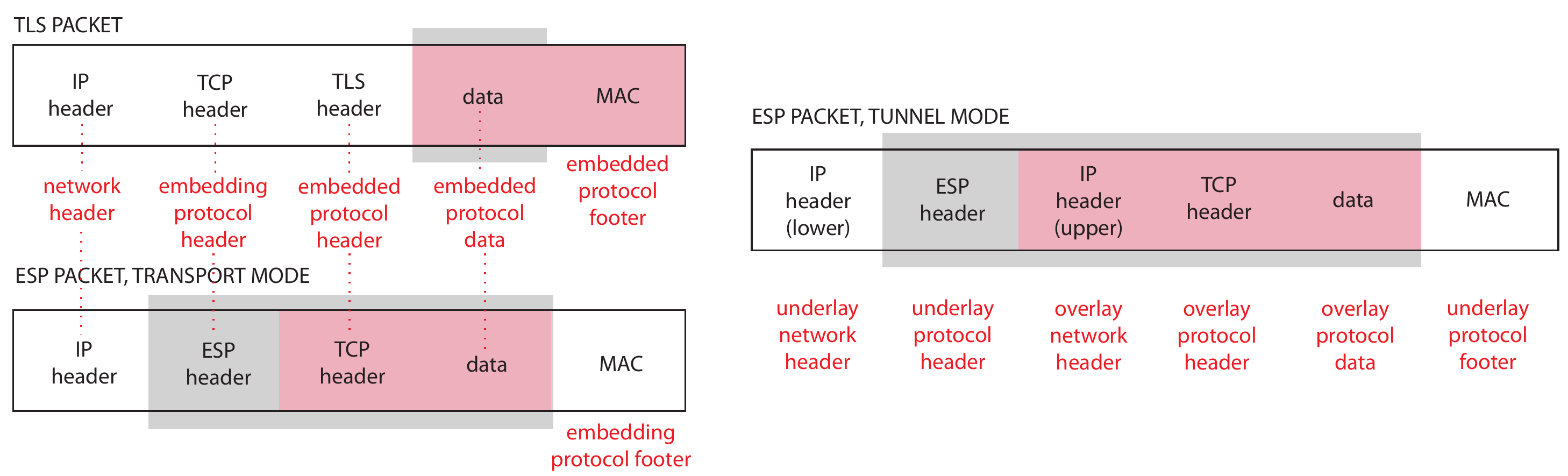} 
\end{center}
\caption
{\small
{Packet formats for cryptographic protocols, slightly simplified.  
The Message Authentication Code (MAC) is a footer that
assists in message authentication.
Pink parts of a packet are encrypted, while gray parts are authenticated.}
\normalsize}
\label{fig:tlsesppackets}
\end{figure*}

When ESP is used in composition with TCP in transport mode, 
TCP is simply embedded in ESP.
In contrast, ESP in composition with TCP in tunnel mode is an instance
of layering (recall \S\ref{sec:Composition-of-networks}).
An entire overlay packet with IP/TCP headers and data is encapsulated
in the data part of an underlay IP/ESP packet.
So the important distinction between ESP transport mode (session-protocol
composition with TCP) and ESP tunnel mode (layering composition with
TCP) is that in tunnel mode there is an upper IP header with a 
completely different destination than in the lower IP header.
Intuitively, the upper destination is the ultimate destination of the
TCP session, while the lower destination is the next hop in the session
path (see \S\ref{sec:Layering4}).

Quic is embedded in UDP, and also uses destination port 443.
When a client accesses an {\tt https://} Web site for the first time,
it should use TLS.
If responses carry an ``I support Quic'' code, subsequent requests
from that client to that server should use Quic, with TLS as a fallback
in case of problems.
\fi

\subsubsection{The setup phase}
\label{sec:The-setup-phase}

In a TLS 1.2 session,
the client and server first have a TCP (control) handshake,
in which they establish the session identifier and other parameters.
They then
begin a TLS 1.2 (control) handshake, which
performs three tasks:
(i) endpoint authentication (\S\ref{sec:Endpoint-authentication}),
(ii) negotiation of a ``cipher suite,'' and
(iii) key exchange (\S\ref{sec:Key-exchange}).
Usually the accepting endpoint is authenticated with a certificate
and the initiating
endpoint is not, because the acceptor is a server and the initiator
is a client.

TLS supports many different methods for exchanging keys, encrypting
data, and authenticating message integrity (see below).
For each of these tasks there are many possible 
algorithms (counting all variations of a few basic algorithms).
A ``cipher suite'' is a collection of algorithms and parameter choices
for doing all the cryptographic tasks within a security protocol.
The most important parameter choices govern key length, because 
key length has a
big effect on the overall security of cryptography.
To negotiate a cipher suite, the initiator sends all the cipher suites
it implements, and the acceptor chooses one that it also implements
and sends back the choice.

The TLS 1.2 handshake adds two round-trip
times for TLS setup on top of the one round-trip for TCP setup.
Slightly simplified, there is one round-trip for authentication and
negotiation, and one for key exchange.
The property of {\it forward secrecy} is achieved because fresh
symmetric keys are computed for each session.

Security in TLS 1.3 is very similar to the security
in Quic.
One difference between TLS 1.2 and Quic 
(or TLS 1.3) is that Quic disallows some older
cipher suites that are known to be insecure, and requires
longer keys.
Another difference is that Quic/TLS 1.3
setups are faster than TLS 1.2 setups.
\ifdefined\bookversion 

\else 
\fi
For faster setups, Quic combines the initial transport handshake
with the initial security handshake.
After this there is one additional round-trip for key exchange.
Further, the key-exchange round-trip can be combined with the
first data round-trip, because the client's first data request
is allowed to use a less-secure symmetric key; the server's first
response and all subsequent data packets are encrypted with the 
final, secure symmetric keys.
Even further, this one-round-trip setup can be eliminated entirely
if the client has saved authentication and negotiation information
from previous contact with the server.
In this ``zero round-trip'' setup,
the first round trip combines data and key exchange as above.

ESP endpoints authenticate each other if required, negotiate
cipher suites, and exchange keys by means of the Internet
Key Exchange (IKE) protocol.
The result is that each ESP endpoint has 
long records called ``security associations''
including choices of cipher suite and actual keys.
Use of full IKE to set up an ESP session
is not always necessary because security associations
can also be introduced into ESP endpoints by configuration,
or saved from previous negotiations.
Needless to say, if perfect forward secrecy is required,
longer-term parts of a security association can be re-used, but
there must be a new key exchange for symmetric keys.

\subsubsection{Data integrity and confidentiality}
\label{Data-integrity-and-confidentiality}

In all three protocols, data and some headers are encrypted with a shared
key by the sender, and decrypted using the same key by the receiver.
A different shared key is used in each direction.
According to the mathematics of symmetric-key cryptography,
encryption satisfies the 
requirement of data confidentiality.

The requirement of data integrity is satisfied by the process of
``message authentication.''
Each packet is sent with a ``message
authentication code'' (MAC) computed from the authenticated data $d$
by appending to the data 
a shared authentication key $k$, and then applying a cryptographic hash 
\ifdefined\bookversion 
function (\S\ref{sec:Endpoint-authentication}) to $d + k$.
\else 
function (\S\ref{sec:Digital-signatures}) to $d + k$.
\fi
The MAC 
$H(d + k)$ is then appended to the data in the packet.
As with encryption keys, all three protocols generate authentication
keys during key exchange, and use a different authentication
key in each direction.
The packet receiver performs the same MAC computation and 
expects it to result in the
same MAC that it received in the packet.
If an attacker inserts or changes packets while they are being
transmitted, it will not be able to compute correct authentication
codes for the packets, and the discrepancy will be detected by the
receiver.

This algorithm alone has the limitation
that an attacker with access to the packet stream
can still delete, re-order, or replay packets, even though it cannot
create new ones.
TLS and ESP require different solutions to this problem, because
of the differences in embedding visible in Figure~\ref{fig:tlsesppackets}.

One might think that this problem would be solved for TLS (both versions)
by the fact
that the enclosing TCP packets have byte sequence numbers.
TCP headers are not encrypted, however, so an attacker-in-the-middle
could alter them to make even an altered TCP byte stream look correct.
The actual
TLS solution is for each endpoint to keep track of packet sequence
numbers as TLS packets are sent and received.
The sequence number is not transmitted directly, but it
is included in the bit string hashed to compute the MAC.
For a packet to be accepted, the receiver must be re-computing its MAC
with the same sequence number that the sender used.
This works because TLS is embedded in TCP, so the authenticated data
and MAC are presented to the authenticator reliably and in sending order.

Message authentication in ESP and Quic must work differently, because 
their packets may not be presented to the authenticator in sending
order.
In these protocols, the headers contain explicit packet sequence
numbers, which are
included in the data on which the MAC is computed.
The authenticator cannot predict the sequence number of the next
packet it will see, so it cannot detect deletion or re-ordering
attacks (which, after all, might not be attacks but flaws in the network).
Rather, authentication
checks only for received packets with sequence numbers that have
already been received, and deletes them.
This is sufficient to defend against replay attacks, which are
part of many man-in-the-middle attacks, because an attacker
cannot change the sequence number of a packet it replays.

\subsubsection{Usage of cryptographic protocols}
\label{sec:Uses-of-TLS-and-ESP}

Almost all Web traffic is now encrypted, at least with TLS 1.2.
Deployments of TLS 1.3 and Quic are both growing rapidly, because of the
motivation of shorter setup times.
TLS is also widely used by other application protocols.
ESP is most commonly used to make ``virtual private networks''
(see \S\ref{sec:Compound-sessions-and-overlays-for-security}).

\ifdefined\bookversion 
Some properties of
the protocols are summarized in Figure~\ref{fig:protocol-table}.
The TLS entry covers both versions.
Quic is like TLS except for message authentication, in which it resembles
ESP.

\begin{figure*}[hbt]
\begin{center}
\includegraphics[scale=0.60]{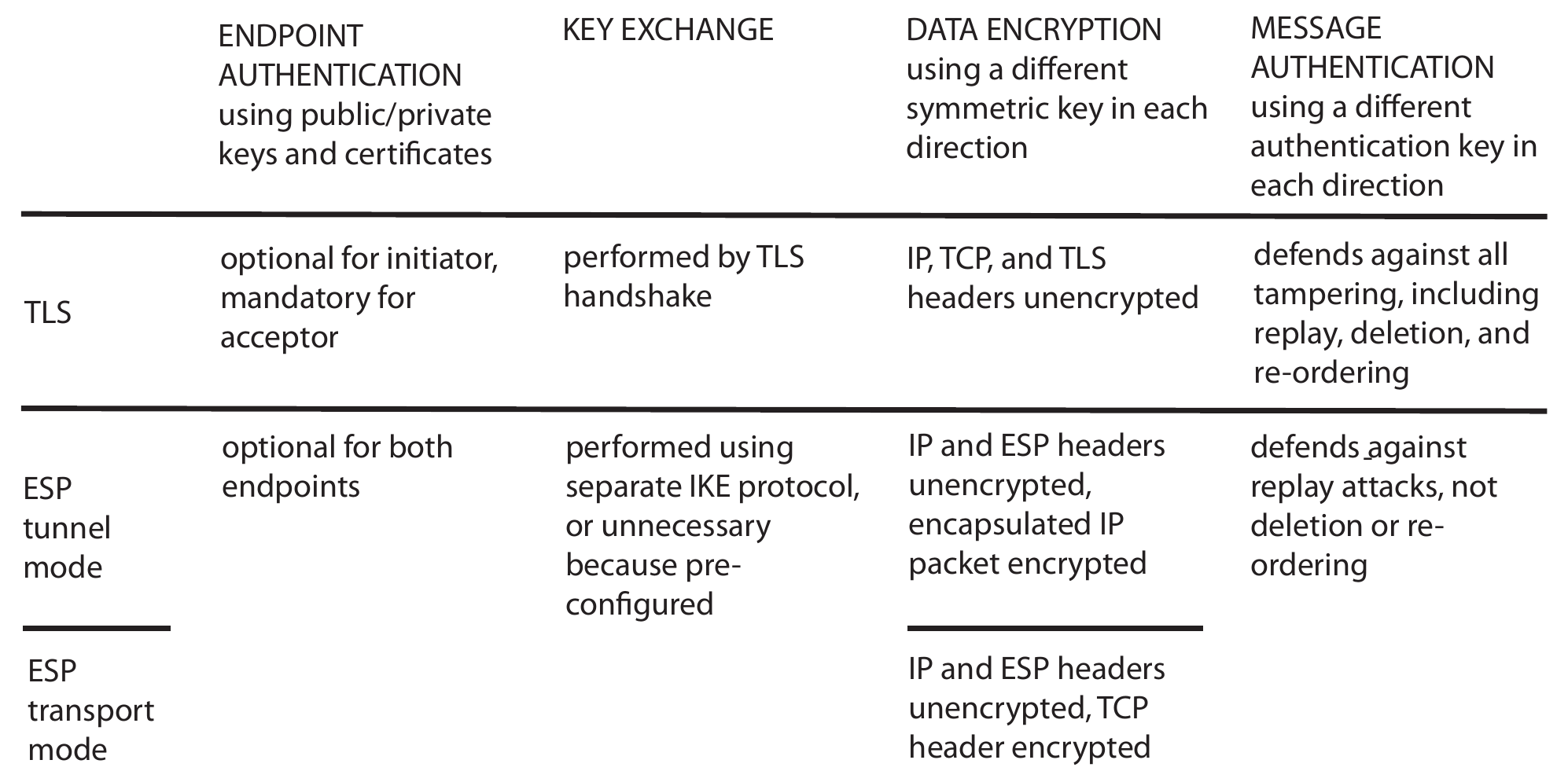}
\end{center}
\caption
{\small
{Summary of protocol properties.}
\normalsize}
\label{fig:protocol-table}
\end{figure*}

Not surprisingly, developers building applications on UDP are
also interested in endpoint security.
For UDP transmission, there is a security protocol called DTLS (Datagram
Transport Layer Security) that is as similar as possible to TLS.
DTLS introduces the notion that a sequence of UDP packets go together
in a session, which is not present in plain UDP.
It should be clear from the previous sections that, because DTLS is not
embedded in TCP, its designers had to solve two problems:
(i) the TLS handshake assumes reliable delivery of the handshake
packets, and (ii) DTLS
message authentication cannot rely on the property that packets
are delivered reliably, in order, and duplicate-free, so that packet
sequence numbers can be computed independently at each endpoint.
DTLS solves the first problem by incorporating packet-loss detection and
retransmission into the DTLS handshake.
DTLS solves the second problem by using explicit sequence numbers,
exactly as ESP does.

Wireless
networks have their own cryptographic protocols based 
on the same principles. 
Security is particularly important for these networks, because
any machine within radio range has physical access to the broadcast
links of the network.
When a new member joins a private wireless network, endpoint
authentication ensures that the new member is authorized.
\else 
\fi

Although cryptographic algorithms and protocols are proved 
mathematically, there is a big difference between mathematical
abstractions and code.
In implementing the algorithms, efficiency is a top priority, and
transformations for efficiency can introduce bugs in addition to
all the other bugs to which software systems are prone.
Advances in processor speeds and the exploitation
of side-channels are making
it easier to crack codes, so that increases in key lengths become
necessary---not even counting the unpredictable disruption that might
be caused by quantum computing.
Cryptographic libraries are improved continually, but each machine
is no more secure than its latest upgrade.
It may even be less secure, when it must use an older software
version to communicate with an infrequently-updated machine.

\subsection{Interactions between cryptographic protocols 
and other aspects of networking}
\label{sec:Interactions-between-cryptographic-protocols} 

Cryptographic protocols have significant interactions with other
security patterns, which will be discussed when the other security
patterns have been presented.
This section is concerned with the interactions of cryptographic
protocols with network architecture and network services other
than security.
\ifdefined\bookversion 
In considering architectural and service interactions, 
we will be looking at multiple composed networks
as well as protocols within a single network.
\else 
\fi

\subsubsection{Layering}
\label{sec:Layering4}

A network with cryptographic session protocols can be layered on top
of one or more networks, as explained in 
\S\ref{sec:Composition-of-networks}.
Because each underlay level can implement an overlay link with a path
of links, forwarders, and middleboxes, users of an overlay network
must accept that its packets can pass through many machines and 
physical links unknown to them.
But cryptographic protocols are designed to work in completely adversarial
environments such as these!
Furthermore, the cryptographic properties of a session can be
assumed to hold for any link that it implements, so
the properties guaranteed by cryptographic protocols
propagate upward through layering.

\subsubsection{Performance}
\label{sec:Performance}

Data encryption and message authentication increase required bandwidth
and computational resources.
The overhead is modest, so it is not a concern in all cases.
It is more likely to be a significant concern for 
battery-operated devices, or for network elements that must decrypt
and re-encrypt at high traffic volumes.

The most direct and significant
performance costs of cryptographic protocols are incurred in the
setup phase, by
endpoint authentication and key exchange, which consume
compute resources and increase latency. 
Even with short round-trip times, a small fraction of
TLS 1.2 setups take 300 ms or more \cite{costTLS}, due to increased
computation time.
We have seen that newer protocols have reduced setup times aggressively,
often by saving and re-using session state, but this causes an
inevitable loss of security
\cite{RFC-7457}.

The performance issue is much more serious in applications for the
Internet of Things (IoT), because
these applications tend to have periodic or irregular short communications
from a large number of networked devices to centralized analysis or 
publish/subscribe servers.
Message Queuing Telemetry Transport, a protocol for IoT
applications, is well-designed from this perspective, because
many short application communications can share the same TLS session. 
\ifdefined\bookversion 
Even so,
group events (such as initialization of a fleet of vehicles) can easily
create spikes in the load on centralized servers \cite{hiveMQTT}.
\else
\fi

For Message Queuing Telemetry Transport and all other application
protocols with short or bursty communications separated by intervals of
inactivity, it is most efficient for many communications to share a
single, long-lived secure channel.
Long-lived Internet channels have been difficult to maintain in the past,
because various components in the path of the channel would time out
and close the channel during intervals of inactivity.
It is easier now---TCP, TLS, and DTLS all have keep-alive options,
sending periodic keep-alive signals to keep long-lived channels open.

\begin{figure*}[hbt]
\begin{center}
\ifdefined\bookversion
\includegraphics[scale=0.60]{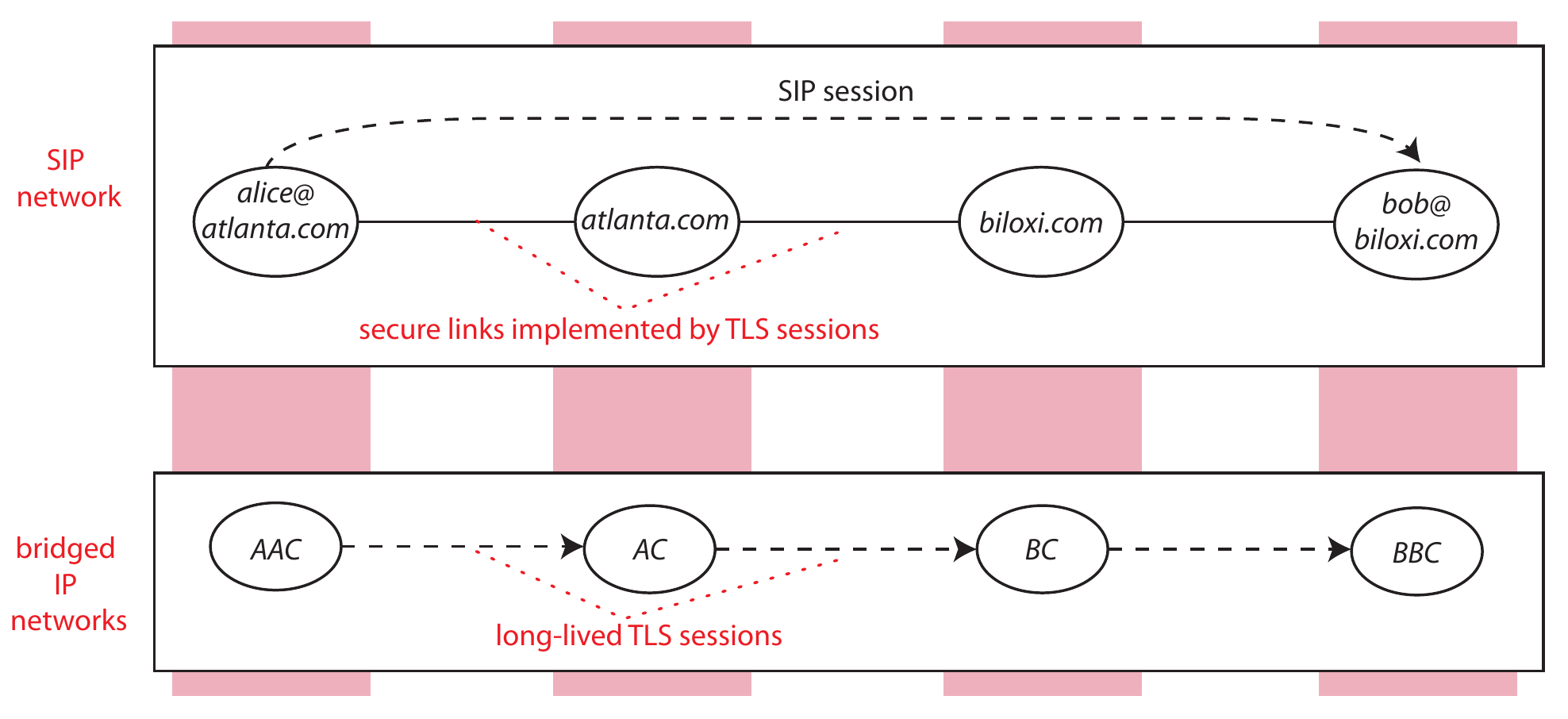} 
\else
\includegraphics[scale=0.45]{fig/newsip.pdf} 
\fi
\end{center}
\caption
{\small
{A SIP application network
layered on bridged IP networks.}
\normalsize}
\label{fig:newsip}
\end{figure*}

Architecturally, there are two ways to implement the optimization
of sharing a secure channel.
The first way is to embed the application protocol in the security
protocol.
For example,
if TLS is the security protocol,
application headers and data would be the data portion
of TLS packets, as shown in Figure~\ref{fig:tlsesppackets}.
\ifdefined\bookversion 

\else 
\fi
Alternatively, an application network could be layered on IP networks,
as shown in Figure~\ref{fig:newsip}.
The Session Initiation Protocol (SIP) is an application protocol
for control of multimedia
applications.
The SIP application network has links that are implemented by
TLS sessions in the IP underlay.
The big difference between this architecture and protocol embedding
is that the TLS sessions have different endpoints (different sources
and destinations) than the SIP application session does.
This makes it more flexible than protocol embedding, for
two reasons:
(i) The application network can insert its own middleboxes into the
path of the application session, as SIP always does.
(ii)
The TLS sessions can last longer than any communication between two
specific
SIP endpoints, and can be shared by communications between many SIP
endpoint pairs.
Recall that embedding and layering
correspond to ESP transport mode and tunnel
mode, respectively.

\ifdefined\bookversion 
\subsubsection{Session protocols}
\label{sec:Session-protocols4}

One significant issue in the use of cryptographic
protocols is their relationship to TCP, because TCP does so many
things: congestion control, reliability, and packet ordering.
We have seen that TLS depends on being embedded in TCP.
This should not be a problem, unless real or perceived implementation
constraints cause designers to make bad choices.
For example, some network architectures use TCP as a session protocol
in an overlay network with secure links implemented by TLS.
(In comparison, in Figure~\ref{fig:newsip}, the overlay network
uses SIP as the session protocol.)
Because of the dependence of TLS on TCP, this design is layering one 
instance of TCP over
another instance of TCP!
This can cause the problem of ``TCP meltdown'' 
\cite{tcp-over-tcp}, as follows.

TCP provides reliability by detecting lost packets by means of a timer,
and requesting retransmission of a packet when it does not arrive in time.
For each session, TCP sets the timeout interval independently and
adaptively.
It can happen that the timeout interval on the upper-level instance
of TCP becomes
shorter than the timeout interval on the lower-level instance.
In this case the lower-level session is experiencing
reduced throughput, because
it is waiting a longer time for each packet.
At the same time, the upper-level session is having frequent timeouts,
making frequent requests for retransmission, and therefore demanding
increased throughput.
This mismatch drastically degrades the end-to-end throughput.

Another significant issue is the relationship between cryptographic
protocols and stateful firewalls in IP networks.
Stateful firewalls record ongoing sessions and use them to filter
packets; for example, firewalls at the edges of private networks
are often configured to allow only two-way sessions initiated from
inside the network.
To do this, the firewall must be able to tell which incoming packets
are in the same session as particular outgoing packets.

The problem in IP networks is that session identifiers are not
standardized across all session protocols.
All firewalls recognize TCP sessions, because the first 32 bits 
of a TCP header consists of two port numbers,
and a session is identified by the IP destination name in each direction
and its corresponding port.
In ESP, on the other hand, the first 32 bits of the protocol header are a
pointer to a security association (see \S\ref{sec:The-setup-phase}),
which is completely different in the forward and reverse directions,
and cannot be used to associate packets in the two directions.
The consequence is that a stateful firewall
will not allow ESP sessions.

In this case protocol composition enables a workaround to the problem.
ESP, whether in tunnel or transport mode,
can be embedded in UDP with well-known port 4500.
(A well-known port for UDP/ESP composition is necessary because
UDP headers have no place for the ``next header,'' as IP and ESP
headers do.)
In this way a two-way sequence of UDP packets forms an identifiable
session, and a stateful firewall
does not see the ESP headers at all.
Quic is already embedded in UDP, and traverses stateful firewalls in
the same way.
\else
\fi

\subsubsection{Mobility} 
\label{sec:Mobility} 

\ifdefined\bookversion 
In its strongest sense, mobility enables a session to persist even though
the network attachment of a device at its endpoint is changing.
This usually means that, at some level of the layering hierarchy,
the network member on the device is changing names
within its network, or dying and being replaced by a member of another
network.
For example, when a mobile phone moves from one cellular
provider's network
to another, its IP name (for data service) must change.
Ideally the data sessions of the phone would persist across such moves,
as its voice sessions do.
\else 
In its strongest sense, mobility enables a session to persist even though
the network attachment of an endpoint device is changing.
This usually means that, at some level of the layering hierarchy,
the network member on the device is changing names
within its network, or dying and being replaced by a member of another
network.
For example, when a mobile phone moves from one cellular
provider's network
to another, its IP name (for data service) must change.
Ideally the data sessions of the phone would persist across such moves,
as its voice sessions do.
\fi

There is usually no interaction between mobility and cryptographic
protocols, because the identity of a mobile machine is at a higher
level than the names that change.
For instance, consider a Web server running on a virtual machine
in a cloud.
Because of failures or resource changes, the virtual machine may
migrate to a different physical machine where it has a different
IP name.
But the identity of the Web server is its domain name, which is at a
higher level and does not change.
Similarly, more than one server can have the same identity,
as when
a Web site of origin delegates its
identity to a content-delivery
server by sharing its certificate and keys.

On the other hand, thinking about mobility brings up 
the possibility of normal mobility in reverse---the  
higher-level identity moves or
changes while the lower-level name remains the same.
This can be a security issue:
after {\it Jane Q. Public} enters her password
(\S\ref{sec:Trust-and-identity}),
she might walk away from her machine, and then any other person who
walks by could retrieve her personal data and request transactions
on her bank account.
For this reason,
secure distributed applications require periodic
re-authentications of the identity of the person using them, especially
after idle periods.

\subsubsection{Infrastructure control protocols}
\label{sec:Infrastructure-control-protocols}

Control protocols are used by network infrastructure
to maintain and distribute network state.
It is important to protect these protocols against subversion
attacks (\S\ref{sec:Subversion-attacks}).

Unsurprisingly, some control protocols incorporate cryptography.
For instance, Border Gateway Protocol Security is a security
extension to BGP that provides 
cryptographic verification of 
packets advertising routes.
Similarly,
Domain Name System Security Extension protects DNS lookups
by returning records with digital signatures.

In many cases, however, it is difficult for control protocols to
rely on cryptography.
An endpoint might not have a certificate or other credential to prove
its identity.
The protocol might require high-speed, high-volume operation.
Or, the protocol might simply be too old to incorporate cryptography,
even if it is feasible.

In these cases there are lighter-weight measures that can help.
Network members that make requests should keep track of their pending
requests and not accept unsolicited replies.
Replies should be checked for credibility, whenever that is possible.
Most effectively, a network member can include a nonce or random
field value
in a session-initiation or request packet.
Subsequent packets of the session must have the same nonce or
random value, so that
no attacker without access to the previous
packets of the session can send packets purporting to be part of it.
Without the nonce, an attacker could do something to trigger a query,
then send a spurious answer to the query.

\section{Traffic filtering}
\label{sec:Traffic-filtering}

{\it Traffic filtering} is performed by forwarders and middleboxes
that are part of a network's infrastructure.
The network's routing ensures that designated traffic passes through
one of these {\it traffic filters,} and the filter examines it
for evidence of flooding attacks, subversion attacks, or policy
violations.
If traffic seems to be part of an attack, the filter takes some
defensive action, most often simply discarding the traffic.

Content-based traffic filtering 
(\S\ref{sec:Content-based-traffic-filtering})
looks at the contents of individual packets or sequences of packets.
Path-based traffic filtering (\S\ref{sec:Path-based-traffic-filtering})
adds to this information about the paths along which traffic has traveled.

As in \S\ref{sec:Cryptographic-protocols}, the usual context of the
discussion in this section will be a single network of any kind, or
a set of similar bridged networks such as
the bridged IP networks of the Internet.
After explaining traffic filtering in individual networks, we
return to the compositional
view (\S\ref{sec:Interactions5}), 
considering how traffic filtering interacts with other network
mechanisms and where it should be placed in a compositional network
architecture.

\subsection{Content-based traffic filtering}
\label{sec:Content-based-traffic-filtering}

\subsubsection{Signature-based-filtering criteria}
\label{sec:Signature-based-filtering-criteria}

{\it Filtering criteria} are predicates used to identify suspected traffic.
Signature-based filtering criteria examine specific header and data fields
of packets.
These criteria are used to detect most policy violations and
subversion attacks.
Often the criteria 
are Boolean combinations of
simple predicates such as {\it destinationPort = 80}
on the values of IP header fields.
For example, suppose that an IP ``firewall'' (traffic filter)
at the edge of a network is enforcing this policy:  
the only external traffic is for Web accesses, which of course
require DNS queries.
The direction of a packet (inbound or outbound) can be determined
from its source and destination names or from the link on which it
arrives.
The firewall might be configured with these four rules:
\begin{enumerate}
\item
Drop all outbound TCP packets unless they have destination port 80.
\item
Drop all inbound TCP packets unless they have source port 80 and 
the TCP ACK bit is set.
\item
Drop all outbound UDP packets unless they have destination port 53.
\item
Drop all inbound UDP packets unless they have source port 53.
\end{enumerate}
In the second rule, the ACK bit indicates that this packet is an
acknowledgment of a previous packet, meaning that it is not a TCP SYN
packet.

These rules are sufficient for the purpose if all packets through the
firewall obey the TCP protocol exactly, but of course an attacker may
not be so polite.
A safer approach would be to make the firewall stateful by having it
maintain a table of all ongoing TCP connections.
Then the second rule above would be replaced by ``Drop all inbound
TCP packets unless their source and destination names and ports
identify them as belong to an ongoing TCP session.''
If a firewall is stateful, it is crucial that all packets of a session
pass through the same firewall.
This property is called ``session affinity.''

\begin{figure*}[hbt]
\begin{center}
\ifdefined\bookversion
\includegraphics[scale=0.60]{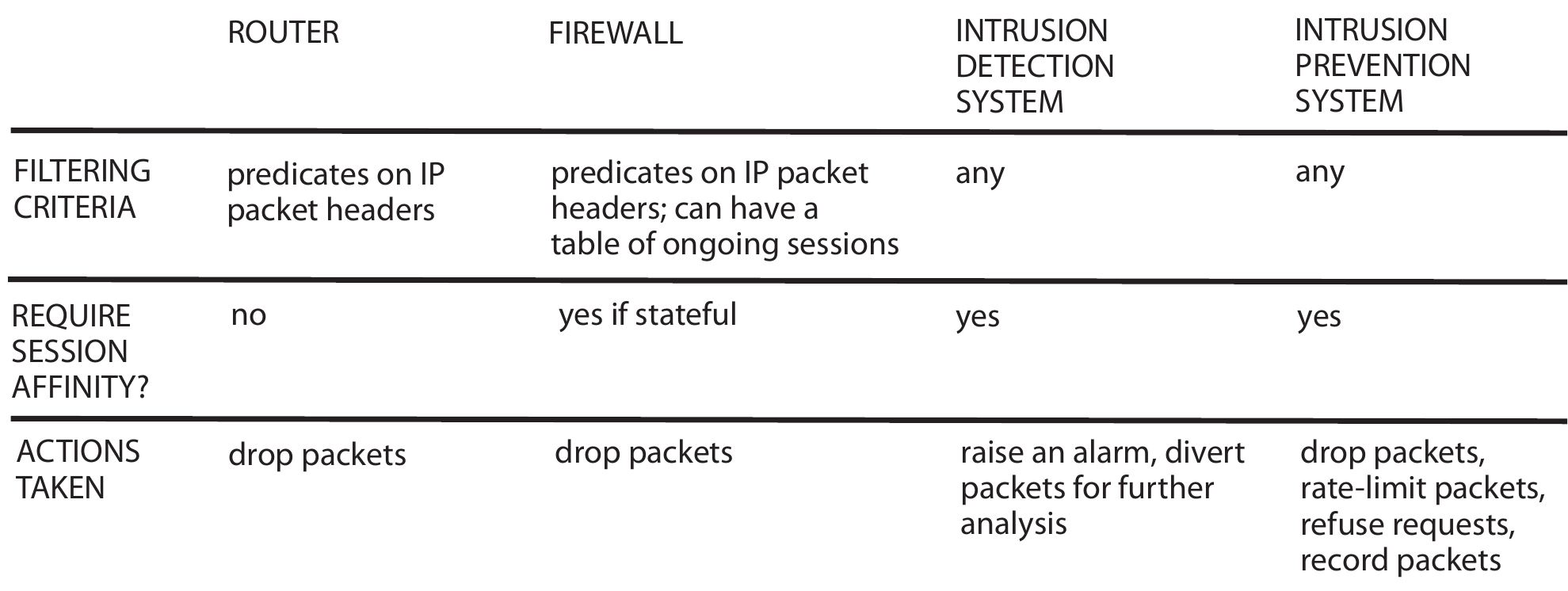} 
\else
\includegraphics[scale=0.50]{fig/filtertbl.pdf} 
\fi
\end{center}
\small
\caption
{\small
Examples of common traffic filters in IP networks.
\normalsize}
\label{fig:filtertbl}
\end{figure*}

For reference throughout \S\ref{sec:Content-based-traffic-filtering},
Figure~\ref{fig:filtertbl} is a table summarizing characteristics
of four common types of traffic filter in IP networks.
The classification is at least as much historical
and marketing-oriented as it is technical!
It is a list of products that have sold well in the past, 
not a prescriptive list
of which options are possible.

IP routers sometimes do dual duty as traffic filters.
To do this, they are configured with predicates on packet
headers, called ``access control lists.''
Routers must work even faster than firewalls, so they
do not perform stateful filtering.

For filtering that looks at packet data as well as headers,
networks often use commercial products known as
``intrusion detection systems'' and
``intrusion prevention systems.''
These filters can use any filtering criteria for any purpose.
Signature-based filters against spam and viruses
look for keywords, sometimes keywords
in specific positions, and other
known attack patterns.
Their criteria can include regular expressions matching fields of
arbitrary length.
They can also be stateful, and check whether protocols are being followed.
These filters can be valuable commercial products because of the
intellectual property in their filtering criteria.
Like all security software, to be effective, they must be kept up-to-date.

In the common case that TCP sessions are being filtered for 
subversion attacks or policy violations,
the filter should reconstruct the correct byte stream (restoring packet
order, replacing lost bytes by retransmitted ones) before filtering.
If there is no reconstruction, attackers can hide attacks simply by
splitting attack data over multiple packets.
Even if there is reconstruction, there may be ambiguities exploitable
by attackers.
For example, if there are missing packets, some bytes may be retransmitted
and received twice.
An attacker can engineer the transmitted stream so that some bytes
will have to be sent twice, 
and place attack bytes only in the second transmission.
The filter might check only the first bytes, and the receiver might use
only the second bytes.
The surest way to avoid all such ambiguities is to have a
``traffic normalizer'' middlebox in the session path, before both
filter and destination, to reconstruct a single unambiguous packet
stream received by both of them \cite{NID}.

One advantage enjoyed by TCP filters
is that attacks require communication in both directions.
Consequently, 
attackers cannot easily hide by giving false source names---if they
did, there would be no two-way commnication.
The sources of flooding attacks, on the other hand,
can hide themselves behind false source names.
This problem is also an opportunity, because having a false source
name is a good indicator that a packet is part of a flooding attack.
Forwarders (and other filters associated with them)
are well-situated for using this as a filtering criteria,
because forwarders have information about routing.
For example,
``ingress filters'' in IP networks check incoming packets to see if
the prefixes of their source names match expectations.
This is an excellent addition to an access network, which may have
detailed knowledge of its user members,
or an Internet service provider's network, which knows the IP prefixes
allocated to each access network bridged with it.
``Unicast reverse path forwarding'' in a forwarder
accepts a packet's
source name as valid only if its forwarding table specifies forwarding
{\it to} the source name on the same 
two-way link on which the packet arrived.
Unfortunately reverse-path checking
cannot be used in the high-speed core of the Internet,
because routes there are not necessarily symmetric.

\subsubsection{Measurement and statistical analysis} 
\label{sec:Measurement-and-statistical-analysis}

Signature-based filtering criteria have two major limitations:
they cannot detect new (called ``zero day'') attacks, and it is
difficult to use them to detect flooding attacks, whose individual
packets look normal (with the exception of false source names).
In response to these limitations, forwarders collect data on
large amounts of traffic, and send it to other network members for
analysis.
Analysis can measure attributes over large collections of packets.
It can then look for known attack patterns, especially of flooding
attacks;
for example, a single destination receiving a large 
number of response packets from many different sources may be 
the victim of a reflection attack (\S\ref{sec:Flooding-attacks}).
Analysis can also detect anomalies, which are new divergences from normal
traffic patterns that may indicate new attacks.
Anomaly detection uses statistical algorithms, including machine learning.

For typical traffic measurement in IP networks, routers collect selected
data and send it to analyzers in some well-known record
format such as NetFlow or IP Flow Information Export (IPFIX).
Data can be collected at multiple locations and
different levels of granularity.
The volume of data can be reduced by 
recording only headers (rather than entire packets), 
by sampling the packets (rather than collecting information about
all packets), or by focusing on 
specific subsets of the packets.
Most importantly,
a {\it flow} comprises a group of packets close together in time that 
have various header fields in common. 
Creating a single record for a flow 
helps reduce the volume of data while still 
providing a timely and detailed view of the network traffic.  

Anomaly detection is a very attractive idea, but it is also very
difficult in practice.
One major reason is that normal Internet traffic is highly
variable, not to mention unusual-but-innocent
occurrences such as congestion
due to failures, or a legitimate flash crowd \cite{pushback}.
The other major reason is that the cost of mistakes (``false positives'')
is high, as many legitimate packets are discarded.
The best use of anomaly detection may be to discover and understand
new attacks,
then turn their characteristics into signatures or
measurable patterns \cite{sommer}.

\ifdefined\bookversion 
\subsubsection{Default-reject filtering}
\label{sec:Default-reject-filtering}

Because the quality of filtering criteria is such a limiting factor,
some of the research on flooding attacks
aims to make filtering criteria precise by recognizing certain packets
as desirable and rejecting all other packets.
We'll call this approach ``default-reject'' filtering, in contrast
to usual filtering with the default behavior of accepting a packet.
In addition to precision, default-reject filtering claims the advantage of
preventing flooding attacks, rather than reacting to them well after
they have begun.

The limitation of default-reject filtering is that it only works in
limited contexts, where desirable packets can be recognized.
For instance, Secure Overlay Services \cite{sos}
is a default-reject
proposal for public emergencies, in which all normal traffic is
suspended and protected servers should be reached by emergency
responders only.
Another example, Ethane \cite{ethane}, is intended for private
IP networks where control software has
very complete knowledge,
including the network's user members, the people who
use the machines, and the peoples' roles in the organization that the
network serves.
With this much information to work with, it is possible to write
very precise rules about which communications should be allowed.
\else 
\fi

\subsubsection{Defensive actions}
\label{sec:Defensive actions}

Obviously, the most common defensive action that a filter can take
is to drop packets, but there are other possibilities.

The only difference between
``intrusion detection systems'' and
``intrusion prevention systems''
is that detection systems only raise alarms,
while prevention systems
automatically take action against suspected attacks.
It might seem that automatic action is always better (it is
certainly faster), but there are good reasons for keeping operators
and enterprise customers in the decision loop.
If a suspected attack is a false positive,
much legitimate traffic may be dropped.
If an operator deploys additional resources on behalf of an enterprise
customer that is under attack, the customer will have to pay for them.
In rare cases, the
defense against a suspected attack may even be a counter-attack,
which is wrong and even dangerous (in a military setting) if not
well-justified.

What actions are normally taken by intrusion prevention systems,
other than dropping packets?
If there is uncertainty about the packets, a filter can
rate-limit them or downgrade their forwarding
priority rather than dropping them.
Rather than dropping 
session-initiation requests, a filter could reply to them with refusals,
which would discourage retries.
A refusal to a TCP SYN (request) is a TCP RST (reset).
A refusal to an HTTP request is an error code.

Finally,
when filtering is being used to defend against policy violations,
sometimes the filter records packets for the purposes of investigation
and legal evidence.
Recorded packets are usually {\it not} dropped but forwarded on to
their destinations, to keep the investigation secret from its targets
until it has been completed.

\subsubsection{Resources and capacity}

Traffic filtering expends a lot of network resources, so the
detailed design of a traffic-filtering 
mechanism must be resource-sensitive.
How does a network ensure that its traffic filters do not
themselves become traffic bottlenecks during flooding attacks?

There are two approaches to providing sufficient filtering capacity.
The first is to host traffic filters on high-capacity machines dedicated
to this purpose.  
The second is to run traffic filters on virtual machines in a cloud.
In this approach it is possible to implement ``dynamic scale-out,''
which means that as the load increases during an attack, more virtual
machines are allocated for the filtering task.
All the filter types in Figure~\ref{fig:filtertbl} have been implemented
with both approaches, although the high-capacity-machine approach is
more often applied to routers and firewalls.

\begin{figure*}[hbt]
\begin{center}
\ifdefined\bookversion
\includegraphics[scale=0.65]{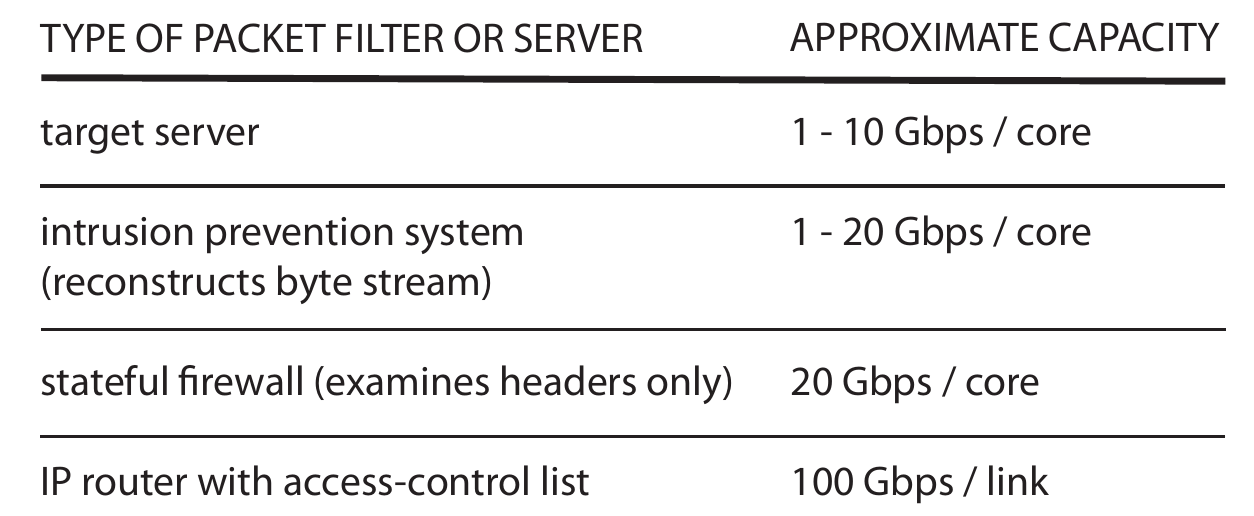} 
\else
\includegraphics[scale=0.55]{fig/capacity.pdf} 
\fi
\end{center}
\caption
{\small
{Data-processing capacities of common traffic filters and servers.}
\normalsize}
\label{fig:capacity}
\end{figure*}

The situation today is fluid, as flooding attacks are becoming
more severe.
A recent flooding attack on a number of
DNS servers \cite{dynAttack}, including
amplification, generated traffic at 10-20 times normal volume,
with bursts up to 40-50 times normal volume, and reportedly
a maximum of 1.2 Tbps (1200 Gbps).
To provide some intuition about the resources needed to handle
such attacks, Figure~\ref{fig:capacity} shows some typical capacities for 
servers and various kinds of traffic filters.

Of course these numbers are subject to frequent change.
On the positive side,
converting an algorithm from software to programmable hardware
increases its speed by a factor of about 10, as does converting
it from programmable hardware to fixed-function hardware.
On the negative side, commercial intrusion-prevention systems 
frequently fall short of advertised capacity.
There are adversarial workloads designed
so that the
TCP byte stream is especially difficult to reconstruct.
More commonly, rule-checking is the performance problem, because it is
more expensive than reconstructing the byte stream; performance
is improved when necessary by dropping rules.

\subsection{Path-based traffic filtering}
\label{sec:Path-based-traffic-filtering}

\ifdefined\bookversion 
Path-based traffic filtering augments filtering criteria based on
the contents of packets, as discussed in the previous section, with
criteria based on the path along which packets traveled.
There are two reasons for introducing path-based filtering.
The first reason is that path information can improve
the precision of filtering criteria, so that fewer good packets are
accidentally included.
For example, say that the overall load on a server suggests a flooding
attack, and intrusion detection proposes a candidate filtering rule
based on packet contents.
If we know that most paths to the server are delivering a trickle of
these packets, and one path's load is dominated by these packets,
there is a good chance that only the packets on the dominated path
are attack packets.

The second reason for path-based filtering
is that it may be possible to filter out
attack packets closer to their sources, which reduces the damage they do.
This section will discuss the trade-offs, why path-based filtering is
not much used today, and why it might be used more in the future.

\subsubsection{Tracing attacks back to their sources}
\label{sec:Tracing-attacks-back-to-their-sources}

In IP networks, because of spoofing, the source name in a packet
is not a reliable indicator of where it came from.
The purpose of a traceback mechanism is to determine the path
along which an attack packet travels.
In other words, traceback associates path meta-data with 
packets.\footnote{The Accountable Internet proposal 
(\S\ref{sec:Endpoint-authentication})
recommends authenticated
source names so that, among other reasons, traceback is not needed.}

From the viewpoint of a victim of a flooding attack, the network is a
directed acyclic graph with many possible packet sources
and a single packet sink.
Often the graph is a tree, with the victim at the root.
The interior of the graph consists of forwarders and middleboxes,
connected by links carrying traffic toward the victim.
Figure~\ref{fig:tree} shows such a graph for attack victim $T$.
In the figure, member names also stand for names of machines.

\begin{figure*}[hbt]
\begin{center}
\includegraphics[scale=0.60]{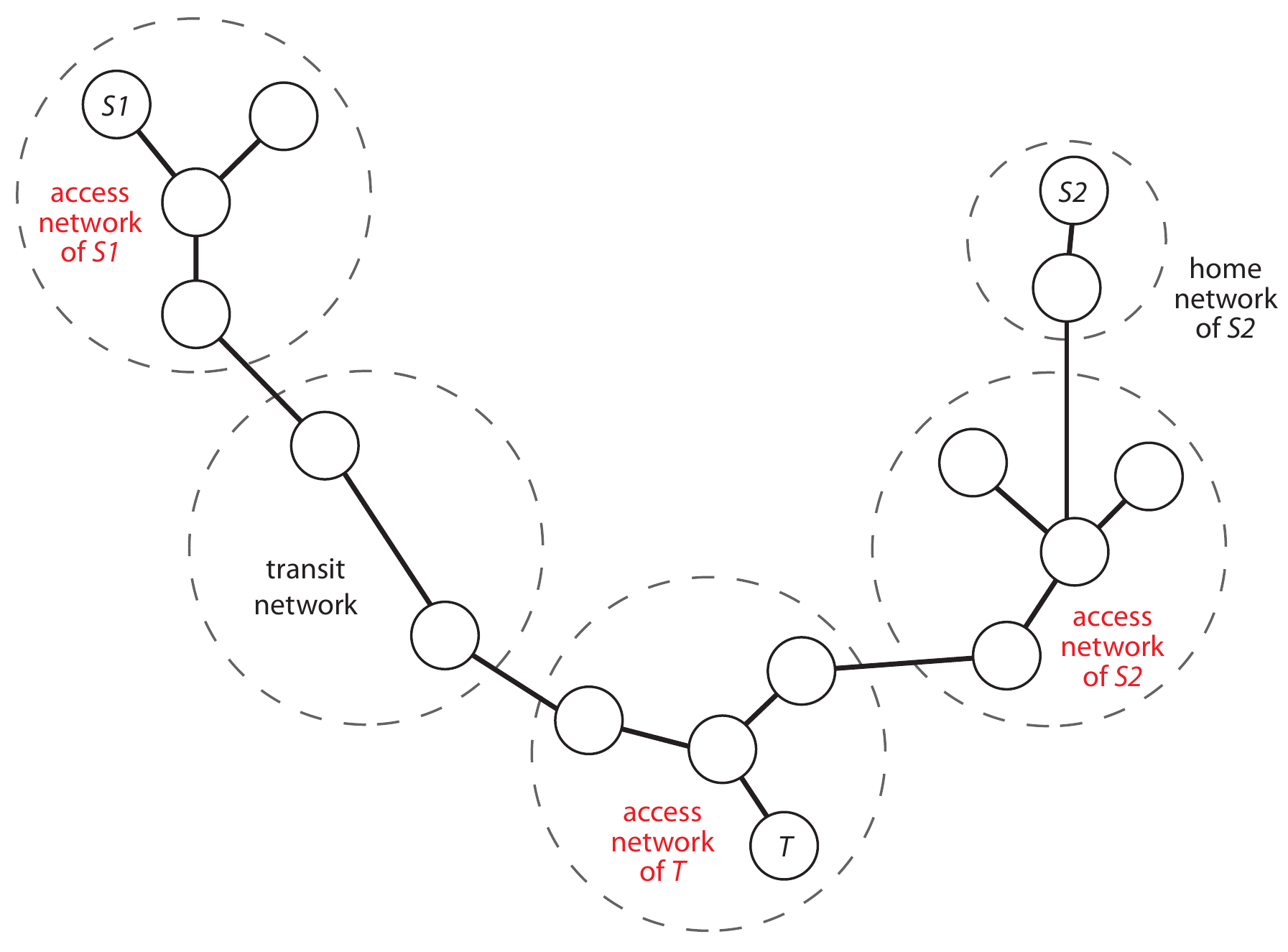}
\end{center}
\caption
{\small
{Paths from packet sources $S1$ and $S2$ to an attack victim $T$.}
\normalsize}
\label{fig:tree}
\end{figure*}

Figure~\ref{fig:tree} illustrates some relevant distinctions.
The {\it access network} of a machine is the first network in all its
outgoing paths whose administrative
authority is different from the owner of the machine
(assuming for simplicity that that there is only one
such network).
The access network of a machine is significant because it is the first
network that is able to filter outgoing packets of the machine.
Often the machine belongs to its access network, as $T$ and $S1$
do.
Sometimes, however, the first network of a machine (for example $S2$)
is a {\it home network} whose administrative authority
is the same as the owner of the machine.
In this case the machine has separate home and access networks.

The simplest traceback mechanism is the Internet Control Message Protocol
Traceback packet.
The idea is that each forwarder samples the packets it is forwarding,
with a very low sampling rate. 
When a packet is chosen,
the forwarder encloses the whole packet, along with the names of 
itself, the preceding forwarder, and the succeeding one,
in a Traceback packet, and sends it to the
packet destination.
The idea is that the victim or a nearby helper
will reconstruct whole paths and
maintain a running view of where its
packets are coming from.
In addition to cumulative overhead, the chief disadvantage of Traceback
packets is that attackers could forge them.
To prevent this, the sources of Traceback packets must be authenticated
with public-key cryptography, introducing significant additional
overhead.

In other traceback mechanisms, forwarders 
mark the packets themselves
with path information as they are forwarded toward the
victim.
The markings allow the victim or a helper near the victim to
reconstruct the path along which packets of an attack
traveled. 

The design of these traceback mechanisms entails many trade-offs.
In the remainder of this section we use representative proposals to
illustrate the issues and indicate some of the
trade-offs, without declaring any particular winners.

Internet measurements indicate that the average-length
path has 10-15 forwarders, and 20-25 is a practical maximum \cite{caida}.
It would take a large amount of space in packets to represent full
paths.
One design allocates space in packets to record only a single
forwarder (32 bits for an IP name), and uses 
probabilistic marking, in which each forwarder marks a packet with
probability $p$, say 0.04 \cite{savage}.
Because forwarders late in the path overwrite the marks
of earlier forwarders when they mark, 
the forwarders in a path can be ordered
by the frequency of their marks.
The disadvantages of simple probabilistic marking are:
\begin{itemize}
\item
A path cannot
be reconstructed until hundreds or thousands of attack packets have been
received.
\item
The attack signature may very well include attack packets from more
than one source, because of botnets and coordinated attacks.
If there are multiple attack paths to a target, then collectively
they will form a tree.
Mark
frequency is not enough information to reconstruct a tree, because
mark frequencies only result in a linear order.
\end{itemize}

The final IP Traceback
design \cite{savage} deals with the problem of multiple attack paths
in simple probabilistic marking
by encoding edges (node pairs representing links)
in the tree rather than nodes.
The resulting doubling of the space needed for marks is dealt with
by very aggressive compression techniques, primarily making each mark field
contain only a fragment of the full-size mark.
This reduces the mark field
to 16 bits, but has the effect that many more packets must be 
received before the path can be reconstructed.
Even so,
simulations show that a path of length 15 can almost always be
reconstructed after the victim or its representative has received
2500 attack packets.
The main disadvantage of this  
is that the reconstructed tree of multiple attack paths becomes
the solution to a large combinatorial puzzle.

The Pi design \cite{Pi} has every forwarder (at least, within a specified
path segment) mark the packet.
Because the mark encodes the entire path (or segment), 
the victim's helper
need only receive one marked packet to have all the information available
about the path.
Deterministic marking
is combined with very aggressive compression of the mark
field, again down to 16 bits.
Primarily, compression in this proposal means that more than one path
can result in the same mark.
The disadvantages of Pi are:
\begin{itemize}
\item
Marks do not have enough information to reconstruct paths, only to
distinguish equivalence classes of paths (all the possible paths that
happen to map to the same mark).
So marks are not helpful in locating or distributing traffic 
filters---all the filtering must be done near the victim.
\item
Marking and filtering require choosing three parameters, each
difficult to choose in general and having interactions
with many other factors,
including the ability of attackers to inject deceptive information.
\end{itemize}

In Active Internet Traffic Filtering \cite{AITF}, every packet is
marked with full names, 
but usually only by the egress forwarder at the edge of the
source's access network, and the ingress forwarder at the 
target's access network.
If a source/destination pair seems to distinguish an attack, a controller
in the target's network
will request both the target-network forwarder and
the source-network forwarder to drop such
packets (see below).  
The source-network forwarder includes a nonce in the mark,
and the controller copies it into the request, which is how
request packets are secured (as in
\S\ref{sec:Infrastructure-control-protocols}).

\subsubsection{Filtering upstream}
\label{sec:Filtering-upstream}

In the Internet, at any given time, there is a relatively small number of 
targets for active flooding attacks.
To defend a target against these attacks, traffic 
filters can conceivably be
placed in the graph of paths downstream, near the target,
or upstream, near packet sources.
There are three main advantages to placing traffic filters upstream:
\begin{itemize}
\item
If filtering is farther from the target, the damage done by attack
traffic is lessened, 
because attack traffic is carried for shorter
distances along fewer links.
Note that the damage of a flooding attack is not limited to the
intended target, because traffic to many other destinations will
also suffer because of congested links.
\item
If a traffic filter is close to sources 
of attack traffic, it may have more information about the sources.
The access network sees all of a suspected source's traffic, so attack
patterns are more likely to be detectable.
An access network may also know more about the type and reputation of
its sources (device type is relevant because some operating
systems and vendor hardware are more
easily penetrated than others).
More precise filtering means less collateral damage.
\item
Very often, attack packets are coming from a botnet, with a large
number of sources well-distributed across the public Internet.
So the total amount of available filtering resources near sources
greatly exceeds the total amount of resources available near a target.
\end{itemize}
A third option, filtering in the topological core of the network,
is never used because the core 
is a region of high-speed links and high-speed forwarders
handling large numbers of packets.
The required speed of filtering, and the potentially large number
of filtering rules to be checked, makes this option infeasible.

Pushback \cite{pushback} is a simple scheme for reducing overall
congestion by pushing filtering upstream.
At a forwarder, congestion on an outgoing link is diagnosed when
there is frequent packet loss (packets must be dropped because there
is no room for them in the link's output queue).
If a particular aggregate of packets is responsible for a significant
portion of the link's traffic, then a predicate describing the aggregate
becomes a filtering criterion.
The forwarder sends upstream, on all its input links carrying packets
in the aggregate, a request to rate-limit these packets.
Upstream forwarders can also request pushback recursively, so
pushback incorporates its own traceback mechanism.
By rate-limiting only specific aggregates along specific paths,
pushback aims to do just enough to protect other traffic, while limiting
collateral damage.

The
Active Internet Traffic Filtering proposal \cite{AITF} employs
upstream filtering because it is particularly
concerned with the botnet case, and with using the many forwarders
in the access networks of bots to help filter.
There are several ways in which its basic idea (above) must be augmented to
make it reliable and secure.
First, the request and acknowledgment packets of the control protocol
itself could be used to flood a network, so they must be rate-limited.
Second, there is a set of mechanisms through which forwarders are
monitored to see if they are keeping their filtering promises,
and through which filtering can be delegated to other forwarders
along an attack path.

\subsubsection{Capabilities}
\label{sec:Capabilities}

In the long struggle to defend public servers against flooding attacks,
researchers have explored an alternative approach based on
``capabilities'' (a capability is an unforgeable record showing the
rights of the bearer).
The idea behind capabilities is that no source should be able to send
Internet packets to a destination unless the destination has already
approved the transmission.

As applied to the protection of a public server accessed through
TCP, the TCP SYN is interpreted as a send request to the destination.
Unless the source is already on a blacklist, the server will grant
permission to send a limited number of packets in a limited period
of time, and reply to the SYN with a capability attesting to the
permission.
The sender includes the capability in subsequent packets, and forwarders
on the path enforce the capability policies.
The destination can grant 
permission for more packets later, if they are needed and the source
has been well-behaved.
Packets with no capabilities, expired capabilities,
or incorrect capabilities {\it may}
be delivered, but with the very lowest priority.

An example of this approach is the
Traffic Validation Architecture \cite{DOSlimiting}.
The architecture includes elaborate mechanisms
to ensure that capabilities
cannot be forged by attackers, and cannot be transferred
to other attackers.
It includes mechanisms to reduce the amount of space needed in packets
for capabilities, which can be considerable.
It also has mechanisms for reducing the amount of state in forwarders
required to implement the security
measures and to track packets sent and time elapsed.

The principal problem with capabilities is session requests, 
which cannot be controlled with capabilities and can be used on their own 
to create flooding attacks.
The Traffic Validation Architecture
handles this problem by rate-limiting request packets to 
5\% of the total volume.
It can be shown, however, that this just turns a flood of request
packets into a denial-of-capabilities attack, in which legitimate
senders cannot get their requests through 
\cite{uglyCapabilities}.

\subsubsection{Filtering downstream}
\label{Filtering-downstream}

The advantages of filtering upstream are
balanced by two major disadvantages:
\begin{itemize}
\item
Upstream 
networks may not have sufficient incentive to use their resources to 
protect targets that are remote from them.
It has been argued that networks under attack might 
be more willing
to accept incoming packets from cooperating upstream networks,
which will give the users of the upstream networks better service
\cite{AITF}.
Historically, however,
cooperation between the networks of different operators has been
scarce \cite{handley}.
\item
Even if source networks are willing to cooperate with target networks,
the necessary coordination is not easy.
Previous sections have illustrated many forms of overhead
and many security vulnerabilities introduced by coordination.
\end{itemize}

The proposals for moving filtering
toward the sources of attack traffic 
date from the early 2000s.
In the 2010s cloud computing advanced so far that 
it altered the evaluation of trade-offs decisively.
Now almost all traffic filtering is performed on behalf of
the access networks of
attack targets. 
It is usually performed in clouds, with virtualized filters
and dynamic scale-out.
Sequenced filters allow simple filters to deflect suspicious
packets to complex filters
for more detailed screening.

The reader might be wondering why we went through
all the detail of
\S\ref{sec:Tracing-attacks-back-to-their-sources} through
\S\ref{sec:Capabilities} if most of it is irrelevant today.
The point is that it was made irrelevant by technology changes
that altered the evaluation of trade-offs,
and technology changes in networking are not finished.
Future changes could easily make old solutions interesting again.
Here are three examples:
\begin{itemize}
\item
If the Internet evolves to offer more paths with reserved bandwidth
for real-time applications, then capabilities might be an excellent
approach to securing the use of reserved paths.
\item
Individual IP networks (under single ownership) now seem to
be growing in size and geographical scope.
If this trend continues, it will become more common for both the upstream
and downstream segments of a path to an attack target to be
controlled by the same operator.
If so, then the administrative barriers to upstream filtering will
disappear.
\item
Most traceback proposals require IP forwarders to perform new 
functions---marking and filtering packets in new ways.
Now that programmable forwarders are coming into more common use,
it will become much easier for network operators
to experiment with traceback and other
such schemes.
\end{itemize}
\else
\fi

\ifdefined\bookversion
\else 
From the viewpoint of a victim of a flooding attack, the network is a
tree with many possible packet sources
and a single packet sink at the root.
From the viewpoint of a packet source, its
{\it access network} is the first network in all its
outgoing paths whose administrative
authority is different from the owner of the machine
(assuming for simplicity that that there is only one
such network).
The access network of a machine is significant because it is the first
network that is able to filter outgoing packets of the machine.
Often the machine belongs to its access network, but the machine
might belong to a home network, which carries its packets to its
access network.

Path-based traffic filtering augments filtering criteria based on
the contents of packets, as discussed in the previous section, with
criteria based on the path along which packets traveled.
There are two reasons for introducing path-based filtering.
The first reason is that path information can improve
the precision of filtering criteria, so that fewer good packets are
accidentally included.
For example, say that the overall load on a server suggests a flooding
attack, and intrusion detection proposes a candidate filtering rule
based on packet contents.
If we know that most paths to the server are delivering a trickle of
these packets, and one path's load is dominated by these packets,
there is a good chance that only the packets on the dominated path
are attack packets.
The second reason is that---with knowledge of where packets came
from---traffic filtering can be moved from its usual downstream
location, near the target, to upstream locations close to packet sources.
Upstream filtering has
three main advantages:
\begin{itemize}
\item
If filtering is farther from the target, the damage done by attack
traffic is lessened, 
because attack traffic is carried for shorter
distances along fewer links.
Note that the damage of a flooding attack is not limited to the
intended target, because traffic to many other destinations will
also suffer because of congested links.
\item
If a traffic filter is close to sources 
of attack traffic, it may have more information about the sources.
The access network sees all of a suspected source's traffic, so attack
patterns are more likely to be detectable.
An access network may also know more about the type and reputation of
its sources (device type is relevant because some operating
systems and vendor hardware are more
easily penetrated than others).
More precise filtering means less collateral damage.
\item
Very often, attack packets are coming from a botnet, with a large
number of sources well-distributed across the public Internet.
So the total amount of available filtering resources near sources
greatly exceeds the total amount of resources available near a target.
\end{itemize}

The principal disadvantage of upstream filtering is that in IP networks,
the source name in a packet
is not a reliable indicator of where it came
from.
Forwarders can attach meta-data to packets so that servers near targets
can reconstruct packet paths, but this is problematic because of the
volume of data involved and the danger of adversarial interference.
All the proposals for ``traceback'' of this kind must make difficult
trade-offs to balance the costs and benefits
\cite{savage,Pi,AITF,pushback}. 
In addition, upstream filtering has two other disadvantages:
\begin{itemize}
\item
Upstream 
networks may not have sufficient incentive to use their resources to 
protect targets that are remote from them.
It has been argued that networks under attack might 
be more willing
to accept incoming packets from cooperating upstream networks,
which will give the users of the upstream networks better service
\cite{AITF}.
Historically, however,
cooperation between the networks of different operators has been
scarce \cite{handley}.
\item
Even if source networks are willing to cooperate with target networks,
the necessary coordination is not easy.
Like traceback, coordination along packet paths invites its own
security attacks.
\end{itemize}

Currently, the net effect of all these factors is that path-based
traffic filtering is uncommon in the Internet.
However, future changes might cause the factors to be weighted
differently.
For example,
individual IP networks (under single ownership) now seem to
be growing in size and geographical scope.
If this trend continues, it will become more common for both the upstream
and downstream segments of a path to an attack target to be
controlled by the same operator.
If so, then the administrative barriers to upstream filtering will
disappear.
\fi

\subsection{Interactions between traffic filtering and other aspects
of networking}
\label{sec:Interactions5}

\subsubsection{Routing}
\label{sec:Routing}

\ifdefined\bookversion 
For a filtering tree or graph (as in Figure~\ref{fig:tree}) to work
\else 
For a filtering tree or graph to work
\fi
correctly,
all packets destined for the protected target must pass through
one or more forwarders or middleboxes acting as filters,
in accordance with the intended design.
This is the province of routing, which populates the forwarding tables
used by forwarders.
Routing is performed in several different ways---sometimes by a
distributed algorithm that forwarders run among themselves, and 
sometimes by a centralized algorithm running in a separate controller.
 
Routing packets through a filtering tree may seem straightforward,
but there is a different tree for each destination, and routing
algorithms are also concerned with reachability,
performance, fault-tolerance,
and other policy constraints.
For this reason, there has been considerable research on 
verifying that forwarding tables are correct, or on generating
them correctly, where the correctness criteria include 
\ifdefined\bookversion 
``waypointing'' constraints about steering packets through filters
\cite{netconfig,propane,fogel,headerspace2,beliefs}.
\else 
constraints about steering packets through filters
\cite{netconfig,fogel,headerspace2,beliefs}.
\fi

Another issue that complicates routing through a filtering tree
is the fact that many traffic filters require session affinity---all
the packets of a session, in both directions,
must go through the same filter.
Wide-area routing frequently creates different paths for packets
traveling in different directions between the same two endpoints.
Even packets traveling in the same direction may be spread across
multiple paths because there has been a failure in one of the paths,
or a need for better load-balancing.
Within a cloud, 
where many virtual machines are running the same filtering software
for scalability, a session can be assigned to any virtual machine.
The assignment must be remembered, however, so that all packets
of the session are steered in the right direction.
Shortcuts such as ``assign a session to one of four virtual machines
based on the last two bits of some identifier'' work well in static
situations, but fail when filtering resources must be scaled up or
down because of fluctuations in load.

\subsubsection{Layering}
\label{sec:Layering5}

Almost always, a packet arriving at a machine is being
transmitted through
multiple layered networks simultaneously, for example an Ethernet
local area network,
an IP network, and an application network.
\ifdefined\bookversion 
Figures~\ref{fig:layering} and 6 
\else 
Figures~\ref{fig:layering} and 4 
\fi
combined illustrate this simple example.
In addition,
layered between the application network and the lowest IP network there is
often a virtual private network 
(\S\ref{sec:Virtual-private-networks}) or other IP network.
If the machine is actually a virtual machine in a multi-tenant cloud,
there is sure to be at least one network between the tenant's IP network
and the Ethernet, 
with the job of sharing cloud resources among all tenants.

The layers are significant because attacks can take place in any
of them.
This is both a challenge and an opportunity.
If only packets in the lower layers are filtered, then many higher-level
attacks will be concealed in the higher-level packets, which are
mere data to the lower-level networks.
For example,
recall that IP intrusion detection systems look into packet
contents for signs of malware at specific locations.
These systems are assuming there are no networks layered between
the filtering network and the application network; if
there are additional networks, then the packet formats will be
different, and the filtering criteria will be useless.

On the other hand, much can be gained by filtering packets 
separately in each network.
Already there are special filters for Web requests and email messages,
which are packets in their own application-oriented networks.
These filters are deployed as middleboxes in these networks.
This idea can be extended to intermediary layers, where each filter
is attuned to the configuration, protocols, and vulnerabilities
of its particular network.
It is often possible to optimize architectures so that filters
at multiple levels can be located on the same machines.

The second interaction between filtering and layering
concerns networks below the filtering network
in the layer hierarchy.
Imagine that you have designed a filtering mechanism within a network,
and convinced yourself that it is correct.
Your argument concerns (among other things) paths in the network to a
potential attack target, and shows that routing places
an appropriate traffic filter in every path.
\ifdefined\bookversion 

\else 
\fi
Whether you remembered to state it or not, your argument that no (or a
limited number of) attack packets reach the target depends on the
assumption that attack packets do not suddenly appear
{\it inside} the perimeter of traffic filters. 
It is easy enough to check that the network members inside the
perimeter are part of the network infrastructure and therefore trusted,
but what about the links?
It must be ensured that no packet is received on trusted link that was not
sent on the link.
If the link is implemented rather than physical, it must be proved
that the implementing network does not inject packets into the
implementation of the link.
\ifdefined\bookversion 

\else 
\fi
This might seem like a fanciful concern, but it is not.
An Ethernet can be penetrated physically, and 
it is easy to make a penetrated Ethernet inject packets
into the links of networks layered on it \cite{kiravuo}.
In a multi-tenant cloud, 
the links of a tenant's network (where the filtering will 
take place) are virtual links implemented by sessions
in the lower-level network
that shares resources among tenants.
If the cloud network 
does not isolate tenants properly, then packets sent by
virtual machines of a different
tenant could be delivered as part of this tenant's sessions.

\subsubsection{Cryptographic protocols}
\label{sec:Cryptographic-protocols5}

There is a profound interaction between cryptographic protocols and
traffic filtering.
If a user session is encrypted, then middleboxes in general, and
traffic filters in particular, cannot read anything
in the session packets beyond their headers.
One response to this interaction is increased interest in filtering
based on traffic attributes that encryption does not change, e.g.,
packet timing and sizes.
It may be too early to tell how effective this will be as a defense,
considering that its recognized successes are spying and tampering
attacks \cite{trafficanalysis}.

\begin{figure*}[hbt]
\begin{center}
\ifdefined\bookversion
\includegraphics[scale=0.55]{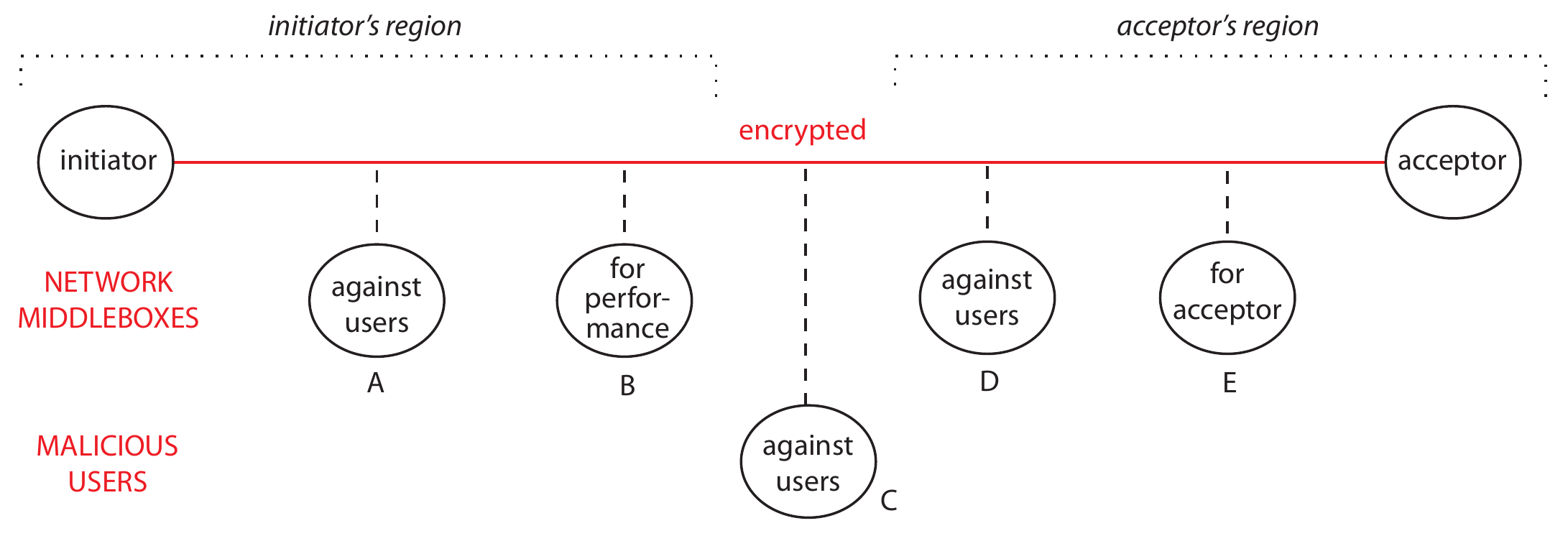}
\else
\includegraphics[scale=0.50]{fig/game.pdf}
\fi
\end{center}
\caption
{\small
{A game: cryptographic protocols {\it versus} traffic filtering.}
\normalsize}
\label{fig:game}
\end{figure*}

In some cases the relationship between users and filters is not
adversarial, and there are three techniques for managing
this interaction in more-or-less cooperative cases.
Before presenting these techniques, we will explain the interested
parties and their powers.
We can think of their interactions as a game, one instance of which
is illustrated by
Figure~\ref{fig:game}.
At the top of the figure we see what the initiating user can do.
It chooses the acceptor of the session, and if the acceptor agrees,
the data of the session will be encrypted end-to-end.

At the second level of the figure, we see what the network can do.
The network has the power to insert middleboxes anywhere in the path
of the user session, simply by routing session packets through them.
The figure shows some common middleboxes, inserted in likely places,
which are often in regions of the session path near the two endpoints.
A middlebox might have the purpose of enhancing performance, for
example by caching or compression ($B$).
For maximum effectiveness, it should be placed near the initiator, as
shown.
A middlebox might be a traffic filter, with the purpose of protecting the 
acceptor from
subversion attacks or policy violations that might damage it ($E$).
This middlebox will probably be placed near the acceptor.
Finally, the network might insert traffic filters that are
working against the interests of the initiator and acceptor,
either by preventing them from violating policies, or by spying on
or tampering with their communication ($A$ and $D$).
These middleboxes might be placed in either region.

At the third level of the figure, we see that other malicious parties
can also insert middleboxes in the path by various techniques such as
wiretapping, for the purposes of spying and tampering ($C$).
Fortunately, physical security and security mechanisms in other networks
constrain such attacks. 
In the illustrated example, 
a malicious party is able to eavesdrop in the middle of the session path,
but not near the endpoints.

If network middleboxes are working on behalf of the user endpoints
of an encrypted session,
and if they need to read data to do their work,
then the cleanest arrangement is to make the middleboxes part of an
application-oriented overlay network.
This is illustrated by a SIP network in Figure~\ref{fig:newsip}.
In the figure, data traveling on the links of the SIP network is
encrypted by TLS in the IP networks, but each middlebox in the SIP
network receives and sends plaintext.

\ifdefined\bookversion 
The second and third techniques can be deployed within the bridged
IP networks of the Internet.
In the second technique, the network introduces another
middlebox that is a proxy. 
The proxy would accept the initiator's TLS session and make a TCP
session between itself and the original acceptor.
The proxy would decrypt packets from the initiator and send their
contents in plaintext packets to the acceptor, so they could be read
by any middleboxes in the path of the TCP session.

If the acceptor is a public server, the limitation on this technique
is that the acceptor must trust and approve of the proxy, so that the
server is willing to lend the proxy its identity and secret keys.
So the only place in Figure~\ref{fig:game} that a proxy could
plausibly be inserted is between $D$ and $E$.

The general idea of proxies that cooperate with endpoints is
developed further in 
Middlebox TLS (mbTLS) \cite{mbTLS}.
In this approach 
all middleboxes must be proxies, and
they create a session consisting of
an end-to-end chain of simple TCP sessions.
Along the chain from initiator to acceptor, there is first a set
of middleboxes inserted on behalf of the initiator, followed by a set
inserted on behalf of the acceptor (see Figure~\ref{fig:mbtls}).

\begin{figure*}[hbt]
\begin{center}
\includegraphics[scale=0.60]{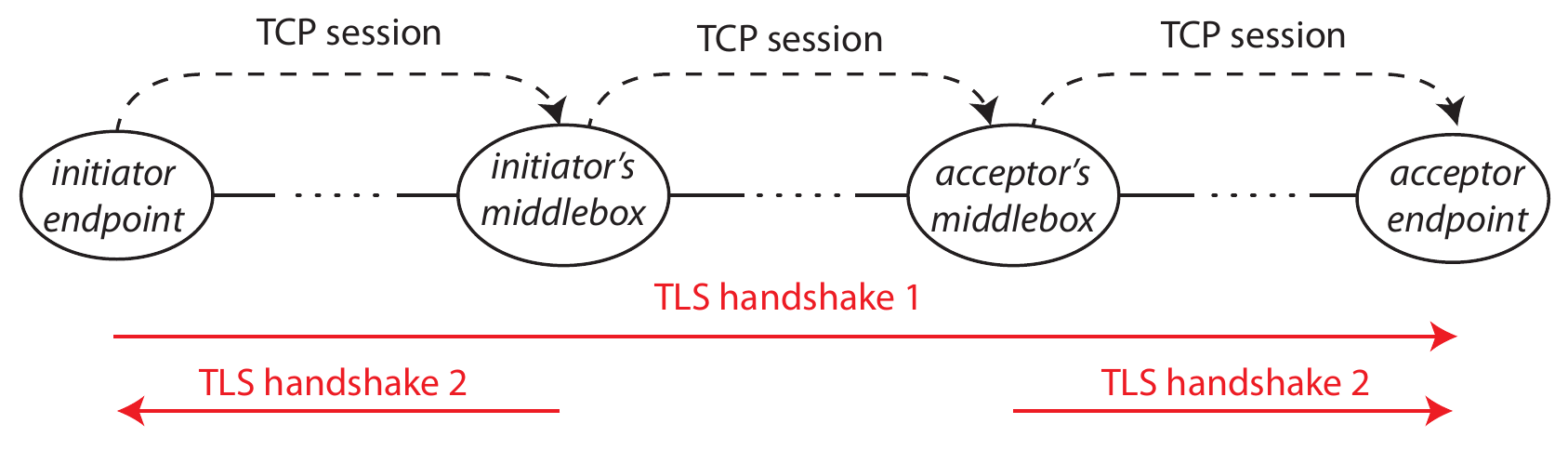}
\end{center}
\caption
{\small
{Control signaling to set up an mbTLS session.}
\normalsize}
\label{fig:mbtls}
\end{figure*}

Within the end-to-end chain of TCP sessions, the initiator and acceptor
first have a normal end-to-end
TLS handshake for endpoint authentication and key
exchange.
Then each middlebox initiates a secondary TLS handshake with the next
element in the direction of its sponsoring endpoint.
For example, if there are two middleboxes {\it M1} and {\it M2} inserted
on behalf of the initiator, {\it M2} initiates a secondary handshake
with {\it M1}, and {\it M1} with the initiator.
The secondary handshakes exchange symmetric and authentication keys for
the individual simple sessions.
They can also perform endpoint authentication of 
middlebox identity (in this case
the responsible owner), software version and configuration, 
security properties of the hardware/software platform, {\it etc.}
This makes sense because the middleboxes associated with each endpoint
are working in cooperation with it, even if they have an adversarial
relationship with the middleboxes of the other endpoint.

After the secondary TLS handshakes, data is transmitted.
In each middlebox data is decrypted, processed as plaintext by
the middlebox application code, then encrypted again for the next 
hop.\footnote{An 
earlier version, Multi-Context TLS \cite{mcTLS}, allowed the endpoints
to place constraints on the read/write access of middleboxes.}
Note that the ``midpoint'' simple session between initiator and
acceptor middleboxes has no secondary handshake;
in this simple session, the keys chosen by the primary TLS handshake
are used.

\else 
The second and third techniques can be deployed within the bridged
IP networks of the Internet.
In the second technique, the network introduces a {\it proxy}---a
middlebox that is a session-protocol endpoint---on behalf of the
acceptor.
The proxy accepts the initiator's TLS session (so it must have
the server's identity and secret keys) and makes a TCP
session between itself and the original acceptor.
The proxy decrypts packets from the initiator and sends their
contents in plaintext packets to the acceptor, so they can be read
by any middleboxes in the path of the TCP session.
In Figure~\ref{fig:game}, the acceptor would prefer a proxy
between $D$ and $E$, and the network would prefer it before $D$.
The general idea of proxies that cooperate with endpoints is
developed further in 
Middlebox TLS (mbTLS) \cite{mbTLS}.
\fi

The third and final technique aims to preserve both middlebox
functionality and user privacy, based on new results in cryptography.
At one extreme,
fully homomorphic encryption \cite{homoencryption} makes it possible
to compute
any function on encrypted data without learning more about the data
than the function's value.
Although fully homomorphic encryption is currently impractical (it is too 
expensive computationally, by orders of magnitude), there are less
capable algorithms for computing functions on encrypted data with
\ifdefined\bookversion 
performance that may be feasible for current use.
\else 
performance that may be feasible for current use \cite{blindbox}.
\fi

\ifdefined\bookversion 
BlindBox \cite{blindbox} is a proposal for allowing middleboxes to
operate on encrypted data.
BlindBox middleboxes can apply detection rules of the
kind commonly used by virus scanners, intrusion-detection systems, and
parental filters.
The scheme also allows a middlebox that has found a keyword or other
suspicious string, as probable cause of a security violation, to
decrypt the entire packet.

BlindBox is implemented as an extension of TLS.
In addition to the basic TLS handshake, endpoints must generate extra
keys.
The data must be sent end-to-end twice (redundantly),
once in the ordinary TLS form and once
in a reformatted and re-encrypted form suitable for the BlindBox
algorithms.

The biggest overhead incurred by BlindBox is due to rule
preparation, because the
middlebox must have the rules themselves encrypted with a 
session-dependent key.
The endpoints must not know the rules (this would make them easier
to evade) and the middlebox must not know the key (otherwise the 
guarantee of data confidentiality would be lost).
So who can encrypt the rules?
For every keyword in every rule, both endpoints must generate and
transmit to the middlebox a special encryption function that
incorporates yet obfuscates the session-dependent key.
The middlebox
must first check the two for agreement (in case one of the endpoints
is insecure) and then apply the encryption function to the rule.
This results in very high performance overhead, which means that BlindBox
is currently practical only for long-lived sessions or small rule sets.

Although both Middlebox TLS and BlindBox are promising efforts,
it seems clear that their complexity and non-uniform communication
among participants are weaknesses.
Complexity itself is a security vulnerability, because it provides
a larger ``attack surface'' for adversaries to probe.
\else 
\fi

\ifdefined\bookversion 
\subsubsection{Session protocols}
\label{sec:Session-protocols5}

There is a general problem affecting all protocols of the IP protocol
suite:  When a new network feature requires additional information in
packets, where can the information be put?
Inventors do not wish to increase packet size, and want their
proposals to have backward compatibility.
So they generally choose to fit their information into some
``unused'' field in current header formats.
For example, both IP Traceback \cite{savage} and Pi \cite{Pi}
squeeze traceback information into the 16-bit identification field
in the IP packet header.
The original use of this field is to group fragments of a fragmented
packet so they can be re-assembled at their destination.
It is declared ``unused'' on the grounds that fragmentation of IP
packets is now rare.

The irony of such proposals is that so many new features and
protocol variations use the same ``unused'' fields.
At the same time, inventors of other new functions have no
compunction in deleting this ``unused'' information from packets
when it is convenient.
For example, a server using
SYN cookies 
(\S\ref{sec:Dynamic-resource-allocation})
effectively drops all optional information 
in TCP SYN packets, which means that any network feature
relying on extensions to TCP is disabled.
Note that many servers using SYN cookies are Web servers, and many
new features relying on extensions
to TCP (such as Multipath TCP \cite{raiciu}, just to name one example) 
have improved Web access as a major use case.

A possible solution to such problems is a generalized
mechanism for composition of session protocols.
This mechanism would build on the structures of 
Figure~\ref{fig:composed-packet}.
If all session protocols could be composed freely, then all new
features requiring space in packets could be introduced---without fear
of interference---as
new, composable session protocols.
For instance, just as TLS is composed with (embedded in) TCP, 
TCP could be composed with (embedded in) Multipath TCP.
TCP would control transmission of byte streams along individual
paths, while Multipath TCP would coordinate the multiple byte streams.
Unfortunately this solution may not be backward-compatible, and also
raises concerns about generating packets
that exceed the maximum packet size of their network.
\else 
\fi

\section{Dynamic resource allocation}
\label{sec:Dynamic-resource-allocation}

Because flooding attacks are resource wars, both network infrastructure
and user members can defend against them by allocating more resources
when they are under attack.
Cloud computing has made it easier for networks to 
scale out traffic filters, and for users to scale out servers.
Even in server-centric defenses the network is usually involved,
for two reasons:
(i) even when server resources are sufficient to absorb an attack,
network bandwidth must also be sufficient to handle the attack, and
(ii) the network provides the service of distributing the load
across servers. 

\begin{figure*}[hbt]
\begin{center}
\ifdefined\bookversion
\includegraphics[scale=0.60]{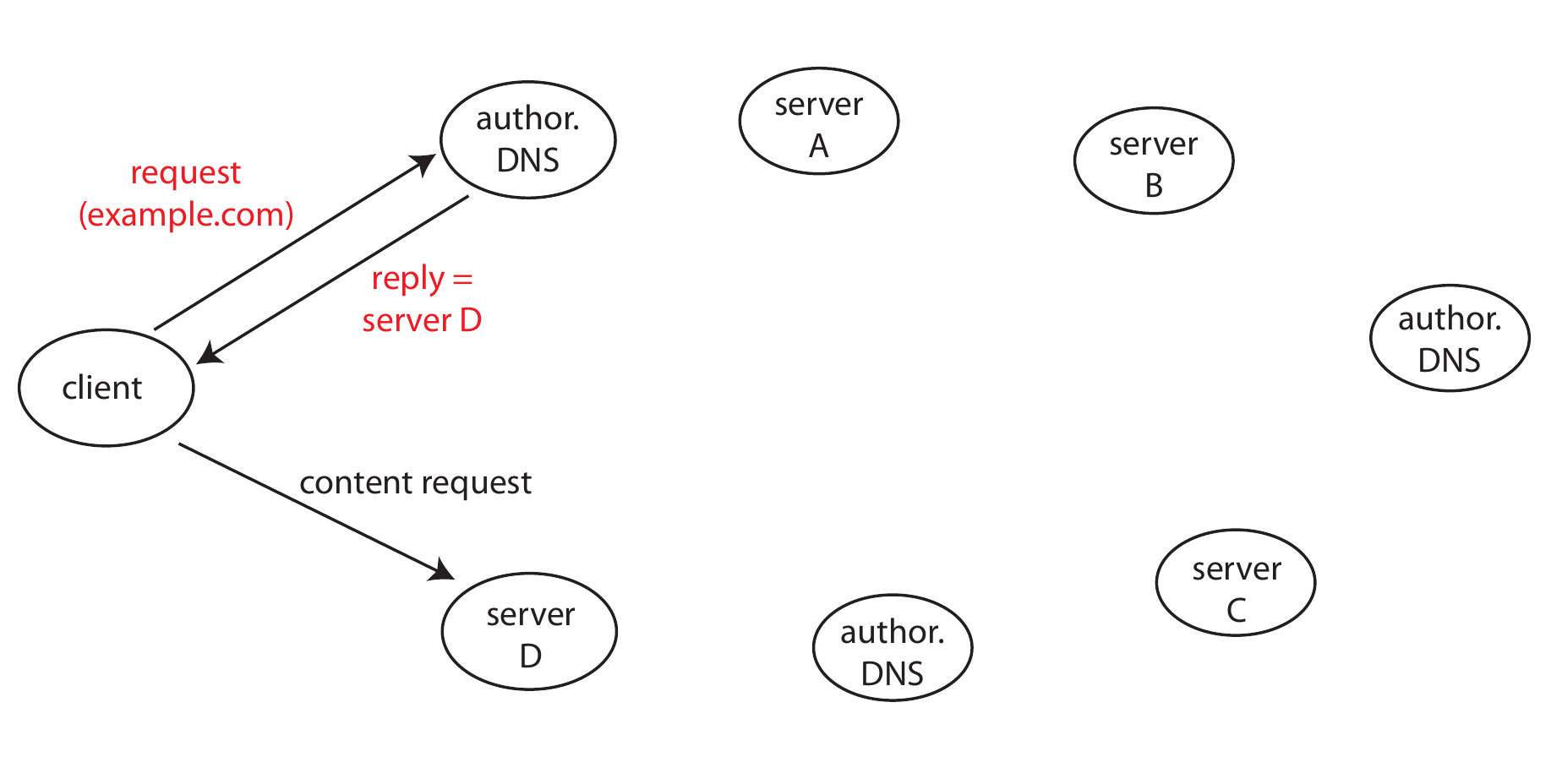}
\else
\includegraphics[scale=0.50]{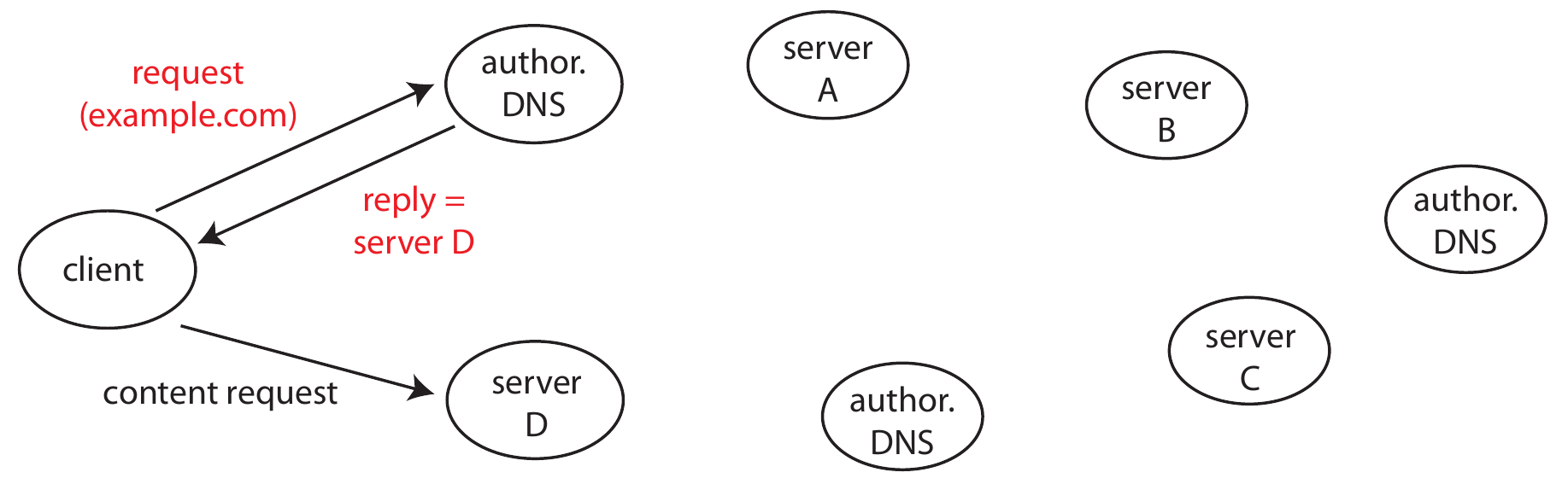}
\fi
\end{center}
\caption
{\small
{Resource replication in a content-delivery network.
\ifdefined\bookversion 
All member labels symbolize IP names, so the three DNS servers have 
the same name.
\else
\fi}
\normalsize}
\label{fig:zakamai}
\end{figure*}

Dynamic resource allocation works better if resource replicas
are geographically distributed, so that some replicas can be reached
when other parts of the network are too congested.
Because attacks on DNS servers are so common and damaging,
it is especially important to have distributed
authoritative DNS servers for popular domain names.
Queries are distributed across the replicas by means of 
IP anycast.
If there are five replicas sharing the load and one has been
overwhelmed by an attack, IP anycast may not be dynamic enough to
redirect queries away from
the failed replica, but at least queries directed by anycast
to the other four will succeed.
In Figure~\ref{fig:zakamai} there are three authoritative DNS servers
for the domain {\it example.com}; 
IP anycast directs the client's query to the closest one.

A ``content-delivery network'' provides many replicas of its customers'
content, geographically distributed so that the latency of
content delivery to each client is minimized.
In Figure~\ref{fig:zakamai} the authoritative DNS servers
for customer domain {\it example.com} are aware that its content is
available at servers A through D, and also maintain information about
location and recent performance of the servers.
So each DNS server can return to a client the IP name of the best
content server for it to contact.

Replication of service resources
is easiest when servers are responding to queries based on
fairly static data.
When queries can update service data, the service implementation
must do extra work to keep the data replicas in some adequate state
of consistency.
(The study of distributed computing has produced many algorithms
for replicated data, satisfying many different definitions of
consistency.)
In some cases dynamic data can be
distributed across multiple sites more easily by sharding, e.g., by
partitioning the keys of a key-value store so that each site is
responsible for a subset of the keys.
No one key-value pair will be replicated, but the total resources
available will be greater.

Instead of dynamically allocating more server resources during attacks,
the same result can be achieved by dynamically reducing the work
per request that servers perform.
For example,
a flood of DNS queries is amplified when servers query other servers.
A very effective defense against these attacks
is longer times-to-live for cache entries,
perhaps 30 minutes,
in recursive and local DNS servers \cite{dns-ddos}.
If local entries are cached longer, there will be fewer queries and
retries made to authoritative servers.
There are many good reasons for DNS cache entries with short times-to-live,
but these can be changed as an adaptive measure during attacks.
\ifdefined\bookversion 
The same idea would work for many other services with caching.
\else 
\fi

SYN floods (\S\ref{sec:Flooding-attacks}) are such a serious problem
that several specialized techniques have been developed for reducing
server work per SYN, and these may be in use at all times rather than
turned on just during attacks.
In a ``SYN cookie'' defense,
the server responds to a SYN with a SYN+ACK packet
having a specially-coded initial sequence number (the cookie).
It then discards the SYN, using no additional resources for it.
If the SYN was an attack, it has caused little damage.
If the SYN was legitimate, on the other hand, it will elicit an ACK
from the initiator with the same initial sequence number incremented by 
one.
By decoding the sequence number, the server can reconstruct the original
SYN and then set up a real TCP connection.

\ifdefined\bookversion 
There is another resource defense against SYN floods that is less efficient
than SYN cookies, but comes with fewer side-effects
(see \S\ref{sec:Session-protocols5}).
This defense uses a middlebox in the path to a Web server that stores
and responds to SYN packets, but does nothing else with a
SYN packet until it receives the ACK that completes the handshake.
On receiving the ACK, the middlebox
forms a new session by sending the SYN to the server, and subsequently
acts as a transparent forwarder between the two sessions.
If the middlebox
does not receive a timely ACK, then the SYN packet was part 
of an attack or the client has failed, so the middlebox drops it.
\else 
\fi

\ifdefined\bookversion 
Viewed globally, dynamic resource allocation is not much different from
static allocation, so its interactions with other aspects of
networking are already understood.
As indicated, services under attack may be available in some locations
and not in others.
\else 
Dynamic resource allocation is not much different from
static resource allocation, so its interactions with other aspects of
networking are already understood.
Services under attack may be available in some places
and not in others.
\fi

\section{Compound sessions and overlays for security}
\label{sec:Compound-sessions-and-overlays-for-security}

\ifdefined\bookversion 
Like cryptographic protocols, compound session and overlays are
mechanisms employed by users to defend themselves against spying
and tampering attacks
(\S\ref{sec:Spying-and-tampering}).
Cryptography alone is not sufficient because it does not conceal 
packet headers.

The focus of this section is on middleboxes.
Traffic filters are middleboxes, inserted by networks into session paths
by means of routing.
In \S\ref{sec:Compound-sessions} we saw that users can also
insert middleboxes into their session paths, by introducing
{\it proxies} that form the endpoints of {\it simple sessions}
in a {\it compound session}.
Packet headers within simple sessions must be correct so that packets
are delivered, but the end-to-end identity of
a compound session is concealed by its multiple headers.

Proxies have a lot of power: they can perform many kinds of computation
on packets, so that 
packets in two associated simple sessions need not correspond
one-to-one.
They can even translate from one session protocol to another, so that
two associated simple sessions use different protocols.
A proxy can get the name of the next
proxy or endpoint in the compound session
by using the session protocol to engage in a
dialogue with the initiator.

Spying and tampering attacks can be launched by other user members
of a network (authorized or unauthorized), and, most importantly,
by infrastructure members of a network performing traffic filtering.
Thus the entire topic of this section can be seen as an interaction
between two patterns, namely compound sessions and traffic filtering.

Compound sessions are useful in many situations, but they have some
limitations.
After covering compound sessions, we will introduce overlays for
security.
These use explicit layering to create implicit compound sessions,
and can do more for users than compound sessions alone.

\else 
Like cryptographic protocols, compound session and overlays are
employed by users to defend themselves against spying
and tampering attacks
(\S\ref{sec:Spying-and-tampering}).
Cryptography is not sufficient because it does not conceal 
packet headers.

Compound sessions and overlays are mechanisms through which users
can insert their own middleboxes into session paths, and use them to
conceal header information.
Thus the entire topic of this section can be seen as an interaction
between two patterns, namely compound sessions/overlays
and traffic filtering.
\fi

\subsection{Compound sessions}
\label{sec:Compound-sessions7}

\ifdefined\bookversion 
\else 
\subsubsection{Definition of compound sessions}
\label{sec:Definition-of-compound-sessions}

A user member initiating a session to some far endpoint can insert
another user member into the session path as a middlebox.
To do this, the initiating user must give the name of the middlebox
as the destination name of its outgoing packets, as shown in
Figure~\ref{fig:appGW}.
The middlebox must learn the initiator's intended far endpoint, 
for example by getting it from some other field of the session-initiation
packet.
Then the middlebox changes the headers of the packets it receives
(source becomes its own name, destination becomes the
initiator's intended) and sends them out.
Recall that a middlebox that behaves in this way is called a {\it proxy}.
Each proxy accepts a session, initiates another
session with a different header, remembers the association
between the two sessions, and relays packets between them.
A {\it compound session} is a chain
of {\it simple sessions} composed
by proxies in this way.

A compound session can have more than one proxy:
the session-initiation packet can contain a list of proxies to
visit, or a
proxy can get the name of the next
proxy or endpoint
by using the session protocol to engage in a
dialogue with the initiator.
Because of the names in forward packet headers, 
return packets naturally pass
through the same proxies in reverse order,
and have their headers re-translated in reverse order.

The principal
security significance of compound sessions is that each simple
session has a different header, so compound sessions can be employed
by users to obscure header information.
In Figure~\ref{fig:appGW}, an observer between $V$ and {\it GW} cannot
observe the true acceptor of the compound session, at least from
packet headers alone,
and an observer between {\it GW} and $P$ cannot observe the true initiator
of the compound session.

Compound sessions are useful in many situations, but they have some
limitations.
After covering compound sessions, we will introduce overlays for
security.
These use explicit layering to create implicit compound sessions,
and can do more for users than compound sessions alone.
\fi

\subsubsection{Proxies in access networks}
\label{sec:Proxies-in-access-networks}

Perhaps the oldest example of a proxy for evading
traffic filtering is an ``application
gateway,'' which is installed in a private IP network for the benign
purpose of evading the too-simple filtering imposed by a firewall.
For example, an enterprise firewall may block all outgoing sessions
except Web accesses.
However, the enterprise may also wish to allow outgoing sessions
of another kind, when they are initiated by specific users.
The firewall cannot enforce this policy because it does not know the
\ifdefined\bookversion 
mapping between internal IP names and users 
(and the mapping may not even be static).
\else 
mapping between internal IP names and users.
\fi

\begin{figure*}[hbt]
\begin{center}
\ifdefined\bookversion
\includegraphics[scale=0.60]{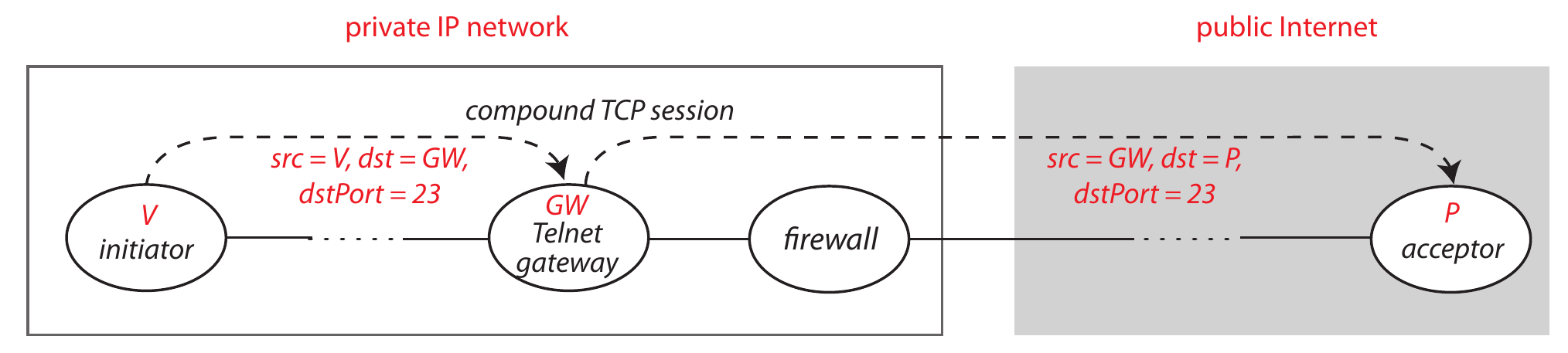}
\else
\includegraphics[scale=0.50]{fig/appGW.pdf}
\fi
\end{center}
\caption
{\small
{A Telnet application gateway inside the access network of $V$.}
\normalsize}
\label{fig:appGW}
\end{figure*}

An application gateway for the application, for instance Telnet,
solves this problem, as shown in Figure~\ref{fig:appGW}.
To use it, a user initiates a Telnet session to the application gateway
inside the enterprise network.
The gateway is a Telnet proxy.
By means of an extension to the Telnet protocol, which is
embedded in TCP, the user supplies
a password to authenticate
himself to the gateway, and also the name of the real Telnet acceptor.
The gateway initiates a Telnet session to the real acceptor outside
the enterprise network, and joins the two simple sessions in a compound
session.
The enterprise firewall allows outgoing Telnet sessions from the 
application gateway only. 

In this example, the operator
of the enterprise network is cooperating with the
user by providing the gateway.
For the operator, it is easier and more efficient
to provide the required user functions
with an application gateway than with a greatly-enhanced firewall.

\ifdefined\bookversion 
In a similarly cooperative situation, the operator of a private
network might provide a proxy (with a public name) that session
initiators outside the network can connect to.
The proxy authenticates the initiator as deserving the
rights of members of the private network.
Then, through the proxy, the initiator can connect to any member
of the private network.
\else 
In another cooperative situation, the operator of a private
network might provide a proxy (with a public name) that
initiators outside the network can connect to.
The proxy authenticates the initiator as deserving the
rights of members of the private network.
Then, through the proxy, the initiator can connect to any member
of the private network.
\fi

\subsubsection{Proxies in transit networks}
\label{sec:Proxies-in-transit-networks}

A user can evade filtering in his access network more easily 
by connecting to a friendly proxy in another network.
This will be
illustrated by the use of a proxy to reach a Web server.
\ifdefined\bookversion 
This kind of proxy
is sometimes called a ``virtual private network.''\footnote{Calling
a proxy a network is a misnomer.  See \S\ref{sec:Virtual-private-networks}
for the real thing.}
\else 
\fi

\begin{figure*}[hbt]
\begin{center}
\ifdefined\bookversion
\includegraphics[scale=0.60]{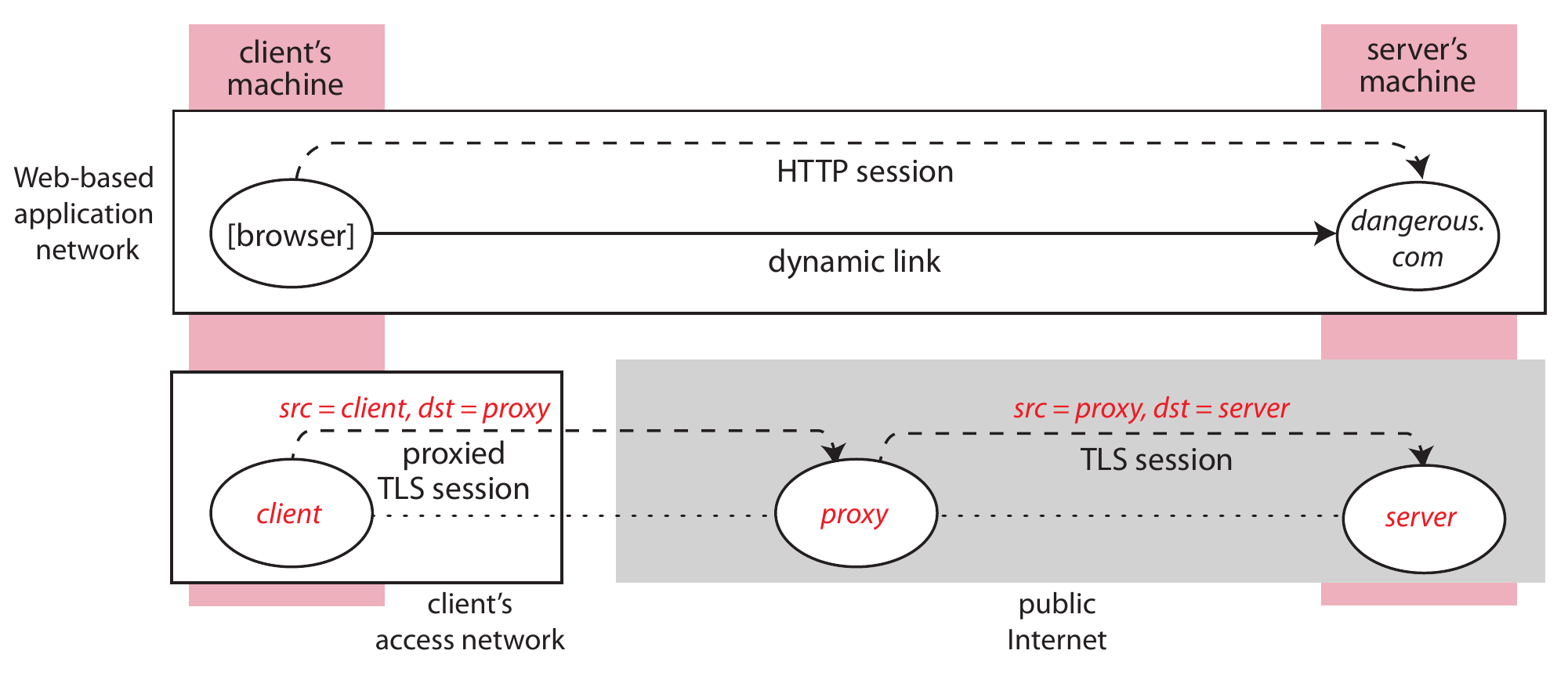}
\else
\includegraphics[scale=0.50]{fig/proxy1.pdf}
\fi
\end{center}
\caption
{\small
{A proxied TLS session protects the client's privacy in his
access network, and provides anonymity at the Web server.}
\normalsize}
\label{fig:proxy1}
\end{figure*}

In Figure~\ref{fig:proxy1} a secure dynamic link in a Web-based 
application network is implemented by a compound TLS session in
bridged IP networks.
First the browser's request causes initiation of a TLS session
with a friendly proxy outside the client's access network.
A proxied TLS session is like a normal TLS session except that:
(i) instead of looking up the domain name {\it dangerous.com} and
using its IP name as the destination of the session, the client's IP
interface uses the
proxy's IP name as the destination of the session;
(ii) the client's IP interface expects and 
verifies the certificate of the proxy, not the Web site;
(iii)
the proxy decrypts the HTTP request in the TLS data, looks up the
domain name, and uses the result of the lookup as the destination of an
outgoing TLS session.
After this the proxy relays packets between the two simple sessions
of the compound TLS session (note that the proxy must decrypt and
re-encrypt the data in each packet, because symmetric keys in the
two simple sessions are different).

Because of the compound session formed by the proxy, 
the client's access network does not know what server the client is
connected to, so the client has privacy from spying and tampering
in his access network.
The client also has anonymity at the server, because the server has
no information about the client.

One disadvantage of this mechanism is that the client has no
privacy from the proxy.
Another disadvantage comes from the fact that the names of
helpful proxies are usually publicly available (so users can find them),
which means that they are available to the user's adversaries as well.
Consequently, if the clients's access network is censoring
the network activity of its users, it can simply block packets
destined for external proxies.
These disadvantages are addressed in subsequent sections.

\ifdefined\bookversion 
The proxy in Figure~\ref{fig:proxy1} is specialized to handle TLS
sessions.
There can be proxies for other session protocols as well.
For example, 
a ``recursive'' DNS resolver is just a proxy for the simple 
request/response session protocol used for DNS lookups.
ODNS \cite{ODNS} is a proposal for improving the privacy of users
doing lookups by introducing a proxy between local DNS resolvers
and authoritative resolvers.

To introduce the proxy, as shown in Figure~\ref{fig:odns},
the user appends---to the domain name it intends
to send in its request---the extra high-level domain name {\tt .odns}.
This will cause a local DNS resolver to send the request to an ODNS
proxy.
To conceal the true desired domain $D$ from the local DNS resolver,
the user generates a symmetric key $k$, encrypts $D$ with $k$, 
and encrypts $k$ with the public key $K$ of ODNS proxies.
A concatenation of these two values is the domain name it sends in its
request.
The ODNS proxy decrypts with its private key to get $k$, and then 
decrypts with $k$ to get $D$.
After getting a response from an authoritative server for $D$, with
an IP name $N$ for $D$, it
encrypts both $D$ and $N$ with $k$, and sends them in a response to
the local DNS resolver.
As a result of this design,
the local DNS resolver will not know what domain name
is being looked up, and 
the ODNS resolver will not know the identity of the user.

\begin{figure*}[hbt]
\begin{center}
\includegraphics[scale=0.60]{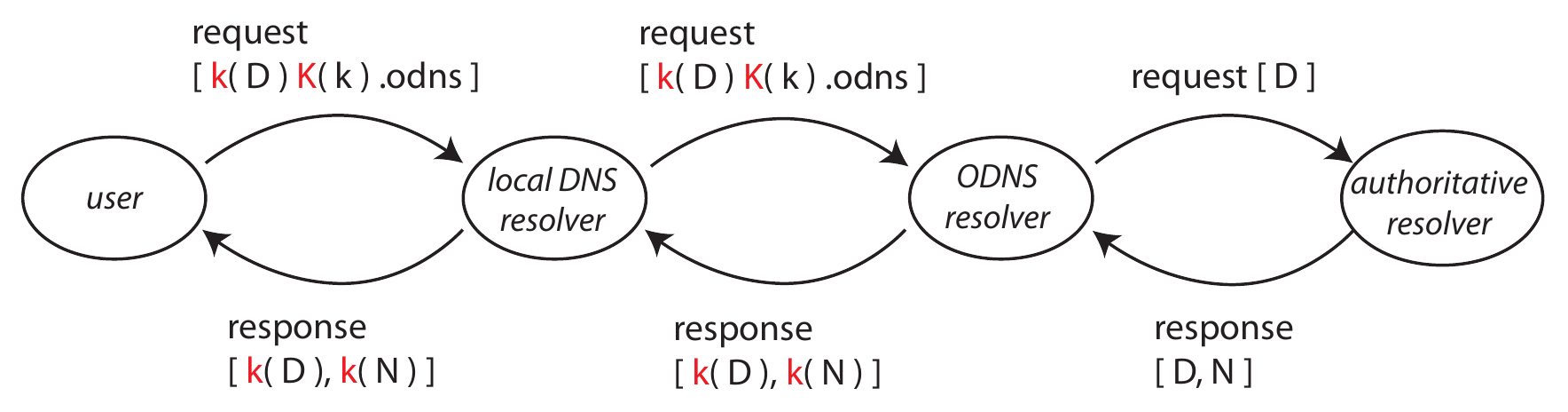}
\end{center}
\caption
{\small
{Oblivious DNS lookup for user privacy.}
\normalsize}
\label{fig:odns}
\end{figure*}
\else 
\fi

\subsubsection{Deflection}
\label{sec:Deflection}

The problem that a censoring access network can block packets to
known proxies has been addressed by 
several similar proposals \cite{cirripede,karlin,telex11}.
They all use proxies,
but in a way that still works despite the blocking.

\begin{figure*}[hbt]
\begin{center}
\ifdefined\bookversion 
\includegraphics[scale=0.60]{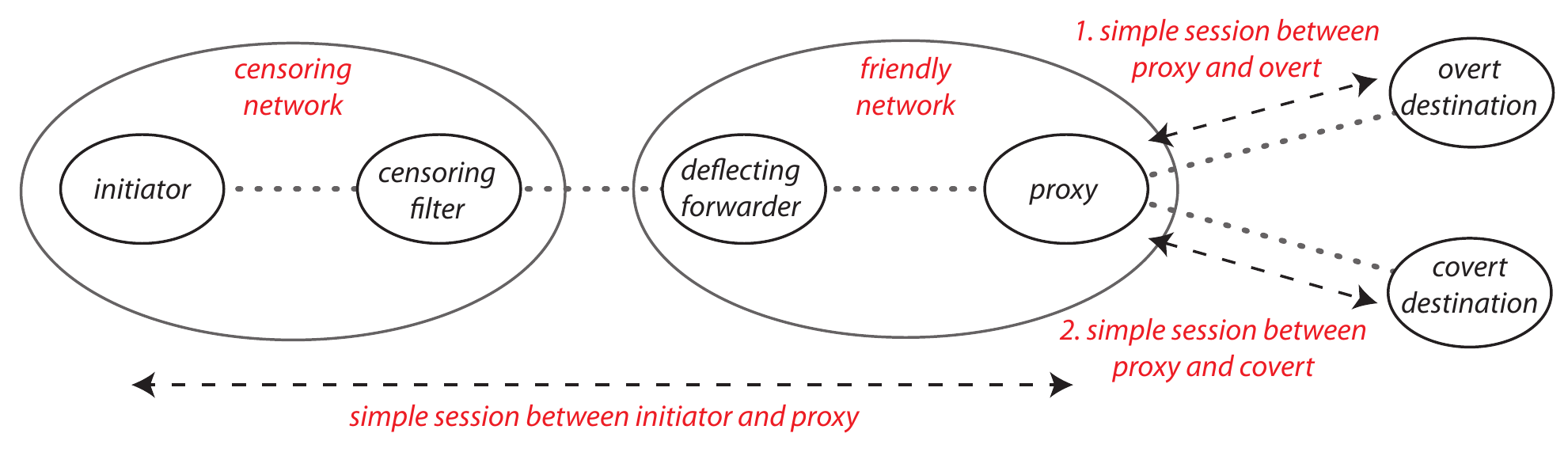}
\else 
\includegraphics[scale=0.50]{fig/cirripede.pdf}
\fi
\end{center}
\caption
{\small
{A deflected session between an initiator and a
covert destination.  In the simple session on the left,
names in the IP header are those of the initiator and overt destination;
packets from the initiator are deflected to the proxy as an exception
to normal forwarding.}
\normalsize}
\label{fig:cirripede}
\end{figure*}

A typical compound session in these proposals is shown in
Figure~\ref{fig:cirripede}.
The access network of the session initiator is
filtering out packets from users to certain destinations,
represented here by the ``covert destination.''
The initiator cannot evade this censorship by using a false source name,
because then
replies from the destination will not be delivered to the initiator
(also, the network may be blocking {\it everyone's} access to the site).
The critical mechanism is that session packets are routed through a
friendly network where a forwarder recognizes that the packets must
be treated specially, and deflects them to a proxy similar to the
proxy in Figure~\ref{fig:proxy1}.

For deflection to work, the initiator must give a hidden signal
to the deflecting forwarder---one that the censoring network is
unlikely to recognize---so the deflecting forwarder knows which
packets to deflect.
In Cirripede \cite{cirripede}, the user registers with the 
friendly network; while the registration is active, all sessions
initiated by the user are deflected.
\ifdefined\bookversion 
In decoy routing \cite{karlin}, this is done on
a session-by-session basis. 
Decoy routing assumes that the initiator and proxy
share a secret key, which is combined by the initiator with time-varying
information to form a hidden detection signal for a session.
The signal goes into a 
pseudo-random field of the first TLS packet.
The decoy-routing system has the same information as all the initiators,
so it can tell the deflecting forwarders which signals to look for
at the current time.
\else
\fi

\ifdefined\bookversion 
In Figure~\ref{fig:cirripede},
the TLS-based session protocol between the initiator and the proxy
is complex.
When the proxy first receives session packets, it initiates a TLS
session to the overt destination.
\else 
In Figure~\ref{fig:cirripede},
when the proxy first receives session packets, it initiates a TLS
session to the overt destination.
\fi
The TLS handshake is completed end-to-end between initiator and
overt destination, so that all packets (including a certificate
in plaintext) look normal to the censoring network.
Once packet data can be encrypted, the proxy 
signals to the initiator that it is in the session path, 
terminates the session to the overt destination,
gets the name of the covert destination from the initiator,
initiates a session to the covert destination,
and relays packets between the client and covert destination. 
During the entire compound
session, the packets seen by the censoring filter will have
the overt destination in their source or destination field.

The final problem to be solved is the placement of deflection forwarders
in friendly networks.
This can be viewed as a game between the censoring network (and its
friends) and the session participants (and their friends).
The administrator of the censoring network would like its outgoing packets
to reach all or most of
the public Internet without passing through a
network with deflection forwarding.
\ifdefined\bookversion 

\else 
\fi
The Cirripede proposal favors deflection forwarders in networks
close to the censoring network, so that many paths from the censoring
network go through friendly networks,
and the censoring network would suffer too much if it stopped bridging
to friendly networks.
The decoy routing proposal favors widespread deflection forwarders,
in particular, in friendly networks close to a variety of important
overt destinations.
This way
an initiator in the censored network can try several overt destinations
until it finds one with deflection in the path, which it knows when
the proxy signals its presence after the TLS handshake.
\ifdefined\bookversion 
In this game BGP inter-network routing helps the endpoints more than
the censoring network, because it 
gives the censoring network only a single route to each destination.
\else 
\fi

The rules of this game may change in the future:
networks may be willing to give some path-selection control to
cooperating networks and even user members, both of which
are recommended by the
SCION project \cite{scion}.
Both now and in the future, when it comes to security contests,
it matters who (and where) your friends are.
Social forces will shape the
Internet in their image, by defining its interest groups and alliances.

\subsection{Overlays}
\label{sec:Overlays}

An overlay is a virtual network layered on top of an
underlay network (\S\ref{sec:Composition-of-networks}).
We will first summarize the differences between overlays and
compound sessions, then show their use in 
\ifdefined\bookversion 
three security designs.
\else 
two security designs.
\fi

\subsubsection{Overlays {\it versus} compound sessions}
\label{sec:Overlays-versus-compound-sessions}

\begin{figure*}[hbt]
\begin{center}
\ifdefined\bookversion
\includegraphics[scale=0.60]{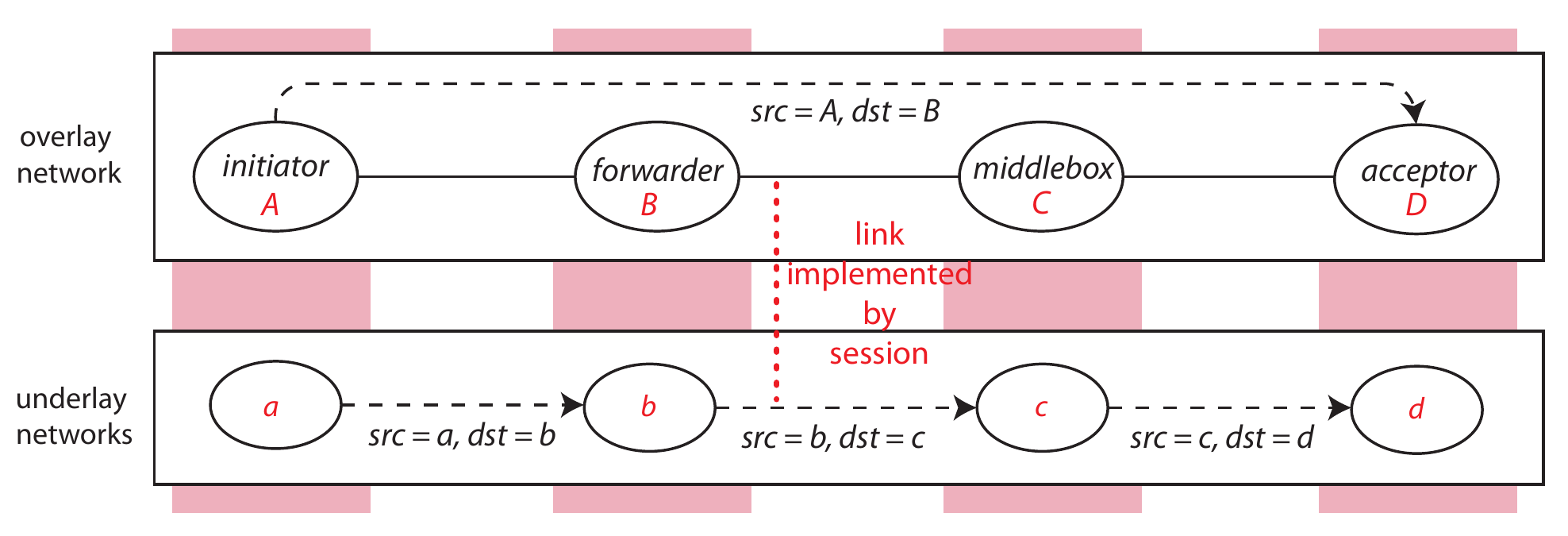}
\else
\includegraphics[scale=0.50]{fig/overlay0.pdf}
\fi
\end{center}
\caption
{\small
{A prototypical overlay session.}
\normalsize}
\label{fig:overlay0}
\end{figure*}

Figure~\ref{fig:overlay0} shows a prototypical overlay session whose
links are implemented by sessions in one or more bridged underlay networks.
All four machines are user members of their underlay networks.
From the viewpoint of the underlay networks, this looks very similiar
to a compound session with three simple sessions connecting user
members.
Yet the sessions in the underlay are completely independent of one
another ($b$ and $c$ are not proxies), and the overlay offers additional
structures that are often useful, as follows:
\begin{itemize}
\item
The overlay has its own namespace.
Overlay names can be the same as in the underlays, but new names
are useful for multiple purposes.
For example, a member of a private IP network with a private, unreachable
name can have a public, reachable name in an overlay.
\item
The overlay has its own routing.
Overlay routing can insert application-specific middleboxes.
In security designs, routing in an overlay is often used to vary and
conceal packet paths.
\item
The overlay has its own session protocols.
We'll see a good use of this in \S\ref{sec:Overlays-for-anonymity}.
\item
The overlay has its own (geographical) span.
It can unite allies in remote underlay networks.
\item
Sessions in the overlay and underlays have different durations.
Overlay links---implemented by underlay sessions---are often long-lived
and reused by many overlay sessions,
which minimizes setup time and computational overhead (as in 
\S\ref{sec:Performance}).
\end{itemize}

\subsubsection{Virtual private networks}
\label{sec:Virtual-private-networks}

\begin{figure*}[hbt]
\begin{center}
\ifdefined\bookversion 
\includegraphics[scale=0.60]{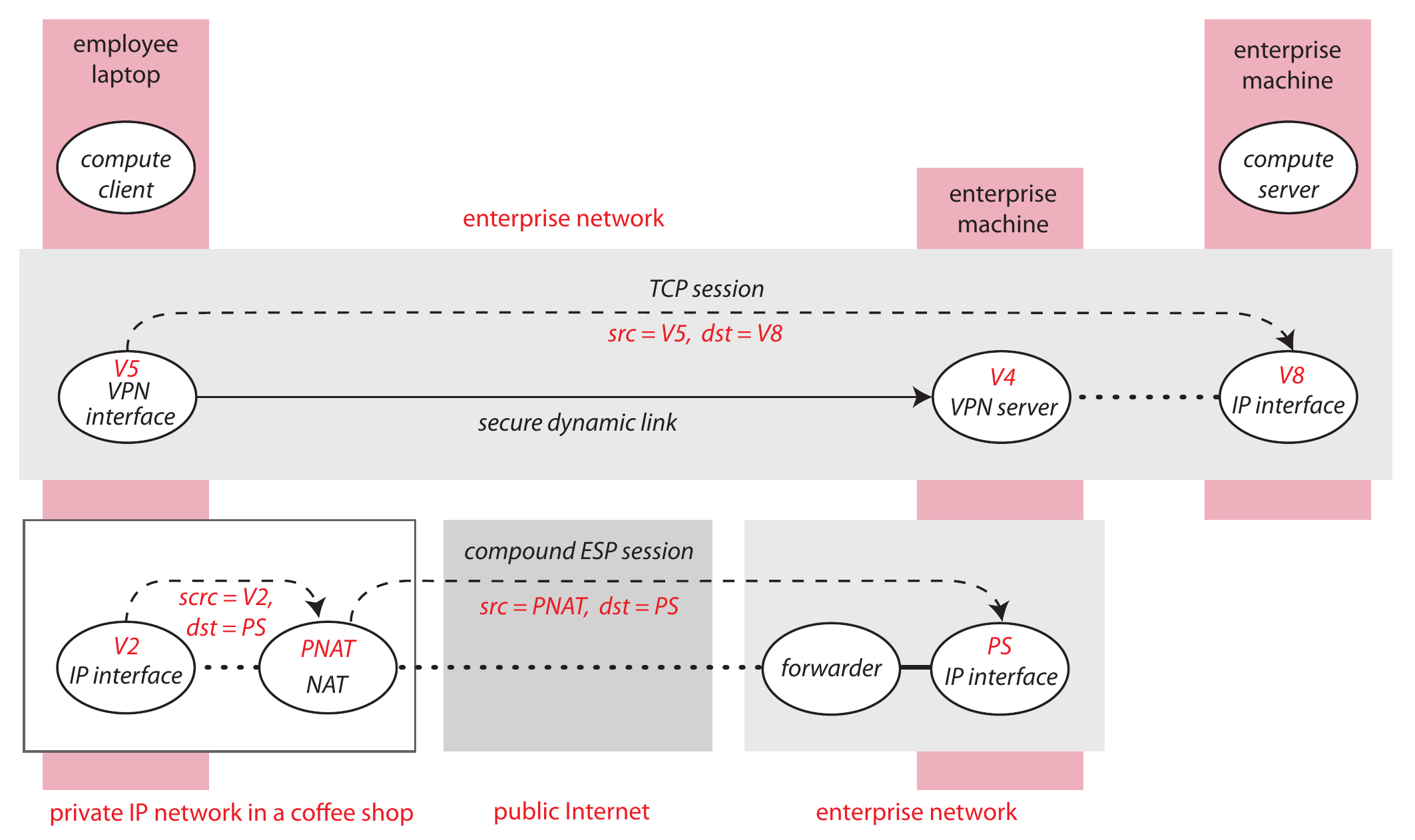}
\else 
\includegraphics[scale=0.50]{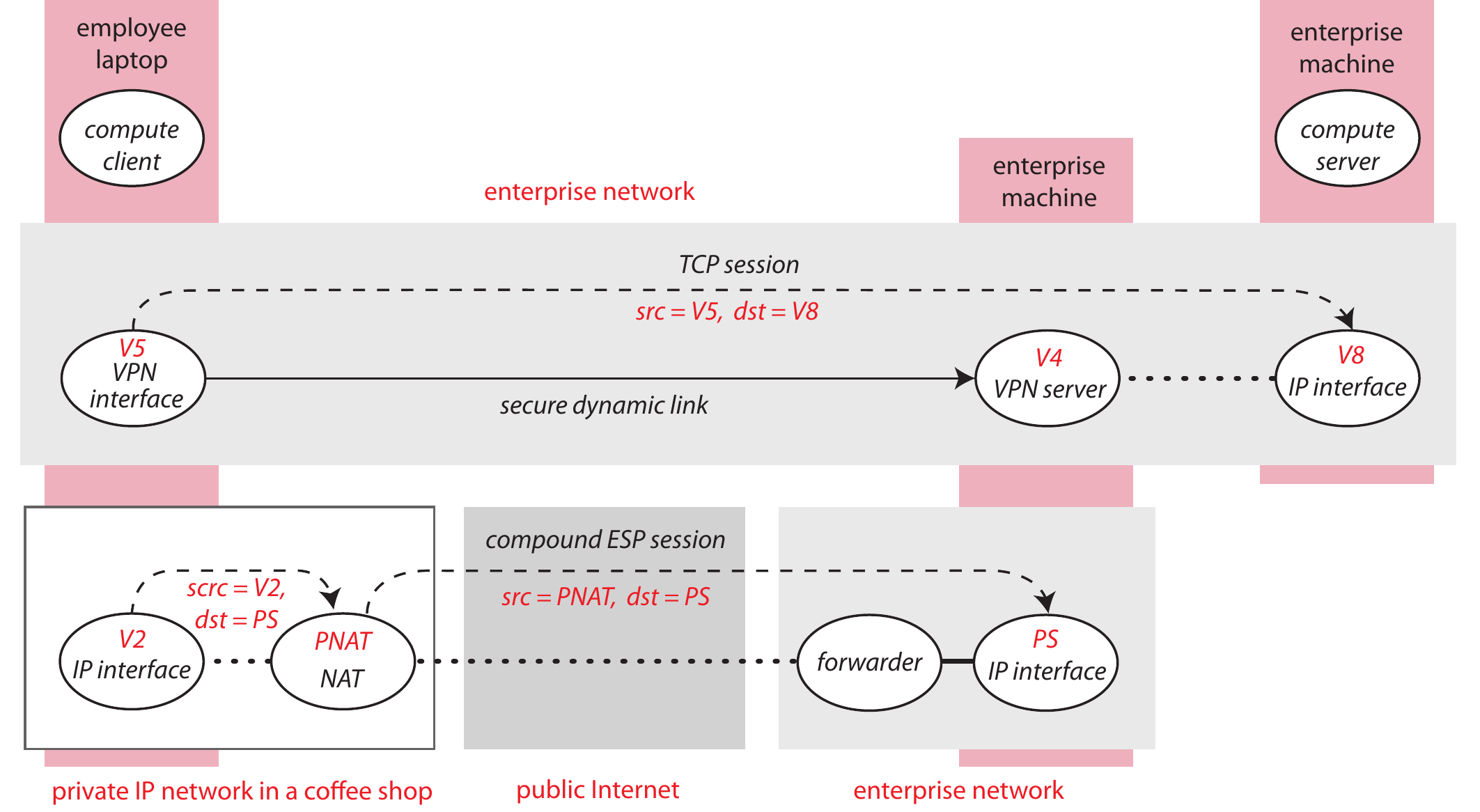}
\fi
\end{center}
\caption
{\small
{An enterprise network using VPN technology.
A secure dynamic link in the enterprise network is implemented by an
ESP session in tunnel mode.}
\normalsize}
\label{fig:vpn3}
\end{figure*}

Strictly speaking, ``virtual private networks'' (VPNs) are not
networks, but rather a technology for
widening the geographic span of a private IP network such as an
enterprise network.
With VPN technology, an enterprise network is composed with
other public and private IP networks in two ways simultaneously: 
(i) as usual,
it is bridged with them, and
(ii) it is layered on them, because some links of the enterprise network
are implemented by sessions spanning other public and private IP networks.
These relationships are illustrated by
Figure~\ref{fig:vpn3}.

In the figure, there is a TCP session between an enterprise machine
and an employee laptop currently located in a coffee shop.
The enterprise-network member on the laptop is described as a ``VPN
interface,'' because it is an IP interface plus VPN client.
Before initiating the TCP session, it must first create a secure dynamic
link to a VPN server in the enterprise network.
To create the dynamic link, the laptop's VPN interface requests
that its IP interface make an ESP session
(\S\ref{sec:Three-IP-cryptographic-protocols})
to public IP name {\it PS}. 
The employee must also enter a password to authenticate his identity
to the VPN server.
The ESP session happens to be compound, because it goes through a
Network Address Translator 
(similar to a proxy) in the coffee shop's private IP network.

Viewed as an overlay network, the enterprise network uses VPN
technology to allow a laptop in an insecure location to participate
fully in the enterprise network.
Most importantly,
the VPN server assigns the laptop's member the name
{\it V5} in the network's
private namespace.
This name can be chosen according to the privileges the laptop's
owner has within the enterprise network.
Consequently, traffic filters in the enterprise network can see from
the source and destination fields of packets which policies should
apply to the laptop's sessions, and enforce them accordingly.

\ifdefined\bookversion 
\subsubsection{Overlays for default-reject filtering}
\label{sec:Overlays-for-default-reject-filtering}

In \S\ref{sec:Default-reject-filtering} 
we introduced default-reject filtering
(filtering in which the default action on a packet is to drop it)
as an interesting technique with limited applicability, because of
the difficulty of finding precise filtering criteria.
Default-reject
filtering is similar in spirit to capabilities, introduced
in \S\ref{sec:Capabilities}, but it turned out that denial of capability
is as much of a problem as denial of service.
Several researchers have explored whether the properties of overlays
can be exploited to make a success of default-reject filtering.

In both Mayday \cite{mayday} and Secure Overlay Services (SOS) \cite{sos},
the initiator of a session is authenticated, and session packets are
not transmitted unless authentication succeeds.
To understand these proposals, it is best to imagine the perspective
of an access network that is trying to protect a Web server within the 
network from flooding attacks.
The obvious problem with authentication at the edge of the access network
is that the authenticators are a limited resource, easily overcome
by denial-of-capability attacks.
Mayday and SOS use overlays to extend the geographical span of the
access network, enlisting allies all over the public Internet to
perform authentication.
The authenticators are the overlay ingress nodes in 
Figure~\ref{fig:overlay1}.
The idea is that there can be enough authenticators, dispersed widely
enough, to resist flooding and denial-of-capability attacks.
This is upstream filtering 
(\S\ref{sec:Filtering-upstream}) in the overlay network.

\begin{figure*}[hbt]
\begin{center}
\includegraphics[scale=0.60]{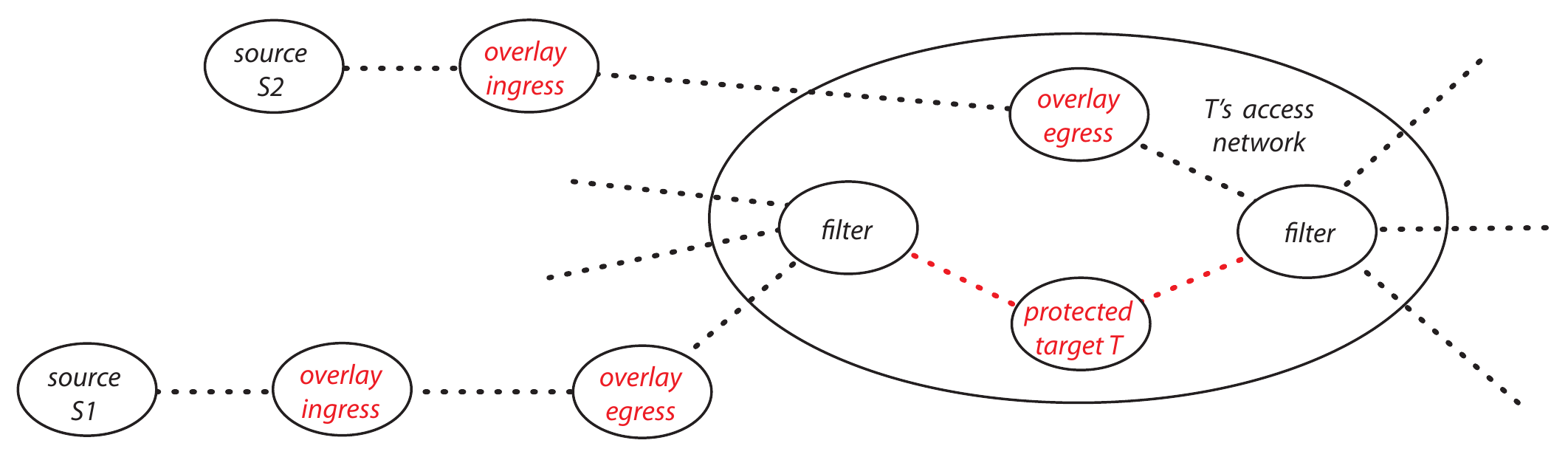}
\end{center}
\caption
{\small
{A network graph of
Internet members involved in overlay-based default-reject filtering.
The members named in red are on overlay machines, i.e., their
machines also have members of the overlay network.  
The paths in red are the {\it only Internet
paths} to the protected target.}
\normalsize}
\label{fig:overlay1}
\end{figure*}

The general idea of Mayday and SOS is that packets of authenticated
sessions travel to the protected target through the overlay as well as
the Internet.
Traffic filters near the protected target can distinguish overlay
packets, and discard all other incoming packets.
Details are given below.

These proposals make use of another overlay property
in addition to flexible span:
because an overlay is a network, it has its own routing.
Overlay routing is used to vary and hide the paths of
packets between ingress members and the target.
This keeps attackers from flooding the paths to the target rather
than the target itself.
For instance, SOS uses a complex routing scheme, with long paths computed
from distributed hash tables, for path secrecy.
Other implications of overlay routing will be discussed further below.

In designing an overlay network for default-reject filtering, there are
three important choices to be made.
SOS makes specific choices, while the Mayday paper points out that there
are other choices, and evaluates some combinations of them.
We now explain the three choices.

\vspace{.075 in}
\noindent
{\it Source authentication}
\vspace{.075 in}

This choice concerns how a session initiator finds an ingress member
and authenticates itself to the overlay
as a source of legimate packets.
SOS is intended for use during an
emergency situation, when networks are so congested that even benign
ordinary traffic must be dropped.
The members of SOS are hosted by a peer group of machines cooperating
to provide emergency services.
The only allowed packets to a given destination come from a few
pre-configured sources used by emergency responders.
So in SOS the source is itself an overlay member,
i.e., it has special software.
SOS source members know the Internet names of many ingress members,
well-distributed so that they cannot all be overwhelmed by flooding
attacks.
It creates a secure link to an ingress member, using ESP
with endpoint authentication.

Mayday emphasizes an authentication
option that is architecturally more complex,
but has broader applicability because the source need not be
an overlay member (both Figures~\ref{fig:overlay1} 
and \ref{fig:overlay2} depict this option).
In this option packets from a source to target name $T$ are routed
to some machine with an ingress member of the overlay.
This can be accomplished by IP anycast, which will route packets
from any source to the destination with name $T$ closest to them.
The underlay IP interface accepts the TCP session and passes control
to the ingress member,
which can then authenticate the source by asking for a user name
and password
associated with the target service.
If the source is authentic,
the ingress member initiates a TCP session {\it through the overlay}
to the target; these two TCP sessions then become two parts of a
compound application session.

\begin{figure*}[hbt]
\begin{center}
\includegraphics[scale=0.60]{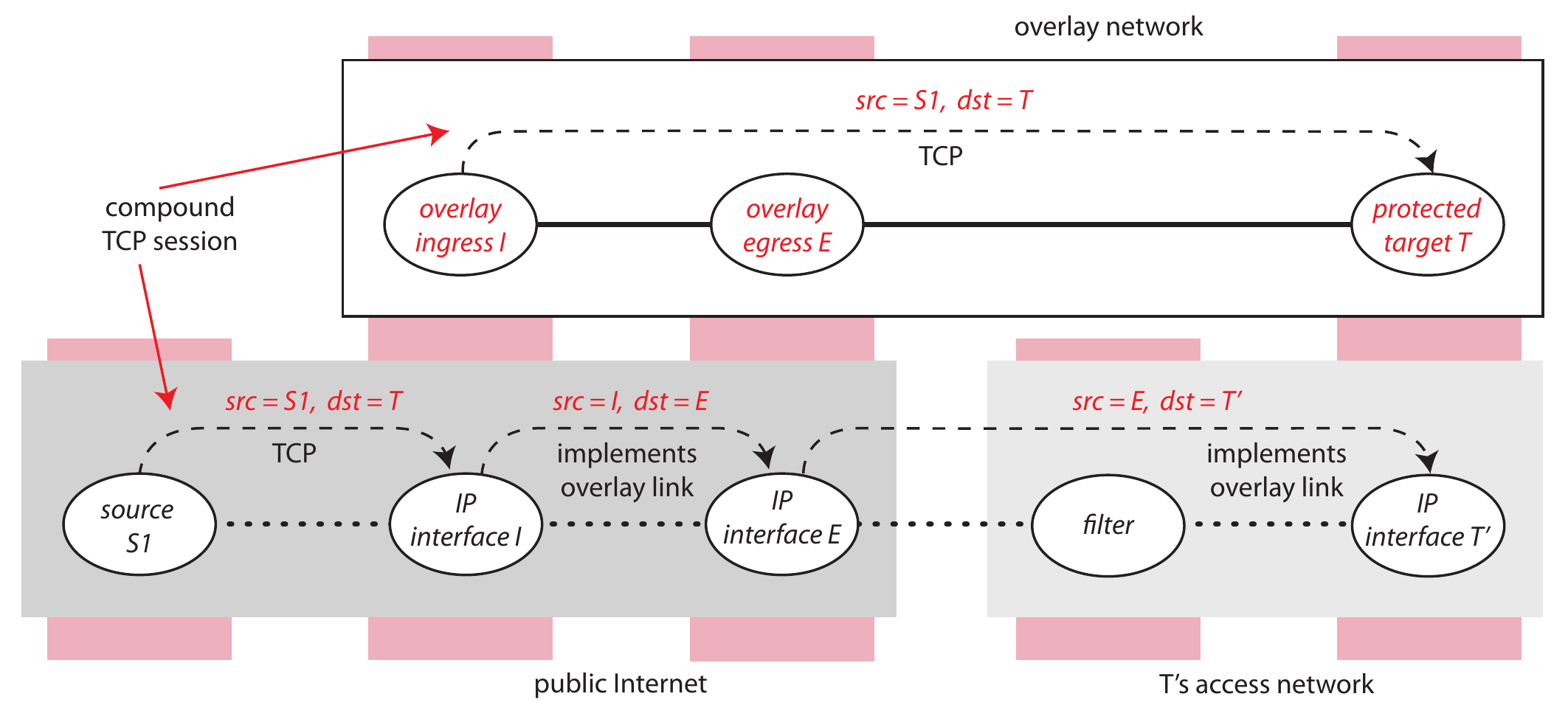}
\end{center}
\caption
{\small
{A session with overlay-based default-reject
filtering, illustrating
the following options: source is not an overlay member, target has
different names in overlay and underlay, routing is singly-indirect.}
\normalsize}
\label{fig:overlay2}
\end{figure*}

Note from Figure~\ref{fig:overlay2} that the target server receives
as source name the underlay name of the initiator.
This means that reverse packets, from the Web server to the initiator,
do not travel through the overlay.
Although the path of return packets is not shown in the figure,
both SOS and Mayday work like this.

\vspace{.075 in}
\noindent
{\it Lightweight authenticator}
\vspace{.075 in}

In addition to the overlay network, a potential target must be surrounded
in the Internet underlay by a ring of ordinary traffic filters.
These ordinary filters, such as firewalls, must have the capacity
to handle flooding attacks, and must be configurable by overlay
machines or by people administering the overlay.

Figure~\ref{fig:overlay2} is a session view of allowed access to 
protected target $T$.
The last overlay hop between an egress member
of the overlay and the target is implemented by an underlay path
that goes through a filter.
The lightweight authenticator is the attribute of underlay packets
from an egress member to the target
that causes the traffic filter to recognize them as overlay packets
and allow them to pass.
The simplest lightweight authenticator is the IP name of the egress
member (here $E$) in the source field of a packet;
this is what SOS uses.
Other authenticators proposed by Mayday include the destination
port, destination name, and other header fields whose contents can
be manipulated by the egress member.

The critical property of a lightweight authenticator is that it must
be a secret---if attackers knew it, they could simply send underlay
packets that match it.
You might think that the destination name is the worst possible
authenticator, but it can be a good one if the underlay name of
the protected target is different from its overlay name,
as shown in Figure~\ref{fig:overlay2}, and if it can be changed
easily and frequently by local control in the target's access network.

\vspace{.075 in}
\noindent
{\it Overlay routing}
\vspace{.075 in}

In addition to hiding whole packet paths, overlay routing
keeps the identities of egress nodes secret, which is
indispensable if the lightweight authenticator is the name of an
egress node.
SOS uses egress names as authenticators; this is safe because of
its elaborate overlay routing.

Mayday takes the position that effective overlay routing can be
much simpler, with options including no routing at all (ingress and
egress nodes are the same), and singly-indirect routing
(one hop between ingress and egress nodes, as in 
Figure~\ref{fig:overlay2}).
The Mayday paper reports on analysis showing that certain combinations
of overlay routing and lightweight authenticator provide ``best cases''
for trade-offs among performance and security.
For example, it says that
designers who want moderate levels of both performance
and security should use singly-indirect routing with any authenticator
other than egress name.
\else
\fi

\subsubsection{Overlays for anonymity}
\label{sec:Overlays-for-anonymity}

In \S\ref{sec:Proxies-in-transit-networks} we showed how proxies
in transit networks can provide session-initiating users
with privacy within their access
networks and anonymity at the accepting endpoint.
The weakness of this mechanism is that the user has no privacy whatsoever
from the proxy.
The purpose of the public service Tor \cite{tor2,tor1} is to
add to the services above a high degree of privacy from the proxies.
\ifdefined\bookversion 
This section describes the second, current Tor design \cite{tor2}.
\else 
\fi

\ifdefined\bookversion 
Tor is an overlay network whose infrastructure members reside on the
machines of volunteers world-wide.
These infrastructure members
are fully connected by long-lived links, each of
which is implemented by a TLS session in the public Internet.
This covers two of the ways Tor makes use of overlay properties:
its membership unites allies across the globe, and
its links are long-lived and reused by many overlay sessions (which
minimizes setup time and computational overhead).

An infrastructure member in Tor acts as a proxy within the overlay.
Users also have Tor members on their machines.
Each proxy has a public key,
which it uses (along with a certificate)
to authenticate itself when setting up links by
means of TLS sessions.\footnote{In Tor terminology, a proxy is an
``onion router'' and a user member is an ``onion proxy.''
But Tor proxies do no routing, and user members do no proxying.} 

\else 
Tor is an overlay network whose infrastructure members reside on the
machines of volunteers world-wide.
They are fully connected by long-lived links, each of
which is implemented by a TLS session in the public Internet.
This shows two of the ways Tor uses overlay properties:
its membership unites allies across the globe, and
its links are long-lived and reused by many overlay sessions (which
minimizes setup time and computational overhead).
An infrastructure member in Tor acts as a proxy within the overlay.
Users also have Tor members on their machines.
Each proxy has a public key,
which it uses (with a certificate)
to authenticate itself when setting up links by
means of TLS sessions.
\fi

Tor is layered between application networks and the public Internet.
Applications use the same interface to get TLS service
from Tor as they would from the public Internet.
User members query Tor directory servers to get lists of available
proxies, each described by its public key, IP name, 
and policies.

\begin{figure*}[hbt]
\begin{center}
\ifdefined\bookversion
\includegraphics[scale=0.60]{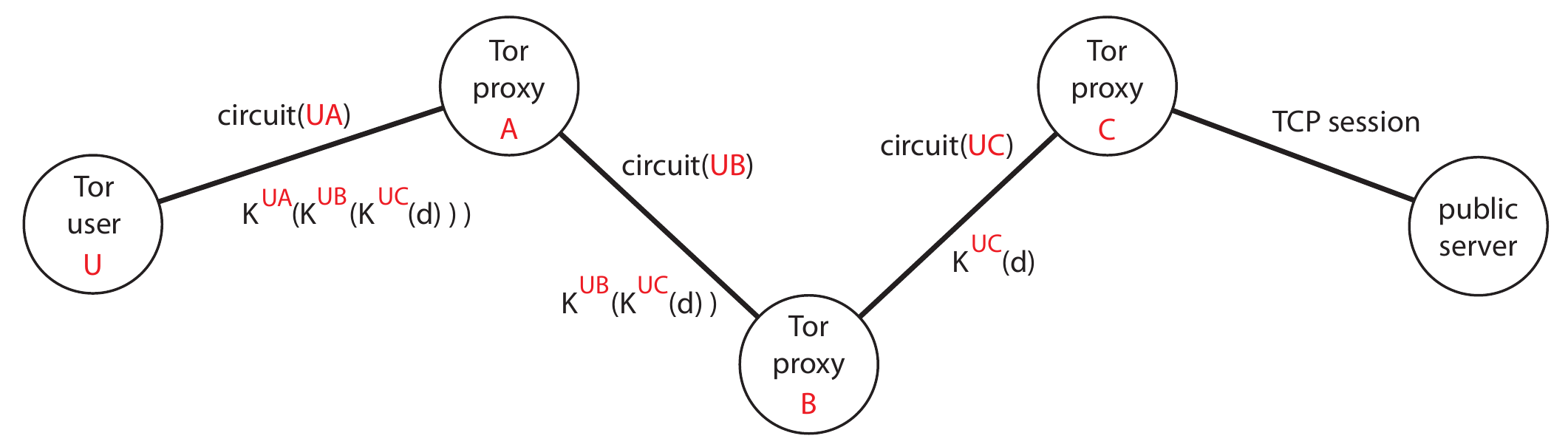}
\else
\includegraphics[scale=0.50]{fig/tor.pdf}
\fi
\end{center}
\caption
{\small
{A compound session made by Tor.  The first three simple
sessions use the Tor circuit protocol, and go through the Tor network.
The last simple session uses TCP, and goes through the public Internet.}
\normalsize}
\label{fig:tor}
\end{figure*}

To make a TLS session for an application (when there is no prior state
in place), a Tor member first chooses a random route through several
Tor proxies
(this is why the proxies themselves do no routing).
As with other overlay routing schemes, this varies and conceals 
packet paths.
Next the user member creates a compound session in Tor that
goes through the chosen proxies, as shown in Figure~\ref{fig:tor}.
The session protocol is the Tor ``circuit'' protocol,
and each simple session is a Tor circuit with its own circuit identifier.

The important thing about circuits is that each one has a unique
security association with the user member that created it.
To make the compound session in Figure~\ref{fig:tor}, the user first
creates a simple session (circuit) to $A$, and executes a key-exchange
protocol with $A$, so that each now knows a shared symmetric key
$K^{UA}$.
Next the user uses {\it circuit(UA)} to send to $A$ an {\it extend}
command telling it to create a new circuit to proxy $B$.
Through the two associated
circuits, $U$ and $B$ execute a key-exchange protocol,
after which each has a shared key $K^{UB}$.
Finally $U$ tells $B$ to {\it extend} the compound session by creating
a new circuit to $C$, with $U$/$C$ key exchange.
Once a compound session has been assembled in Tor, it can be used
to carry many TLS application sessions.
In the background, the compound session
is reconfigured piece-by-piece about once a minute,
to confuse adversarial observers who are analyzing traffic
patterns.\footnote{Spies know that Tor sessions are deliberately
concealed, so they have reason to analyze side-channels such as
packet timing and sizes.  
These attacks can be successful in
correlating packet streams coming into and out of Tor.
Note that the adversarial observers can be Tor proxies, too.}

Tor users use the security associations to conceal packet data from 
all except the last Tor proxy.
The data transmitted on each circuit is multiply-encrypted
as shown in the figure.
When $A$ receives a packet from $U$, it decrypts it before 
forwarding it to $B$, but it cannot read the packet because it is 
doubly encrypted with keys $K^{UB}$ and $K^{UC}$ that are unknown to $A$.
For a similar reason, $B$ cannot read it either.

\begin{figure*}[hbt]
\begin{center}
\ifdefined\bookversion
\includegraphics[scale=0.55]{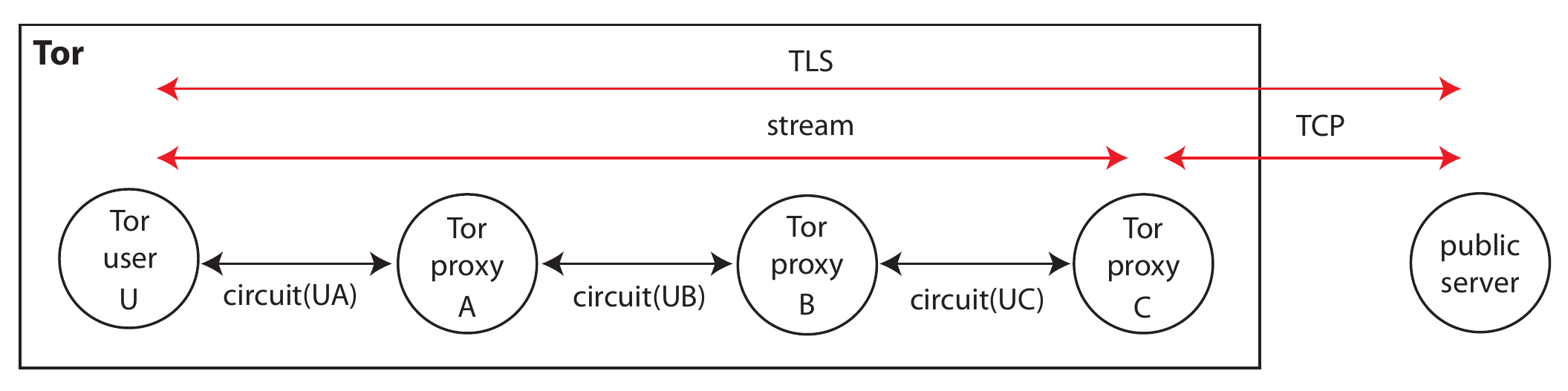}
\else
\includegraphics[scale=0.45]{fig/tor2.pdf}
\fi
\end{center}
\caption
{\small
{Session protocols and their embeddings, for a single TLS
application session made through Tor.  Sessions of the circuit
protocol last longer than application sessions.}
\normalsize}
\label{fig:tor2}
\end{figure*}

\ifdefined\bookversion 
To understand the rest of the Tor design, it is necessary to consider
TLS as a separate protocol embedded inside TCP.
Tor has a second session protocol, the stream protocol, which is embedded
in the circuit protocol.
Figure~\ref{fig:tor2} shows all the session protocols, with protocols
above embedded in protocols below, used for a single TLS application
session through Tor.
\else 
To understand the rest of the Tor design, it is necessary to consider
TLS as a separate protocol embedded in TCP.
Tor has a second session protocol, the stream protocol, embedded
in the circuit protocol.
Figure~\ref{fig:tor2} shows all the session protocols, with protocols
above embedded in protocols below, used for a single TLS application
session through Tor.
\fi

The stream protocol substitutes for TCP within Tor;
there is a one-to-one correspondence between external TCP sessions and
Tor streams, and TCP data is simply reformatted for streams.
There are two reasons for using streams instead of TCP inside Tor:
(i) if the data sent on Tor circuits were TCP packets, then proxy $C$
would see their source and destination fields in plaintext;
(ii) the reliable, ordered packet delivery of TCP is not required
within Tor, because all of its links are implemented by TLS, and already
have these properties.

When $C$ has received enough stream packets to carry an HTTPS
request with a domain name,
it can complete the compound session end-to-end.
It sends a TCP SYN packet with its own IP name as source and the
IP name of the domain name as destination.
After the TCP handshake, it continues converting data packets between
the TCP and stream-protocol formats, and forwarding them in both
directions.
The TLS handshake between $U$ and the server goes end-to-end, so that
$U$ can validate the server's certificate, and so that even $C$ does
not see plaintext packet data.

\ifdefined\bookversion 
Tor is a complex network in its own right.
We have seen that it has its own session protocols, compound sessions, and
routing. 
It also has internal mechanisms for denial-of-service protection and 
congestion control.
Rate-limiting can be managed on circuits or on streams.
\else 
\fi

Unfortunately the Tor design for privacy has one serious deficiency,
which is the fact that the final acceptor of the TCP session can
know that Tor is being used, 
because there are readily accessible lists of Tor nodes.
Fraudsters, spammers, and other criminals are big users of Tor, along
with law-abiding people in need of privacy.
Consequently an increasing number of services are rejecting or
otherwise discriminating against Tor users \cite{torDifferential}.
Tor protects the reputation of its volunteer machines by allowing
them to restrict their exiting TCP sessions or refuse to be exit proxies.
Some volunteers must shoulder this burden, however, or the service
will not be available to those who really need it.

\ifdefined\bookversion 
Although the Tor-exit problem appears to have
no solution in today's Internet,
privacy is so valuable that some researchers propose to replace much of IP
with the concepts of Tor \cite{tor-insteadof-ip}.
If the use of Tor became much more widespread as this proposal argues
it should, then the stigma of using Tor would fade away.
\else 
\fi

\section{Conclusion}
\label{sec:Conclusion}

\ifdefined\bookversion 

Modeling and security are tightly intertwined.
Given a rigorous model of a network, security attacks, and defenses,
we can reason rigorously or even formally that the defenses will prevent
the attacks---or at least mitigate them.
Where there are gaps in the model, i.e., possible real-world
behaviors that the model does not describe, there are possible attacks
against which the defenses are useless.

As networks have become increasingly important in most aspects of
daily life, their complexity has grown in proportion, and the early
models have become increasingly inadequate.
In this tutorial we have used a new, compositional model, which
emphasizes that there are many networks in an overall network
architecture, each one being a microcosm of all the basic aspects
of networking.
Because all networks are similar in important ways, they can be
modeled with common structures and common compositional interfaces.
This modularity greatly expands the range of network behaviors that
can be modeled without overwhelming complexity \cite{cacm}.

In this tutorial, the new model has enabled us to find a new and useful
classification of security attacks, and to explain all common
defenses by means of just four patterns.
There is a clear relationship between the attack categories and the
defense patterns, because the categories are based on which agents
are the attackers, defenders, and potential victims,
and some defenses are only available to some defenders.

The model has also helped us understand how the patterns interact with
each other and with other aspects of networking,
which is a dimension of security that has received little prior
attention.
Nevertheless it is essential, because networks must satisfy many
goals simultaneouly, and designers cannot succeed if they are ignorant
of the consequences of their decisions.

The modeling and defenses in this tutorial are obviously not complete,
yet we believe that any progress toward organized thinking about
network security will be helpful for building defenses 
and---ultimately---making verifiable claims about network security.
It has been shown that real software can be made secure with
modularity and verified components \cite{klein},
and the world is sure to demand this level of assurance from more
and more computer systems, including networks.

\else 

Modeling and security are tightly intertwined.
Given a rigorous model of a network, security attacks, and defenses,
we can reason rigorously or even formally that the defenses will prevent
the attacks---or at least mitigate them.
Where there are gaps in the model, i.e., possible real-world
behaviors that the model does not describe, there are possible attacks
against which the defenses are useless.

As networks have become increasingly important in most aspects of
daily life, their complexity has grown in proportion, and the early
models have become increasingly inadequate.
In this tutorial, a new model has enabled us to find a new and useful
classification of security attacks, and to explain all common
defenses by means of just four patterns.
There is a clear relationship between the attack categories and the
defense patterns, because the categories are based on which agents
are the attackers, defenders, and potential victims,
and some defenses are only available to some defenders.
The model has also helped us understand how the patterns interact with
each other and with other aspects of networking,
which is a dimension of security that has received little prior
attention.
The modeling and defenses in this tutorial are obviously not complete,
yet we believe that any progress toward organized thinking about
network security will be helpful for building stronger defenses.
\fi 

\bibliographystyle{abbrv}
\ifdefined\bookversion 
\bibliography{all}
\else 
\bibliography{compsurv}

\begin{thebibliography}{10}

\bibitem{mayday}
D.~G. Andersen.
\newblock Mayday: Distributed filtering for {I}nternet services.
\newblock In {\em Proceedings of the 4th {USENIX} Symposium on Internet
  Technologies and Systems}, 2003.

\bibitem{aip}
D.~G. Andersen, H.~Balakrishnan, N.~Feamster, T.~Koponen, D.~Moon, and
  S.~Shenker.
\newblock Accountable {I}nternet {P}rotocol {(AIP)}.
\newblock In {\em Proceedings of ACM SIGCOMM}, 2008.

\bibitem{arbor}
{Arbor Networks}.
\newblock {NETSCOUT} {A}rbor's 13th annual worldwide infrastructure security
  report.
\newblock
  \url{https://pages.arbornetworks.com/rs/082-KNA-087/images/13th_Worldwide_Infrastructure_Security_Report.pdf}.

\bibitem{AITF}
K.~Argyraki and D.~R. Cheriton.
\newblock Active {I}nternet traffic filtering: Real-time response to
  denial-of-service attacks.
\newblock In {\em Proceedings of the {USENIX} Annual Technical Conference},
  2005.

\bibitem{uglyCapabilities}
K.~Argyraki and D.~R. Cheriton.
\newblock Network capabilities: The good, the bad, and the ugly.
\newblock In {\em Proceedings of the 4th Workshop on Hot Topics in Networks},
  2005.

\bibitem{scion}
D.~Barrera, L.~Chuat, A.~Perrig, R.~M. Reischuk, and P.~Szalachowski.
\newblock The {SCION} {I}nternet architecture.
\newblock {\em Communications of the {ACM}}, 60(6):56--65, June 2017.

\bibitem{netconfig}
R.~Beckett, A.~Gupta, R.~Mahajan, and D.~Walker.
\newblock A general approach to network configuration verification.
\newblock In {\em Proceedings of ACM SIGCOMM}, 2017.

\bibitem{propane}
R.~Beckett, R.~Mahajan, T.~Millstein, J.~Padhye, and D.~Walker.
\newblock Don't mind the gap: Bridging network-wide objectives and device-level
  configurations.
\newblock In {\em Proceedings of ACM SIGCOMM}, 2016.

\bibitem{end-to-end2}
M.~S. Blumenthal and D.~D. Clark.
\newblock Rethinking the design of the {I}nternet: The end-to-end arguments vs.
  the brave new world.
\newblock {\em ACM Transactions on Internet Technology}, 1(1):70--109, August
  2001.

\bibitem{canetti}
R.~Canetti.
\newblock Universally {C}omposable {S}ecurity: A new paradigm for cryptographic
  protocols.
\newblock \url{https://eprint.iacr.org/2000/067.pdf}, 2019.
\newblock Accessed 15 October 2019.

\bibitem{ethane}
M.~Casado, M.~J. Freedman, J.~Pettit, J.~Luo, N.~McKeown, and S.~Shenker.
\newblock Ethane: Taking control of the enterprise.
\newblock In {\em Proceedings of SIGCOMM}. ACM, August 2007.

\bibitem{caida}
{Center for Applied Internet Data Analysis}.
\newblock Archipelago monitor statistics.
\newblock \url{https://www.caida.org/projects/ark/statistics/}.
\newblock Accessed 29 November 2019.

\bibitem{philo}
D.~D. Clark.
\newblock The design philosophy of the {DARPA} {I}nternet protocols.
\newblock In {\em Proceedings of SIGCOMM}. ACM, August 1988.

\bibitem{clark-book}
D.~D. Clark.
\newblock {\em {{\it Designing an Internet}}}.
\newblock Information Policy Series, MIT Press, 2018.

\bibitem{tussle}
D.~D. Clark, J.~Wroclawski, K.~R. Sollins, and R.~Braden.
\newblock Tussle in cyberspace: Defining tomorrow's {I}nternet.
\newblock {\em IEEE/ACM Transactions on Networking}, 13(3):462--475, June 2005.

\bibitem{tor2}
R.~Dingledine, N.~Mathewson, and P.~Syverson.
\newblock Tor: The second-generation onion router.
\newblock In {\em Proceedings of the 13th USENIX Security Symposium}, 2004.

\bibitem{dynAttack}
Dyn analysis summary of {F}riday {O}ctober 21 attack.
\newblock
  https://dyn.com/blog/dyn-analysis-summary-of-friday-october-21-attack/.
\newblock Accessed 10 November 2018.

\bibitem{fogel}
A.~Fogel, S.~Fung, L.~Pedrosa, M.~Walraed-Sullivan, R.~Govindan, R.~Mahajan,
  and T.~Millstein.
\newblock A general approach to network configuration analysis.
\newblock In {\em Proceedings of the 12th USENIX Conference on Networked
  Systems Design and Implementation}, 2015.

\bibitem{dangerous}
M.~Georgiev, S.~Iyengar, S.~Jana, R.~Anubhai, D.~Boneh, and V.~Shmatikov.
\newblock The most dangerous code in the world: Validating {SSL} certificates
  in non-browser software.
\newblock In {\em ACM Conference on Computer and Communications Security},
  2012.

\bibitem{homoencryption}
S.~Goldwasser, Y.~Kalai, R.~A. Popa, V.~Vaikuntanathan, and N.~Zeldovich.
\newblock Reusable garbled circuits and succinct functional encryption.
\newblock In {\em Proceedings of Symposium on Theory of Computing}. ACM, 2013.

\bibitem{trustSurvey}
T.~Grandison and M.~Sloman.
\newblock A survey of trust in {I}nternet applications.
\newblock {\em IEEE Communications Surveys and Tutorials}, 3(4):2--16, 2000.

\bibitem{handley}
M.~Handley.
\newblock Why the {I}nternet only just works.
\newblock {\em BT Technology Journal}, 24(3):119--129, July 2006.

\bibitem{NID}
M.~Handley, V.~Paxson, and C.~Kreibich.
\newblock Network intrusion detection: Evasion, traffic normalization, and
  end-to-end protocol semantics.
\newblock In {\em Proceedings of the 10th {USENIX} Security Symposium}, 2001.

\bibitem{hiveMQTT}
How does {TLS} affect {MQTT} performance?
\newblock https://www.hivemq.com/ blog/how-does-tls-affect-mqtt-performance/.
\newblock Accessed 19 September 2018.

\bibitem{tcp-over-tcp}
O.~Honda, H.~Ohsaki, M.~Imase, M.~Ishizuka, and J.~Murayama.
\newblock {TCP} over {TCP}: Effects of {TCP} tunneling on end-to-end throughput
  and latency.
\newblock In {\em Proceedings of SPIE}, volume 6011, pages 138--146.
  International Society for Optical Engineering, 2005.

\bibitem{cirripede}
A.~Houmansadr, G.~T.~K. Nguyen, M.~Caesar, and N.~Borisov.
\newblock Cirripede: Circumvention infrastructure using router redirection with
  plausible deniability.
\newblock In {\em Proceedings of the ACM Conference on Computer and
  Communications Security}, 2011.

\bibitem{osi}
ITU.
\newblock Information {T}echnology---{O}pen {S}ystems
  {I}nterconnection---{B}asic {R}eference {M}odel: The basic model.
\newblock {ITU-T} {R}ecommendation {X}.200, 1994.

\bibitem{karlin}
J.~Karlin, D.~Ellard, A.~W. Jackson, C.~E. Jones, G.~Lauer, D.~P. Mankins, and
  W.~T. Strayer.
\newblock Decoy routing: Toward unblockable {I}nternet communication.
\newblock In {\em Proceedings of the USENIX Workshop on Free and Open
  Communications on the Internet}. USENIX, 2011.

\bibitem{headerspace2}
P.~Kazemian, M.~Chang, H.~Zeng, G.~Varghese, N.~McKeown, and S.~Whyte.
\newblock Real time network policy checking using {H}eader {S}pace {A}nalysis.
\newblock In {\em Proceedings of the 10th USENIX Conference on Networked
  Systems Design and Implementation}, 2013.

\bibitem{sos}
A.~D. Keromytis, V.~Misra, and D.~Rubenstein.
\newblock {SOS}: {S}ecure {O}verlay {S}ervices.
\newblock In {\em Proceedings of SIGCOMM}. ACM, August 2002.

\bibitem{torDifferential}
S.~Khattak, D.~Fifield, S.~Afroz, M.~Javed, S.~Sundaresan, V.~Paxson, S.~J.
  Murdoch, and D.~McCoy.
\newblock Do you see what {I} see? {D}ifferential treatment of anonymous users.
\newblock In {\em Proceedings of the Network and Distributed Security
  Symposium}. Internet Society, 2016.

\bibitem{kiravuo}
T.~Kiravuo, M.~Sarela, and J.~Manner.
\newblock A survey of {E}thernet {LAN} security.
\newblock {\em {IEEE} Communications Surveys \& Tutorials}, 15(3):1477--1491,
  2013.

\bibitem{klein}
G.~Klein, J.~Adronick, M.~Fernandez, I.~Kuz, T.~Murray, and G.~Heiser.
\newblock Formally verified software in the real world.
\newblock {\em Communications of the ACM}, 61(10):68--77, October 2018.

\bibitem{Quic}
A.~{Langley {\it et al.}}
\newblock The {QUIC} transport protocol: Design and {I}nternet-scale
  deployment.
\newblock In {\em Proceedings of ACM SIGCOMM}, 2017.

\bibitem{delegationLogic}
N.~Li, B.~N. Grosof, and J.~Feigenbaum.
\newblock Delegation logic: A logic-based approach to distributed
  authorization.
\newblock {\em ACM Transactions on Information and System Security},
  6(1):128--171, February 2003.

\bibitem{tor-insteadof-ip}
V.~Liu, S.~Han, A.~Krishnamurthy, and T.~Anderson.
\newblock Tor instead of {IP}.
\newblock In {\em Proceedings of the 11th Workshop on Hot Topics in Networks},
  2011.

\bibitem{beliefs}
N.~P. Lopes, N.~Bjorner, P.~Godefroid, K.~Jayaraman, and G.~Varghese.
\newblock Checking beliefs in dynamic networks.
\newblock In {\em Proceedings of the 12th USENIX Conference on Networked
  Systems Design and Implementation}, 2015.

\bibitem{pushback}
R.~Mahajan, S.~Bellovin, S.~Floyd, J.~Ioannidis, V.~Paxson, and S.~Shenker.
\newblock Controlling high bandwidth aggregates in the network.
\newblock {\em Computer Communication Review}, 32(3):62--73, July 2002.

\bibitem{greatCannon}
B.~Marczak, N.~Weaver, J.~Dalek, R.~Ensafi, D.~Fifield, S.~McKune, A.~Rey,
  J.~Scott-Railton, R.~Deibert, and V.~Paxson.
\newblock An analysis of {C}hina's {``Great Cannon''}.
\newblock In {\em {USENIX} Workshop on Free and Open Communications on the
  Internet}, August 2015.

\bibitem{NRL}
C.~Meadows.
\newblock The {NRL} protocol analyzer: An overview.
\newblock {\em Journal of Logic Programming}, 26(2):113--131, February 1996.

\bibitem{dns-ddos}
G.~C.~M. Moura, J.~Heidemann, M.~Muller, R.~de~O.~Schmidt, and M.~Davids.
\newblock When the dike breaks: Dissecting {DNS} defenses during {DDoS}.
\newblock In {\em Proceedings of the ACM Internet Measurement Conference},
  2018.

\bibitem{trafficanalysis}
M.~Nasr, A.~Houmansadr, and A.~Mazumdar.
\newblock Compressive traffic analysis: A new paradigm for scalable traffic
  analysis.
\newblock In {\em {ACM} Conference on Computer and Communications Security},
  2017.

\bibitem{costTLS}
D.~Naylor, A.~Finamore, I.~Leontiadis, Y.~Grunenberger, M.~Mellia, M.~Munafo,
  K.~Papagiannaki, and P.~Steenkiste.
\newblock The cost of the `{S}' in {HTTPS}.
\newblock In {\em Proceedings of ACM CoNEXT}, 2014.

\bibitem{mbTLS}
D.~Naylor, R.~Li, C.~Gkantsidis, , T.~Karagiannis, and P.~Steenkiste.
\newblock And then there were more: Secure communication for more than two
  parties.
\newblock In {\em Proceedings of ACM CoNEXT}, 2017.

\bibitem{mcTLS}
D.~Naylor, K.~Schomp, M.~Varvello, I.~Leontiadis, J.~Blackburn, D.~Lopez,
  K.~Papagiannaki, P.~R. Rodriguez, and P.~Steenkiste.
\newblock Multi-context {TLS} (mc{TLS}): Enabling secure in-network
  functionality in {TLS}.
\newblock In {\em Proceedings of ACM SIGCOMM}, 2015.

\bibitem{raiciu}
C.~Raiciu, C.~Paasch, S.~Barre, A.~Ford, M.~Honda, F.~Duchene, O.~Bonaventure,
  and M.~Handley.
\newblock How hard can it be? designing and implementing a deployable
  {Multipath TCP}.
\newblock In {\em Networked Systems Design and Implementation}, 2012.

\bibitem{tor1}
M.~G. Reed, P.~F. Syverson, and D.~M. Goldschlag.
\newblock Anonymous connections and onion routing.
\newblock {\em IEEE Journal on Selected Areas in Communications},
  16(4):482--494, May 1998.

\bibitem{savage}
S.~Savage, D.~Wetherall, A.~Karlin, and T.~Anderson.
\newblock Practical network support for {IP} traceback.
\newblock In {\em Proceedings of SIGCOMM}. ACM, 2000.

\bibitem{ODNS}
P.~Schmitt, A.~Edmundson, A.~Mankin, and N.~Feamster.
\newblock Oblivious {DNS}: Practical privacy for {DNS} queries.
\newblock arXiv:1806.00276v2, 2018.

\bibitem{RFC-7457}
Y.~Sheffer, R.~Holz, and P.~Saint-Andre.
\newblock Summarizing known attacks on {T}ransport {L}ayer {S}ecurity ({TLS})
  and {D}atagram {TLS (DTLS)}.
\newblock {I}nternet {E}ngineering {T}ask {F}orce {R}equest for {C}omments
  7457, 2015.

\bibitem{blindbox}
J.~Sherry, C.~Lan, R.~A. Popa, and S.~Ratnasamy.
\newblock {BlindBox}: Deep packet inspection over encrypted traffic.
\newblock In {\em Proceedings of SIGCOMM}. ACM, 2015.

\bibitem{sommer}
R.~Sommer and V.~Paxson.
\newblock Outside the closed world: On using machine learning for network
  intrusion detection.
\newblock In {\em Proceedings of the IEEE Symposium on Security and Privacy},
  2010.

\bibitem{sidechannels}
R.~Spreitzer, V.~Moonsamy, T.~Korak, and S.~Mangard.
\newblock Systematic classification of side-channel attacks: A case study for
  mobile devices.
\newblock {\em IEEE Communications Surveys \& Tutorials}, 20(1):465--488, 2018.

\bibitem{vertesi}
J.~Vertesi.
\newblock My experiment opting out of big data made me look like a criminal.
\newblock \url{https://time.com/83200/privacy-internet-big-data-opt-out}.
\newblock Accessed 15 October 2019.

\bibitem{telex11}
E.~Wustrow, S.~Wolchok, I.~Goldberg, and J.~A. Halderman.
\newblock Telex: Anticensorship in the network infrastructure.
\newblock In {\em Proceedings of the 20th {USENIX} Security Symposium}, 2011.

\bibitem{Pi}
A.~Yaar, A.~Perrig, and D.~Song.
\newblock Pi: A path identification mechanism to defend against {DDoS} attacks.
\newblock In {\em Proceedings of the Symposium on Security and Privacy}. IEEE,
  2003.

\bibitem{DOSlimiting}
X.~Yang, D.~Wetherall, and T.~Anderson.
\newblock A {DoS}-limiting network architecture.
\newblock In {\em Proceedings of SIGCOMM}. ACM, August 2005.

\bibitem{cacm}
P.~Zave and J.~Rexford.
\newblock The compositional architecture of the {I}nternet.
\newblock {\em Communications of the ACM}, 62(3):78--87, March 2019.

\bibitem{ndn14}
L.~Zhang, A.~Afanasyev, J.~Burke, and V.~Jacobson.
\newblock Named data networking.
\newblock {\em ACM SIGCOMM Computer Communication Review}, 44(3):66--73, July
  2014.

\end{thebibliography}
\fi

\end{document}